\documentclass[11pt,a4paper]{article}
\usepackage{jheppub}

\usepackage{mathrsfs}
\usepackage{tabularx}
\usepackage{booktabs}

\usepackage{braket}

\usepackage{pdfpages}

\def\be{\begin{equation}}
\def\ee{\end{equation}}
\newcommand{\nn}{\nonumber}
\newcommand{\pt}[1]{\left(#1\right)}
\newcommand{\pq}[1]{\left[#1\right]}
\newcommand{\pg}[1]{\left\{#1\right\}}

\newcommand{\num}{\addtocounter{equation}{1}\tag{\theequation}}
\newcommand{\pmat}{\begin{pmatrix}}
\newcommand{\fpmat}{\end{pmatrix}}
\newcommand{\eq}{\begin{equation}}
\newcommand{\feq}{\end{equation}}
\newcommand{\cas}{\begin{cases}}
\newcommand{\fcas}{\end{cases}}

\newcommand{\mbf}[1]{\mathbf{#1}}

\newcommand{\eqarray}{\begin{eqnarray}}
\newcommand{\feqarray}{\end{eqnarray}}

\newcommand{\diag}[1]{\operatorname{diag}\pt{#1}}

\newcommand{\lagr}{\mathcal{L}}

\newcommand{\Imm}{\,\text{Im}\,}

\def\bea{\begin{eqnarray}}
\def\eea{\end{eqnarray}}

\newcommand{\al}{\alpha}

\newcommand{\ve}{\epsilon}

\newcommand{\La}{\Lambda}
\newcommand{\m}{\mu}
\newcommand{\n}{\nu}
\newcommand{\om}{\omega}

\newcommand{\si}{\sigma}

\newcommand{\Si}{\Sigma}

\newcommand{\half}{\frac{1}{2}}

\newcommand{\quar}{\frac{1}{4}}

\begin{document}

\includepdf[pages=-, offset=75 -75]{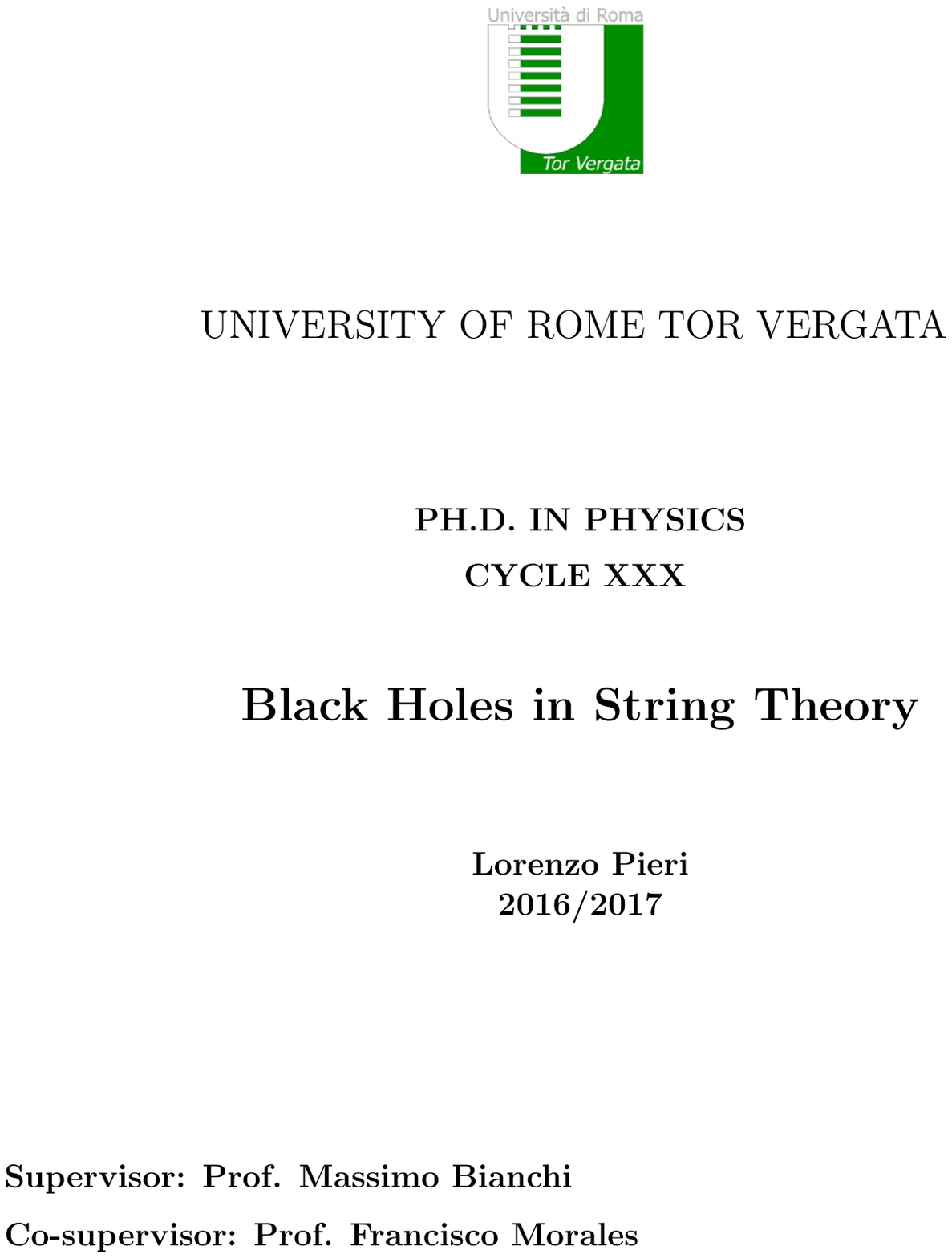}

\tableofcontents

\pagebreak

\section{Introduction}
 
Black holes are classical solutions of Einstein's equations with very peculiar characteristics:  the presence of a singularity and of an event horizon. For a long time these solutions were seen as mere mathematical curiosities, but slowly the paradigm has shifted and indeed nowadays we have gathered an enormous amount of both theoretical and experimental evidences supporting their existence. The apex has been reached recently, with the detection of the gravitational wave fingerprint of a binary black hole merger \cite{Abbott:2016blz,Abbott:2016nmj,Abbott:2017vtc}, or to be more precise with the detection of an event compatible with the general relativistic description of a black hole.

This pedantic clarification is in fact the motivation for writing this thesis and the triggering factor for thousand of papers in the last decades. The underlying idea is that General Relativity (GR) cannot be the ultimate theory of gravity, since it fails to give sensible predictions in key scenarios like black holes and the birth of the universe. To fully describe the physics of black holes we are forced to use both gravity and the quantum theory at the same time, far away from of any possible perturbative regime. We need a theory of quantum gravity, an ultimate theory able to describe whatever can be defined as physical (and maybe a bit more). 

A striking issue is that we don't have access to energy scales at which Quantum Gravity (QG) is expected to be relevant and we cannot probe the interior (if any) of black holes. Perhaps this is a unique situation in the history of mankind, in which we know that something must be there for fundamental reasons, but we are not able to probe it, so that we are forced to rely on the consistency of our theory to make progress in the quest for QG. Many researchers believe that the main fundamental inconsistency is given by the so called "Information Loss Paradox". Every serious attempt to write down a QG theory should answer to what happen to a pure state entering into a black hole. As we will see this is a devilish problem in which we are required to employ QG in a regime in which classical gravity should be enough.

One proposal for the a consistent QG theory is given by string theory, which introduce extended fundamental objects like strings and branes. Like everything else in the theory, black holes are therefore constructed from these primitive building blocks. Many calculations suggest that string theory is indeed able to describe black holes, giving a deep statistical meaning to the black hole entropy by identifying the quantum microstates of the system. Moreover string theory is able to recover the Bekenstein-Hawking formula \cite{Sen:1995in,Strominger:1996sh} in the classical (super)gravity limit. The main limitations of the theory comes from his complexity, forcing us to work in simplified and (super)symmetric unphysical settings.

A great deal of work has been carried on microstates counting, for various systems of different dimensionality, asymptotic and charges. Nevertheless, due to the intractability of strong gravity and the fact that the non perturbative formulation of the theory is, to say the least, incomplete, very little is known about the description of the microstates in the black hole regime. Many results have been  in fact obtained by clever exploitation of weak-strong coupling dualities or the use of protected quantities like supersymmetric indices. 

We will take the working hypothesis that string theory is the correct QG theory and we will try to answer to the following question: what is the description of the black hole microstates in supergravity? Indeed supergravity (SUGRA) is the low energy limit of string theory and we know that at some distance from the horizon it must  be the right description. Moreover even though the description of the generic black hole will likely require the full theory of quantum gravity and all the non perturbative technology, possibly already  at the horizon, it's still possible that some very special states, some sort of coherent states, can be described fully in supergravity. One possibility is given by the fuzzball proposal \cite{Lunin:2001jy,Lunin:2002iz,Mathur:2003hj,Lunin:2004uu,Giusto:2004id,Giusto:2004ip,Mathur:2005zp, Skenderis:2008qn, Mathur:2008nj}, in which every quantum microstates is associated to a smooth and horizonless solution of SUGRA, getting rid of the pathological features of the GR black hole and therefore solving the information paradox.

In this thesis we will jump back and forth the string theory and SUGRA description of a black hole\footnote{Notice that here and in many other places, by black hole we really mean "the SUGRA-string solution associated to the corresponding GR black hole". This solution can be in fact be horizonless and without singularities, so it is not a black hole in any sense! Of course from far away it, in the regime in which we cannot probe the horizon or the interior, it must resemble arbitrary well a GR black hole, in order to be experimentally significant.} and in particular we will focus on four BPS dimensional black holes. The fundamental stringy picture of the microstates is given by open strings stretched between intersecting branes. By studying the back reaction of this system, one can retrieve information on the closed string signature arising from a given open string configuration. Concretely one can compute the stringy open-closed scattering amplitude  of a Neveu Schwarz-Neveu Schwarz (NSNS) or Ramond-Ramond (RR) field with the open strings binding the branes. Remarkably a large class of SUGRA solutions can be linked in this way to the underlying string description, establishing a  dictionary since, apart subtleties, a particular choice of the string polarizations gives rise to a particular gravity solution \cite{Bianchi:2016bgx}. The SUGRA solutions associated to the four dimensional black holes will be constructed in the U-dual frame with only D-branes: 4 stacks of D3 branes or 1 stack of D6 and 3 stacks of D2.

Despite this relationship between the two picture will explicit show which fields in the SUGRA solutions are turned on (among metric, B field, dilaton and C forms),  much freedom remain at the two extremes after the scattering amplitude computation. First of all,  the supergravity solutions are actually a family of solutions valid for some generic harmonic functions, like usual when working with extremal solutions. This implies that many (the majority) of these solutions will be singular as the classical GR solutions. If the fuzzball conjecture is correct there must exist some choices of the harmonics resulting in smooth and horizonless solutions. We will show that such solutions can indeed be found therefore linking stringy microstates with regular gravity solutions \cite{Bianchi:2017bxl}. We will mainly use the formalism of 5-dimensional SUGRA, since the solutions will typically be regular only when the additional compact dimensions of string theory are taken into account.

The second freedom in the stringy-SUGRA dictionary comes from the stringy side. Indeed the information contained in  the open string binding the branes can be summed up into a vacuum expection value. The  value of this vev doesn't explicit arise from the string amplitude computations and remain an arbitrary constant; this is due to simplified assumptions in the calculations done so far in literature, but in general the vevs are not arbitrary and  can be determined once we determine all the normalizations of the string fields and the dependence on closed string moduli. To fix these vevs one can study the Supersymmetric Quantum Mechanics (SQM) arising on the worldvolume of the branes. It is a quantum mechanics, since the 4 stack of branes forming the black hole intersect in a point, leaving only the time direction as common direction. In the following we will study the SQM associated to the purely D-brane system analyzed previously, in the regime of small brane charges, in order to count the number of supersymmetric vacua and compute the vevs appearing in the string computations.

The plan of the thesis is as follows: in section 2 we  review classical black holes in GR and remind how the addition of quantum physics result in the infamous information loss paradox. We  briefly outline many proposed solutions, both in string theory and in other QG approaches.  In section 3 we  focus on black holes in string theory, including how they are built and how microstates counting is performed; then we describe the fuzzball proposal starting from the pioneering two charge system and concluding with the  3 and 4-charges systems, the ones relevant to  describe a black hole with non zero area in the SUGRA limit.  Starting with chapter 4 we present material published during the course of this thesis. In chapter 4 we build the dictionary between open strings and SUGRA solutions. We first review the computation for 2 charge systems and then focus on the 4 charge system, in a frame with all D3 branes. In chapter 5 we work exclusively in SUGRA, to find some examples of smooth solutions living in the same SUGRA family studied in chapter 4. In chapter 6 we analyze the SQM on the worldvolume of the branes. We close the thesis with remarks on the future of black hole physics and quantum gravity.

\pagebreak

\section{Classical Black Holes and the Information Paradox}

\subsection{Black Holes in General Relativity}

Consider a strongly asymptotically predictable spacetime $M$. A black hole $\emph{B}$ is defined as:

\bea
\emph{B} = [ M -J^-(\mathscr{I}^+)]
\eea

In words, in a spacetime $M$ with an open region conformally related to a globally hyperbolic spacetime, a black hole region is defined as the closure of the complementary of the spacetime $M$ with respect to the causal past of future null infinity. The black hole horizon is the boundary of this region.

In simple words: a black hole is a region of space in which once entered, nothing can escape to infinity anymore. Nothing includes light, that's why they are black. A nice introduction to the subject is given in \cite{Wald:1984rg,Carroll:2004st,Poisson, Hartle, Frolov, Townsend, Schutz,Misner:1974qy} and we will relegate precise definitions to the appendix \eqref{grDef}.

The archetypical example is given by the Schwarzschild black hole; in spherical coordinates $(t,r,\theta,\phi)$ the metric is:

\bea
ds^2=-\left(1-\frac{2M}{r}\right)dt^2+\left(1-\frac{2M}{r}\right)^{-1}dr^2+r^2(d\theta^2+\mathrm{sin}^2\theta d\phi^2)
\eea

where in these units the gravitational constant $G$ is set to one.

This spacetime wonderfully illustrate the key points which are associated will all the solutions dubbed as black holes: a singularity and an event horizon concealing it. By inspecting the metric, two point seems to be troublesome: $r = 0$ and $r=2m$. The latter is the location of the horizon and in fact it's an harmless coordinate singularity, that is we are simply choosing a bad set of coordinates for an observer willing to probe that region of space. A better frame is given by the Eddington-Finkelstein coordinates $(v,r,\theta,\phi)$:

\bea 
v &=& t + r + 2M \ln \left(\frac{r}{2M}-1\right) \nn \\
ds^2 &=& -\left(1-\frac{2M}{r}\right)dv^2+2 dr dv+r^2(d\theta^2+\mathrm{sin}^2\theta d\phi^2)
\eea

here we can see that as $r \rightarrow 2M $ the metric remains regular and invertible, since $g = - r^4 sin^2\theta$. Another confirmation comes from the curvature invariants, that are well behaved at this point.

\begin{figure}
\begin{center}
\includegraphics[width=8cm]{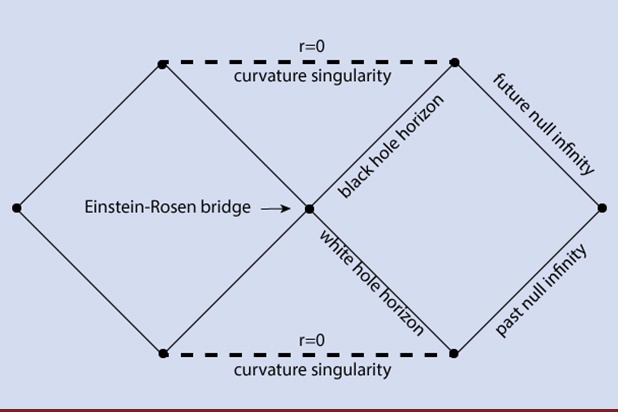} 
\end{center}
\caption{Penrose diagram of the maximally extended Schwarzschild spacetime, in which our universe is the square on the right. Since light moves at $45°$ degrees, once entered into a black hole horizon everything is forced to fall into the singularity. }
\end{figure}

Nevertheless something odd is happening, as witnessed by computing the light cones for an observer  radially falling into the black hole. By setting $ds^2 = 0 $ and fixing the angles we can find out the boundaries of the cones: 

\begin{equation}
    \frac{dv}{dr}=
    \begin{cases}
      0 & (infalling) \\
      2 \left(1-\frac{2M}{r}\right)^{-1} & (outgoing)
    \end{cases}
  \end{equation}
  
  For $r < 2M $ the light cones tilt inside the black holes, meaning that every future trajectory will never cross the boundary $r = 2M$ again. In summary the horizon is a global concept, related to the causal structure of the spacetime, and locally nothing strange happen to the infalling observer, at least according to this classical picture.

\begin{figure}
\begin{center}
\includegraphics[width=8cm]{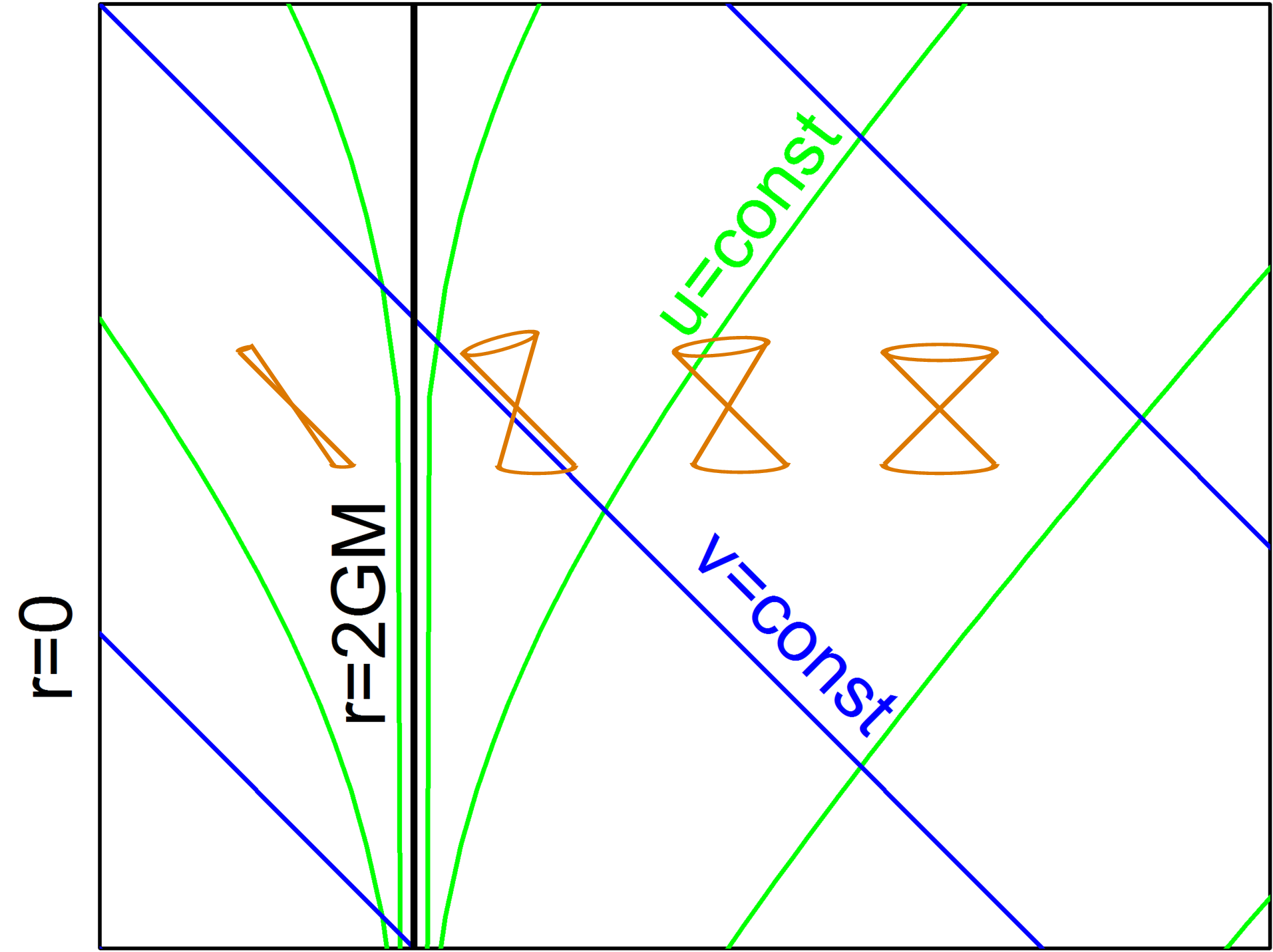} 
\end{center}
\caption{Tilting of the light cones as the event horizon is approaching.}
\end{figure}

  The point $r=0$ is instead a genuine curvature singularity, since the Kretschman invariant $K$ blows up there:
  
\bea
 K= R^{\mu \nu \rho \sigma}R_{\mu \nu \rho \sigma} = \frac{ 48 M^2}{r^6}
\eea

An observer probing the singularity will experience extremely high tidal forces. This can seem reasonable, since astrophysically we  consider black holes as the last stage of evolution of certain collapsing stars or the remnant of supernova explosions, therefore we can naively think at super dense objects (notice that the  Schwarzschild black is a vacuum solution). The issue is that the black hole is actually infinitely dense at the singularity and the tidal forces are divergent. This is not a physical insight, but a signal that our theory is breaking down. It's analog to consider a model of an massive charged pointlike particle in classical physics:  gravitational and electomagnetic potential energy are infinite.

Singularities comes in many different flavors \cite{Ellis:1977pj}, indeed it's not even required for the curvature invariants or the Riemann tensor to be divergent. A pretty satisfactory definition is the following: a spacetime is singular, it has a singularity, if it possesses at least one incomplete geodesic. An incomplete geodesics has finite affine length (range of the affine parameter) in at least one direction. Intuitively the geodesics it's not free to be extended to infinity or at least to circle on itself (but that would give closed time like curves (CTCs)). One may hope that the theory will make the formation of singularities impossible, while in fact they seem to be a very generic feature of every gravitational collapse. These considerations are formalized in the Singularity Theorems, which  states that under reasonable conditions like absence of CTCs, validity of energy conditions and some other technical assumption, the presence of a trapped surface imply a singularity in the spacetime. The boundary of a region filled with trapped surfaces is called apparent horizon and a trapped surface is a surface inside which the expansion $\theta$ of a congruence of null geodesics is negative both for outgoing and ingoing rays. It formalize the concept of "tilting of the light cones".  For instance, consider a sphere of fixed radius emanating shells of photons in Minkowski, both inward and outward. Here some shells will go to infinity and other will collapse at the center; instead the same scenario inside $r < 2M$ in Schwarzschild would result in both shells going to decreasing values of the radius. For Schwarzschild apparent and event horizon are the same hypersurface, but in general they can be different.

To cover these monstrosity it has been conjectured that the gravitational collapse of a body always result in a singularity covered by an event horizon, the so called Cosmic Censorship Conjecture. A naked singularity would be an intolerable problem for the predictivity of the theory, the space would be not globally hyperbolic,  even though no conclusive arguments has ever been made on their absence. All in all, considering singularity theorems and cosmic censorship, black holes are very generic predictions of general relativity.

To reiterate how much paradoxical  black holes are, please notice that only after an infinite amount of time we can really say to be outside the black hole. The mechanism for the horizon growth is strongly non causal and non local, indeed the event horizon will expand in response to all the infalling matter that will ever cross the horizon, including the one in the future of a given observer at a certain time. To concretely work with black holes  in numerical relativity one usually employs milder concepts than the black hole horizon, see \cite{Booth:2005qc} for more details.

\subsection{No-Hair Theorem}

The gravitational field generated by a planet  is influenced by every detail of the emitting objects, for instance by the distribution of every grain of sand on the surface. The multipole expansion of this field is characterized in general by an huge number of indipendent coefficients\footnote{A counterexample is given by a perfectly round planet, which in the exterior is completely described by the Schwarzschild solution due to the Birkhoff Theorem. Planets without a grain of sand displaced from exactly spherical symmetry are not so common though.}. 

Remarkably this property is not shared by stationary black holes, which appears to be so simple that an handful of parameters are enough to describe all of them. This property is formalized in the so called No Hair Theorem that, apart from subtleties,  roughly states the following:

\begin{quotation}\small
Stationary, asymptotically flat black hole solutions to GR coupled to electromagnetism that are non singular outside the horizon are fully characterized by mass, charge and angular momentum. 
\end{quotation}

Here we assume that electromagnetism is the only long range non gravitational field; the charge  can in principle be both electric and magnetic.

This theoreom is very powerful since ultimately every black hole will settle to a stationary configuration since every departure from stationarity will rapidly be emitted as gravitational waves. All the stationary black holes of GR are inside the Kerr-Newmann family of solutions, that in the Boyer-Lindquist $\lbrace t,  r,\theta,\phi \rbrace$ frame can be written as follow:
 
   \bea
 d s^{2}&=& -\left(1-\frac{2Mr}{ R^2}+\frac{(Q^2+P^2)}{R^2}\right) d t^{2}-   \frac{2a  \sin^{2}\theta}{ R^2}   \left[ 2\,M\,  r - (Q^2+P^2)\right]d t\, d\varphi  \nn  \\
&&+\frac{R^2}{\Delta}d  r^{2}+R^2 d\theta^{2}+\frac{sin^2\theta}{ R^2}\left[ (  r^2+a^2)^2-a^2 \Delta sin^2\theta\right] d\phi^{2} \nn\\
A_{\mu} dx^{\mu} &=& -{4 \over  R^2}  ( Q\,  r+P\, a\, \cos\theta)\,  dt+\frac{4}{R^4} ( Q\, a \, r\, \sin^2\theta+P\, ( r^2+a^2) \cos\theta )\, d\phi
\eea
where
    \bea
R^2( r,\theta)= r^2+a^2 cos^2(\theta) \qquad \Delta( r)= r^2-2M\,  r+a^2+(Q^2+P^2) \nn \\
t \in (-\infty,+\infty) \quad\quad  r \in [0,+\infty) \quad\quad \theta \in [0, \pi] \quad\quad \phi \in [0,2 \pi]
\eea
 The parameter $M$ represents the mass, $Q$ and $P$ the electric and magnetic charges and $J=Ma$ the angular momentum. 
   If
  \bea
  M^2\geq Q^2+P^2+a^2 
  \eea
 the solution represents a black hole with horizons located at
 \bea
  r_{\pm}=M \pm \sqrt{M^2-a^2-Q^2-P^2}
 \eea
   and a curvature singularity at $ r=0$. 
  The area of the black hole horizon is $A=4\pi ( r_+^2+a^2)$ and the asymptotic in the large $r$ expansion is given by: 
\bea
 d s^{2}=-\left(1-\frac{2M}{ r}+o(r^{-2})\right) d t^{2}-\left(4\frac{J\sin^{2}\theta}{ r}+o(r^{-2})\right)d t\, d\varphi  \nn  \\
+\left(1+\frac{2M}{r}+o(r^{-2})\right) d r^2 + r^2 \left(d \theta^2+\sin^2\theta d\varphi^{2}\right)
\eea

Important subcases are Reissner-Nordstom ($J=0$) and Kerr ($Q=P=0$). 
\subsection{Black Holes Thermodynamics}

Black holes are black, at least classically. Consider a Schwarzschild black hole: since nothing is emitted and it depends only on the mass, every object falling inside will have the effect of increasing the area. In fact this result is very general and it turns out that under reasonable assumptions every possible process results in a net increase of the event horizon. The appearance of a never decreasing quantity is reminiscent of the second law of thermodynamics, and indeed a complete analogy with all the four laws of thermodynamics was uncovered:

\begin{itemize}

\item \textbf{0 Law:} The surface gravity $\kappa$ of a stationary black hole is uniform over the whole horizon.

\item \textbf{I Law:} The following is true:
\bea 
dM= \frac{1}{8 \pi} \kappa dA + \Omega_H dJ
\eea

where $\Omega_H$ is the angular velocity of the hole, $A$ is the area  and we have not included contributions due to the charge. It's apparent the identification $ E \leftrightarrow M$, $ T \leftrightarrow C \kappa$, $ S \leftrightarrow \frac{1}{8\pi C}A$ for some constant $C$.

\item \textbf{II Law:} The surface area of a black hole can never decrease:

\bea 
dA > 0
\eea

This result comes mainly from two ingredients: the observation that the horizon is a null hypersurface generated by null geodesics with no future endpoint and the focusing theorem. This  theorem make use of the Raychaudhuri equation for a congruence of null geodesics together with the null energy condition, therefore exotic fields can in principle violate this law.

\item \textbf{III Law:} If the stress energy tensor is bounded and it satisfies the weak energy condition, then the surface gravity cannot be reduced to zero by any process in a finite amount of (advanced, $v = t+r$) time. Basically, extremal black holes cannot be reached starting from  non-extremal black holes.

\end{itemize}

\textcolor{white}{i}

\begin{table}[]
\begin{center}
\begin{tabularx}{\textwidth}{lXXX}
 Law      & \textbf{Thermodynamics}       & \textbf{Black Holes } \\
 \toprule
\textbf{Zeroth} & $ T$  constant trough body in thermal equilibrium   &  $\kappa$  constant over (Killing) horizon of stationary BHs  \\
\midrule
\textbf{First} & $ dE= T dS - p dV$ & $ dM= \frac{1}{8 \pi} \kappa dA + \Omega_H dJ$\\
\midrule
\textbf{Second} & $ \delta S \geq 0$  in any process  &  $\delta A \geq 0$  in any process \\
\midrule
\textbf{Third} & Impossible to achieve $ T=0$ by a physical process   &  Impossible to achieve $\kappa=0$ (extremality)  by a physical process 
\end{tabularx}
\caption{Analogy between old fashioned thermodynamics and black hole thermodynamics.}
\end{center}
\end{table}

\textcolor{white}{i }

Restricting to classical physics, this analogy is merely formal, since a black hole doesn't emit anything while a temperature is associated to the dynamic equilibrium between a system an its surrounding. Considering quantum physics will fill the gap.

\subsection{Quantum Field Theory in Curved Space}

Consider a scalar field coupled to gravity in a two dimensional Schwarzschild spacetime (without angular coordinates), with action:

\bea
S[\phi]=\frac{1}{2} \int d^2 x \, \sqrt{-g}  \, g^{\mu \nu} \phi_{, \mu}\phi_{, \nu} 
\eea

This toy model will be useful to illustrate the main features of Quantum Field Theory (QFT) in curved space applied to black holes and the case for GR in $3+1$ dimensions will be a generalization of this one; see \cite{Mukhanov:2007zz,Birrell:1982ix} for details.

Since the action is conformally invariant we can write the solution to the scalar equation of motion in terms of different coordinates systems. First of all, we choose a frame that is singular on the horizon, for instance the lightcone coordinates $(\tilde{u},\tilde{v})$ obtained from $(t,r)$ trough the tortoise coordinate: 

\bea
\tilde{u}=t-r^* \qquad \tilde{v}=t+r^* \qquad dr=\left( 1-\frac{2M}{r} \right) dr^*
\eea

 Secondly, we pick a frame that is regular everywhere apart from the central singularity, such as lightcone Kruskal-Szekeres $(u,v)$ defined by 

\bea
u=-4 M e^{-\frac{\tilde{u}}{4M}} \qquad v = 4M e^{\frac{\tilde{v}}{4M}}
\eea

An observer at rest located far away will associate particles with positive frequency modes $\Omega$ with respect to the time coordinate t. The expansion of the scalar fields will be (ignoring the left moving part for simplicity):

\bea
\hat{\phi}= \int_{0}^{\infty} \frac{d\Omega}{(2\pi^{1/2})} \frac{1}{\sqrt{2\Omega}} \left[e^{-i\Omega \tilde{u}} \hat{b}_{\Omega}^-+e^{i\Omega \tilde{u}} \hat{b}_{\Omega}^+\right]
\eea

where the annihilation operator defines the Boulware vacuum:

\bea
\hat{b}_{\Omega}^-  \Ket {0_B} =0
\eea

Therefore this vacuum contains no particles from the point of view of the far away observer. Nevertheless this vacuum it's not physical since it's singular on the horizon and this would require an infinite amount of energy to actually prepare such a state. For the Kruskal coordinates:

\bea
\hat{\phi}= \int_{0}^{\infty} \frac{d\omega}{(2\pi^{1/2})} \frac{1}{\sqrt{2\omega}} \left[e^{-i\omega u} \hat{a}_{\omega}^-+e^{i\omega u} \hat{a}_{\omega}^+ \right]
\eea

\bea
 \hat{a}_{\omega}^-  \Ket {0_K} =0 \eea

The Kruskal vacuum (also known as Hartle-Hawking vacuum) is well behaved on the horizon and suitable to be a physical state. Notice that it contains particle from the point of view of the far away observer, indeed on can relate the two set of creation-annihilation operators:

\bea
\hat{b}_{\Omega}^-= \int_{0}^{\infty} d\omega [\alpha_{\Omega \omega} \hat{a}_{\omega}^- -\beta_{\Omega \omega}  \hat{a}_{\omega}^+] \,\,
\eea

via the Bogolyubov coefficients, resulting in $\hat{b}_{\Omega}^-  \Ket {0_K}  \neq 0$. In summary, a far away observer will see a spectrum of scalar  particles coming from the black hole:
\bea
\langle \hat{N}_{\Omega} \rangle \equiv \Bra {0_K}\hat{b}_{\Omega}^+ \hat{b}_{\Omega}^- \Ket {0_K}= \int d \omega |\beta_{\Omega \omega}|^2
\eea

\bea
n_{\Omega}=\frac{\langle \hat{N}_{\Omega} \rangle }{V}= \frac{1}{e^{\frac{2\pi \Omega}{\kappa}}-1} \qquad T_{BH}=\frac{\kappa}{2 \pi} = \frac{1}{8 \pi M}
\eea

for a finite volume $V$. The black hole it's not black anymore and ironically is now emitting with a black body spectrum! The radiation is dubbed Hawking-Zel'dovich radiation or more commonly Hawking radiation. Since the radiation stress energy tensor doesn't satisfy the null energy condition the second law is violated and the black hole loses mass and it shrinks. Pictorially we can imagine pairs of positive/negative particles emitted near the event horizon: the positive energy particle escapes to infinity as radiation, while the negative energy one is captured and the black hole loses mass. We can appreciate the striking parallel of the whole analysis with the phenomenon of Unruh radiation, in which an accelerated observer (analog of the far away observer) in flat space measures a thermal spectrum of particles.

\begin{table}[]
\begin{center}
\fbox{\begin{minipage}{30em}
Accelerated  Observer  \qquad \qquad  \, \, \, Schwarzschild spacetime\\
Minkowski Vacuum  $\Ket {0_M}$ \qquad \qquad    Kruskal Vacuum $\Ket {0_K}$\\
Rindler Vacuum  $\Ket {0_R}$  \qquad \qquad  \, \,\, \, Boulware Vacuum $\Ket {0_B}$\\
Acceleration $a$   \qquad  \qquad   \qquad \qquad \,\, Surface gravity $\kappa=(4M)^{-1}$ 
\end{minipage}}
\end{center}
\caption{Parallel between Unruh radiation on the left and Hawking radiation on the right.}
\end{table}

\textcolor{white}{i }

This calculation allowed to completely determine the Bekenstein-Hawking relation:

\bea
 S_{BH} = \frac{1}{4 \hbar}A
 \label{SA}
\eea

This entropy can be put on the same conceptual level of the usual thermodynamical entropy $S$ and the second law in presence of black holes can be generalized to:

\bea
dS' > 0 \qquad S' = S_{BH} +  S
\eea

Notice that without quantum physics, even this generalized law would be violated, for instance by considering a black hole in a thermal bath with $T < T_{BH}$, thereby transferring heat from a cold body to a hotter body.

Still, some conceptual issues remain. We know that thermodynamic is a macroscopic coarse-grained description of an underlying microscopic theory. In particular the entropy can be associated to the information contained in the macroscopic state or equivalently\footnote{The entropy of a system, for a given observer, can be defined as a measure of the ignorance of the exact microstates of the system. In thermodynamic, the ignorance of the precise position and velocity of every particle constituting a gas give rise to the associated entropy. Every single microstate has zero entropy, since we have complete information about it. } with the number of microstates $N_{micro}$ as:

\bea
S  = k_B \ln(N_{micro})
\eea

Unfortunately such microstates are not there in GR. In fact due to the No Hair theorem, for a single black hole with given asymptotic charges, no matter how many encyclopedia have fallen inside,  we have:

\bea
S  = k_B \ln(1) = 0
\eea

Another issue will be the subject of  the next section. 

\subsection{Information Paradox}

 Consider an information carrying pure state falling into a macroscopic black hole. We will make the following assumptions:
 
 \begin{itemize}
 \item A) General Relativity is the correct theory of gravitation at low energy. In particular:
 
 A1) the equivalence principle applies at the horizon of a macroscopic black hole and an infalling observer doesn't experience anything special.
  
 A2) the No-Hair theorem is correct. The region around the horizon is information-free.
 
 \item B) Quantum Field Theory is the valid theory at low energy. Therefore\footnote{Loss of unitarity and loss of information are often treated  as equivalent, while in principle they can happen independently, see \cite{Mathur:2009hf} for an example. In the information paradox they both show up.}:
 
 B1) the theory is unitary (a pure state cannot evolve into a mixed state). 
 
 B2) there is no loss of information.
 
 B3) the theory is local.
 
 \item C) A semiclassical treatment is appropriate to describe the physics of an evaporating black hole until the black hole is macroscopic. Moreover this implies that:
 
 C1) There is no need to invoke quantum gravity before reaching some high energy scale, usually taken to be the Planck scale.
 
 \end{itemize}
 
At least one of the above is incorrect: this is the  information loss paradox. It's a paradox since all of the assumptions are believed to be true in our universe, if taken  singularly. There can be some redundancy in these conditions, but we will stick to these to be as explicit as possible.
 
 Let's follow an encyclopedia, or any other information carrying object that you would like to throw in, falling into the horizon to have a grasp of the situation. In the Hilbert space of our quantum theory it's possible to describe the combined state of the encyclopedia plus the black hole as a pure state. As stated in the hypothesis, nothing special happens at the horizon crossing and the encyclopedia sink into the central singularity. The black hole emits a spectrum of Hawking quanta that is thermal, therefore mixed. This is not a problem, since the emitted quanta are entangled with quanta falling inside the black holes and the overall system is still pure, but an issue arises when the black hole eventually evaporates completely, leaving only radiation entangled with nothing and in a completely mixed state. The net process is a pure state evolving into a mixed state, violating the basic principles of the quantum theory.
 
 A key role has been played by the No-Hair theorem: when the encyclopedia falls in, an external observer is able to measure only the mass, the charge and the angular momentum and it's impossible for him to differentiate between a book written in english and one written in  extra-terrestrial language (given the same mass!). Moreover the emitted pairs of quanta don't talk to each other, since their are generated from the vacuum, i.e. from an information free region, independently from previously emitted pairs \cite{Mathur:2008nj}. This explain why burning an encyclopedia don't give rise to a paradox: even though the information is de facto lost, it can in principle be recovered by the subtle correlation between the emitted quanta, since every emission starts in an information-full region, namely the surface of the ardent pages, and not from the vacuum.
 
\begin{figure}
\begin{center}
\includegraphics[width=5cm]{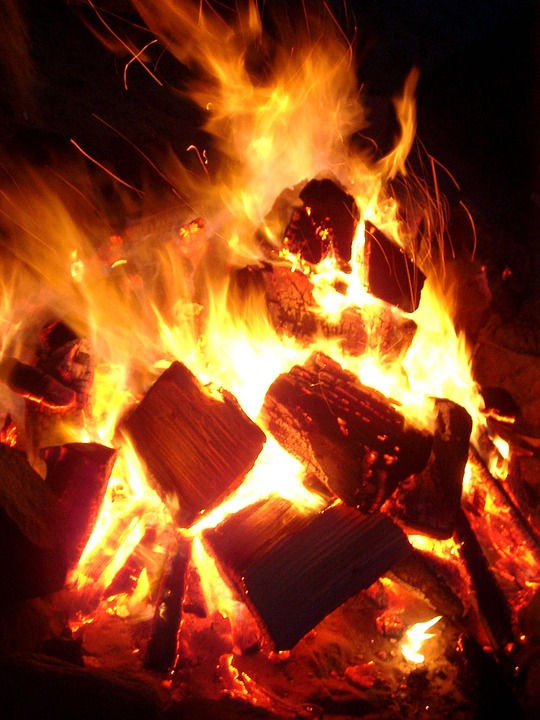} 
\end{center}
\caption{ Lighting a fire doesn't destroy information.}
\end{figure}
 
 The information paradox involves many ingredients, so it's not difficult to come up with plausible loopholes; here we follow  \cite{Preskill:1992tc,Polchinski:2016hrw,Chakraborty:2017pmn,Marolf:2017jkr} to argue that in fact there are no obvious loopholes and it's likely that we will need to give up one of more assumptions of the paradox. Here we list some proposals not violating A,B or C and discuss the implications.  
 
 \begin{itemize}
 \item \textbf{The information is in the radiation:} A possible solution would be that the radiation is not really thermal or equivalently the information is encoded in correlations between the quanta. 
 The original  Hawking's  calculation  showed that the radiation is thermal and therefore carry no information, but one can wonder if there are subtle quantum effects spoiling the calculation and accumulating over time, restoring a unitary evaporation. 
  Intuitively the answer seems negative, since the No-Hair theorem implies some sort of factorizability of the Hilbert space black hole-radiation. This belief can be made rigorous using the No hiding Theorem \cite{Braunstein:2006sj}: assuming a bipartite system black hole-radiation, unitarity and Hawking's semiclassical analysis of the radiation, no information can be carried by the radiation or in correlations between the radiation and the black hole. A way out is to consider a tripartite system, composed by the black hole and the early and late radiation. The tripartite possibility has been analyzed extensively recently, particularly  in the context of Black Hole Complementarity \cite{Susskind:1993if} and the Firewall argument \cite{Almheiri:2012rt}, but apparently a modification of physics violating A,B or C it's nonetheless required.

 In  \cite{Mathur:2009hf}  it was shown that given some "niceness conditions", believed to be true in the evaporation process for a big black hole,  and the absence of remnants, for the radiation to be pure it's necessary to alter the evolution of low energy modes at the horizon by order one, ruling out any "small correction" way out.
 
 \item \textbf{Remnant and baby universes:} if the evaporation process ends up with a remnant carrying all the information there is no paradox since the radiation remain entangled with the remnant; of course the formation of the remnant will require the full theory of quantum gravity to be properly described \cite{Aharonov:1987tp}. Here we restrict to microscopic remnants, since macroscopic ones would violate hypothesis C anyway.
 
 Unfortunately a remnant could bring more problems than it solves:  since the initial black hole can have arbitrary high mass, the remnant must be able to carry an arbitrary amount of information. Therefore there must be an huge number of stable remnant with mass comparable to the Planck scale and indistinguishable from an effective field theory viewpoint, so these remnant should appear inside quantum loops and evaporation process of other black holes. Even though the production of a single remnant will be highly suppressed, the huge phase space implies that the probability of generating a remnant is not negligible in a conventional process \cite{Preskill:1992tc,Giddings:1993km}. As a minor point, the entropy of the remnant cannot be accounted by the Bekenstein-Hawking formula. 
Despite these difficulties, the  possibility of stable remnants remain open, see \cite{Chen:2014jwq} for an extensive review. 
 
 The idea of baby universes can be traced back to the possibility of gluing external black hole geometries with internal non trivial geometries, see \cite{Corvino:2003sp} for an application to Kerr. The fundamental problem with this approach is that we don't know how model the evaporation from the point of view of an external observer. One can speculate that being created from a quantum fluctuation, a closed baby universe will carry exactly zero energy away and by the uncertainty principle it will be completely delocalized in spacetime. This means that a baby universe is more similar to a global property of our spacetime, and it's not clear how to measure in which superselection sector we live \cite{Preskill:1992tc,Giddings:1988cx,Coleman:1988cy}. There is little doubt  that a baby universe it's far from being a natural and economic solution to the paradox.

 \item \textbf{All the information comes out at the end:} A fundamental result on evaporating black holes is the Page's theorem \cite{Page:1993df}, roughly stating that for a unitary process the information starts leaking out at the Page time (when the black hole has already radiated half of its entropy), which can be very large for macroscopic black holes. This theorem make use of a bipartite system, so we can still hope to find a model in which the information comes out entirely due to quantum gravity effects at end. This possibility partially overlap with the previous one, since the endpoint of the evaporation of a black hole with mass $M$ would be some sort of metastable remnant, with an estimated life time of the order $\sim M^4$, compared to an evaporation time $\sim M^3$ \cite{Carlitz:1986ng,Giddings:1995gd}. Intuitively, the remnant takes a long time to evaporate, due the small energy involved ($ M_{remn} \sim M_{pl}$) and the time-energy uncertainty principle.   These are de facto   long-lived remnants, incurring in the issues outlined previously. Again, this possibility cannot be completely ruled out, but there are no convincing proposal at the moment.
 \end{itemize}

\subsection{Solving the Paradox}

 In this section we outline some of the possible resolutions to the information paradox, in order to appreciate different ways of relaxing at least one of the initial assumptions. We will limit to some  selected proposals,  whereas an overview of all the  resolutions to the paradox will probably require more than one human lifespan to be written; at least part of the material omitted can be found in \cite{Chakraborty:2017pmn}.  It's worth mentioning that some authors still support the simplest possible solution, namely that the original calculation is correct and the information is indeed lost \cite{Unruh:2017uaw}. At the cost of stating the obvious, no proposed solution so far is universally accepted or even favored by the community.

\subsubsection{Black Hole Complementarity and Firewalls}

Consider a slicing in Cauchy surfaces of an evaporating black hole. In particular we pick a surface $\Sigma$ before the black hole forms, a surface $\Sigma_S =\Sigma_{out} \times \Sigma_{bh} $ intersecting outside and inside and written as union of two disjoint surfaces, and finally $\Sigma'$  after the evaporation. Given a pure quantum state $  |\psi (\Sigma)\rangle $ on $\Sigma$ and if the whole process is unitary and no information is lost, then:

 \bea
   |\psi (\Sigma')\rangle  =U | \psi (\Sigma)\rangle 
   \label{noLoss}
 \eea

is another pure state, where $U$ is a unitary operator enforcing the evolution governed by the Schr\"{o}dinger equation.  

Since $\Sigma_{out}$ is causally disconnected from $\Sigma_{bh}$, and considering that $\Sigma'$ is causally connected to $\Sigma_{out}$ only, we can write respectively that:

 \bea
   |\psi (\Sigma_{S})\rangle  = | \psi (\Sigma_{out})\rangle  \otimes  | \psi (\Sigma_{bh})\rangle 
 \eea
 
\bea
   |\psi (\Sigma')\rangle  =U | \psi (\Sigma_{out})\rangle 
 \eea
 
One can show that  linearity and unitarity of the quantum theory lead to puzzling conclusion. Indeed if both $ |\psi (\Sigma')\rangle$ and $ |\psi (\Sigma_{out})\rangle$  depend linearly on  $|\psi (\Sigma)\rangle$, we must conclude that $| \psi (\Sigma_{bh})\rangle $ cannot depend on  $|\psi (\Sigma)\rangle$ , otherwise  $ |\psi (\Sigma_{S})\rangle$ will not depend linearly on the initial slice. In summary, if we impose no information loss, as in \eqref{noLoss}, we find that the black hole carry no information about what have fallen inside. 

The black hole complementarity proposal is an attempt to restore the interior of the black hole, while retaining a unitary evaporation \cite{Susskind:1993if}. Substantially, we relax the assumptions B, by modifying the quantum theory. The core observation is that no observer can probe  $| \psi (\Sigma_{bh})\rangle $ and $| \psi (\Sigma_{out})\rangle $ at the same time, therefore one can imagine a complementary picture in which the infalling observer experiences nothing special entering the horizon, while an external observer views a "stretched horizon" encoding all the information fallen inside. The two pictures are incompatible, but since there is no way for two different observer to communicate their findings the whole process is consistent.

Actually, a careful analysis reveals that the postulates of complementarity  black holes  old enough cannot coexist with the equivalence principle; a proposed way out is to introduce an hard wall of quanta, a firewall, at the horizon  
\cite{Braunstein:2009my,Almheiri:2012rt}. Divide the whole system in three parts: the black hole, the early radiation $R$ and the late radiation $R'$. If we suppose that the evaporation is unitary, then the entanglement entropy\footnote{Given a bipartite system A and B, the entanglement entropy of the region A is given by $S_A^{ent} = -Tr(\rho_A \log \rho_A)$, with $\rho_A^{ent} = Tr_B(\rho)$, where $\rho$ is the density matrix of the whole system. See \cite{Harlow:2014yka} and reference therein for other definitions and results.} of the total radiation will be $S_{R R'}= 0$. Consider an infalling observer Alice and a static outer observer Bob. Bob will be able to see both early and late radiation, with the latter coming from two region: a region SH, the stretched horizon (roughly a surface at planck distance from the horizon), and a region near the horizon that we call NH populated by modes that we call B. A mirror region inside the horizon is populated by modes that we call A. If B is sufficiently weakly interacting with the rest of the black hole, it can be purified by some early radiation $R_B$, i.e.

 \bea 
 S_{BR_B} = 0
 \eea 
 
 This relation can be verified by both Bob and Alice, provided that Alice wait long enough to be able to measure the early radiation. Alice can do more, since at the horizon she will find nothing special due to the equivalence principle, and measure that
  \bea 
S_{AB}=0
 \eea 
 This it's true, since the Rindler wedges are maximally entangles if the global state is Minkowski vacuum, the two wedges being A and B. The double relation of B  violates a basic theorem regarding the monogamy of maximally entanglement, namely the strong sub-additivity of the entropy. As a way out, a firewall would  prevent  Alice from jumping inside, therefore getting rid of the conflict with the strong sub-additivity.

\subsubsection{Non Locality and Wormholes}

Violating locality has always been met with skepticism, considering that everything that we see around us is well described by local theories. Nevertheless entanglement is a  non local property of quantum mechanics well tested experimentally, even though probably not fully understood. The universal attractive behavior of gravity has an interesting parallel with the ubiquity of entanglement, and indeed at a very fundamental level gravity could be an emergent concept related to the entanglement of an underlying quantum theory, see \cite{VanRaamsdonk:2016exw} for more details. An interesting idea, dubbed $EPR=ER$, links the entanglement (Einstein-Podolsky-Rosen) with wormholes (Einstein-Rosen bridges) \cite{Maldacena:2013xja}. In the strongest form this conjecture states that to every entangled pair we can associate a wormhole. Applied to the black hole paradox, this means that  the external radiation is non locally connected to the interior of the black hole, therefore the evaporation process is potentially unitary.

\subsubsection{Soft Hairs}

 Relaxing the No-Hair theorem is an economic and simple possibility, and indeed it has been tried many times in the past with particular emphasis on scalar hairs, considering extension to general relativity and different matter content \cite{Herdeiro:2015waa}. Remarkably, a recent proposal suggest that even in pure GR we overlooked an infinite number of soft hairs related to the asymptotic structure of the spacetime \cite{Hawking:2016msc,Hawking:2016sgy}.
 
 Apart from Poincare' symmetries, it's well know that an asymptotically flat spacetimes enjoy an infinite number of diffeomorphisms changing the physical data at past and future null infinity while preserving the asymptotic; this enlarged infinite dimensional group is called  BMS group \cite{Bondi:1962px,Sachs:1962wk}. These symmetries allow to define an infinite number of conserved charges, in particular a  suitable antipodal combination of  generators on past and future infinity is an exact symmetry of gravitational scattering \cite{Strominger:2013jfa}, actually an analog of the   earlier discovered soft theorems for gravitons in QFT \cite{PhysRev.140.B516}.

At the quantum level, the Minkowski vacuum it's now infinitely degenerate, since different vacua are labeled by a different number of soft gravitons, which can be seen as Goldstone bosons of the spontaneously broken supertraslation symmetry. Notice that each vacuum it's physically inequivalent, even though they have the same energy. These features influence the information paradox in two ways: first of all the final state of the evaporation it's not a unique vacuum state anymore, since the vacuum is degenerate and gravitational process can induce transition between vacua. Secondly, there an infinite number of soft hair, in principle capable of storing the information of what falls inside. 

\pagebreak

\section{String Theory and the Fuzzball Proposal}

Most of the resolutions to the information paradox and descriptions of the black hole's quantum structure  introduce fancy machinery to solve a very specific issue. While there is nothing wrong with this, it would be nice to have a description embedded in a bigger framework of quantum gravity. A strong candidate to describe our universe is given by string theory, a quantum gravity framework in which the main novelties are the existence of fundamental extended objects and extra space-time dimensions. Everything in this theory must be constructed using the basic building blocks, namely strings and branes, and black holes made no exception. In the following sections we are going to discuss how black holes can emerge in string theory and how they appear in the corresponding low energy limit, that is supergravity, focusing in particular on the fuzzball proposal. The literature on string theory and supergravity is ginormous, a good introduction to strings and branes can be found in 
\cite{Zwiebach:789942,Blumenhagen:2013fgp,Polchinski:1998rq,Polchinski:1998rr,Kiritsis:2007zza,Becker:2007zj,Tong:2009np,Green:1987sp}, while for supergravity we recommend \cite{Ortin:2015hya,Freedman:2012zz,Tanii:2014gaa}.

\subsection{Black holes in String Theory and Supergravity}

The greatest achievement in string theory's black hole physics is the confirmation of a statistical interpretation for the Bekenstein-Hawking relation \cite{Sen:1995in,Strominger:1996sh}. The early calculations identified the quantum structure of the black holes with an exponential number of microstates, and remarkably the log of the number of these microstates reproduce  the area-entropy relation \eqref{SA}  in the thermodynamical limit. The seminal example in \cite{Strominger:1996sh} was performed for a supersymmetric black hole in five dimensions; in this section we are going to review how to build the corresponding black hole in gravity and how to count the microstates in the quantum setting. Notice that in the original derivation there is no mention to the appearance that these microstates have in the gravity picture; we will show an explicit occurrence of them when talking about fuzzballs.

\subsubsection{The Gravity Solution}

BPS black holes obtained by dimensional reduction on a tori from $D=10$ superstring theory or $D=11$ M-theory can be constructed by superposition of single BPS constituents, like Dp-branes and strings. Looking at supergravity solutions, without taking into account stringy corrections, the majority of these black holes have zero area in the supergravity limit. Indeed it can be shown that the only black holes with macroscopic finite area are the ones with 3 charges in five dimensions and 4 charges in four dimensions \cite{Klebanov:1996un}. 

A systematic procedure do build supergravity solutions, and therefore  black holes, is given by the "harmonic function rule": in the string frame the metric factorize in a product structure and simply superposes the harmonic function , with an exponent factor $-1/2$ for directions in which the brane is extended and $1/2$ otherwise. This ansatz imposes that the harmonic functions can depend only on the transverse coordinate to the whole bound state \cite{Peet:2000hn}.
As an example, we can build the $D1-D5$ system, a two charge system that will be useful later when talking about the fuzzball proposal. But before doing this, let's briefly review the expression for a Dp-brane solution.

Dp-branes are fundamental non perturbative entities of superstring theory (in particular they appear in IIA,IIB and I superstring theories), related by a web of dualities to every other object in the theory, including strings and branes of different dimensionality. From a string theory perspective, they are the hypersurface on which the fundamental string end points lie and they are $1/2 $ BPS solutions charged under the Ramond-Ramond fields. Finally, they extend in a odd number of spatial dimensions in IIB superstring theory and and in an even number in IIA. On the other end they can be seen as semiclassical solution of supergravity, the low energy limit of string theory, provided that the energies involved are far below the string scale, otherwise higher powers of the space time curvature must be added to the action. The relevant bosonic effective action in the 10d string frame, neglecting B field and the topological Chern-Simons term, is given by:

\bea
S_{D_p} =\frac{1}{2 \kappa^2_{10}} \int \, d^{10}x \sqrt{-g} \left[ e^{-2\phi} (R + 4 (\nabla\phi)^2 )-
\frac{1}{2(p+2)!} F^2_{p+2}\right] 
\eea

with the Ramond field strength given by $F_{p+2} = dC_{p+1}$ and the gravitational coupling is related to the string length and coupling via $2\kappa^2_{10} = (2\pi)^7 l_s^8 g_s^2 $. The equation of motion for the metric, the dilaton and the C-fields are given by:

\bea 
R_{\mu \nu} + 2 \nabla_{\mu}\nabla_{\mu} \phi = \frac{e^{2\phi}}{2(p+1)!} (F^2_{\mu \nu} - \frac{g_{\mu \nu}}{2(p+2)}F^2) \nn \\
d^* F_{p+2} = 0 \qquad R= 4 (\nabla \phi)^2 - 4 \square \phi
\eea

A p-brane solution is given by \cite{Kiritsis:2007zza}:

\bea 
ds_{10}^2 &=& (H_p(r))^{-1/2} (-f(r) dt^2 + \vec{dy} \cdot \vec{dy}) + (H_p(r))^{1/2} (f(r)^{-1} dr^2 + r^2 d\Omega_{8-p}^2)  \nn \\
e^{2\phi} &=& g_s^2 H_p^{(3-p)/2} \nn \\
F_{01...pr} &=& - \sqrt{1+\frac{r_0^{7-p}}{L^{7-p}}} \frac{H_p'}{H_p^2} \nn \\
H_p(r) &=& 1 + \frac{L^{7-p}}{r^{7-p}} \qquad f(r) = 1 - \frac{r_0^{7-p}}{r^{7-p}} \nn \\
L^{7-p} &=& \sqrt{(\frac{2 \kappa_{10}^2 T_p N}{(7-p)\Omega_{8-p}})^2 + \frac{1}{4}r_0^{2(7-p)}} - \frac{1}{2}r_o^{7-p}  \qquad \Omega_n = \frac{2\pi^{(n+1)/2}}{\Gamma((n+1)/2)}
\label{dp}
\eea

where $r$ is the radial transverse direction, namely if the wordvolume of the brane spans $t,y^i$ with $i = 1, \dots , p$, then $r^2 = x_{p+1}^2 + x_9^2$. Moreover $d\Omega^2_{8-p}$ is the line element of the $(8-p)$-sphere of unit radius, $N$ is an integer equal to the number of branes and finally $T_p$ is the $D_p$ brane tension:

\bea
T_p = \frac{1}{(2\pi)^p l_s^{p+1} g_s}
\eea 

The factor $f(r)$ is something called the blackening factor, and for extremal branes $f(r)=1$, i.e. $r_0 = 0$. The extremal metric assumes a particularly simple and elegant form:

\bea
ds_{10}^2 &=& H_p^{-1/2}  \eta_{\mu \nu} dx^{\mu}dx^{\nu} + H_p^{1/2}  dx^a dx^a 
\label{extremeDp}
\eea

where $\mu$ runs over the worldvolume directions and $a$ over the transverse directions. The point $r=0$ is a genuine  singularity for $p<7$, with the only exception of  $p=3$.

Given a single $D_p$ brane solution we can make superpositions to find new solutions for composite objects. If we want to remain inside a supersymmetric ansatz we need to assure that the zero point energy of the open strings ending on the branes add up to an half integer number, so that we satisfy level matching. Concretely, for pure $D_p$ brane bound states, this implies a number $n = 0 \, mod(4)$ of $ND$ directions. The D1-D5 solution perfectly fits in this constraint, therefore using the harmonic superposition rule and restricting to extremality we can write:
 
 \bea
 ds^2_{D1,D5} &=& (H_1 H_5)^{-1/2} (-dt^2+dy^2_{5})+(H_1 H_5)^{1/2} (dx^2_{1}+\dots  dx^2_{4})+H_1^{1/2} H_5^{-1/2} (dy^2_{6}+\dots  dy^2_{9})  \nn \\
  F_{0 5 r} &=& \partial_r H_1^{-1}  \qquad  F_{056789r}= \partial_r H_5^{-1} \qquad e^{\phi}=g_s H_1^{1/2}H_5^{-1/2}
  \label{naive2}
 \eea

\begin{table}[]
\begin{center}
\begin{tabular}{|c|c|c|c|c|c|c|c|c|c|c|}
\hline
    Brane     &  t       & $x_1$   &  $x_2$  & $x_3$   & $ x_4 $ & $ y_5$    & $ y_6 $ & $ y_7$ & $ y_8 $ & $ y_9$        \\
\hline
 $ D1 $       &  $-$      &$.$    & $.$   &$.$    & $.$  &$-$   &$.$    &$.$     & $.$   & $.$     \\
$ D5$           &  $-$      &$.$    & $.$   &$.$    & $.$  &$-$   &$-$    &$-$     & $-$   & $-$     \\
\hline
\end{tabular}
\caption{D1-D5 brane configuration. Neumann and Dirichlet directions of the D-branes are indicated by lines and dots respectively.}
\end{center}
\end{table}

where $x_1, \dots x_4$ are the extended dimensions, so $r^2 = x_1^2 + \dots + x_4^2$, and $y_5, \dots , y_9$ are compact.
The harmonic functions are given by:

\bea 
H_1 = 1 + \frac{q_1^2}{r^2} \qquad H_5 = 1 + \frac{q_5^2}{r^2}
\eea

This expression  respects \eqref{dp} for the D5, while the D1 doesn't match with our expectation for a single D1; actually the D1 has been delocalized, or "smeared", over the ND directions, see \cite{Mohaupt:2000gc,Johnson:2003gi} for details. We can check that the horizon area in the 5d Einstein frame is zero: we first shift from String to Einstein frame in ten dimensions:

\bea
g_{\mu \nu}^{E,10} = e^{-\frac{1}{2}\phi_{10}} g_{\mu \nu}^{S,10}
\eea

and then dimensionally reduce on the torus $T^4$ of the ND directions $y_6, \dots , y_9$ and on the $S^1$ labeled by $y_5$:

\bea
g_{\mu \nu}^{E,5} = g_{(T^4 \times S^1)}^{\frac{1}{3}} (g_{\mu \nu}^{E,10})_{(t,x_i)}
\eea

where we have isolated the non compact 5d piece of the 10d metric $g_{\mu \nu}^{E,10}$. We end up with:

\bea
ds^2_{E,5} = -dt^2 (H_1H_5)^{-\frac{2}{3}}+ (dr^2 + r^2 d\Omega_{3}^2) (H_1H_5)^{\frac{1}{3}}
\eea

Since $A \propto r_H^3$ in 5d, it's enough to compute the radial size of the horizon, located at $r=0$: 

\bea
\lim_{r \rightarrow 0} r_H \sim r (H_1H_5)^{\frac{1}{6}} \sim r^{1/3}
\eea

therefore the area is vanishing in this limit. Moreover the Ricci scalar goes like $R \sim r^{-2/3}$, therefore there is a naked singularity.$  $

The simplest non trivial black hole than can be constructed using BPS constituents is the 5d Reissner-Nordstrom and it's an extension of the D1-D5 solutions just seen, again obtained by compactifying IIB superstring theory on $T^4 \times S^1$, but  with $Q_1$ D1 branes, $Q_5$ D5 branes and $Q_P$ units of momentum along the $S^1$. As visible from \eqref{extremeDp}, going near horizon the p-branes spacetime  tend to shrink the  longitudinal directions and expand the transverse space. A generic superposition of branes will therefore deform the space in a non homogeneous fashion, as happen in the D1-D5 solution in the compact $S^1$. Indeed while the ND directions are stabilized by the combined action of expansion and shrinkage respectively of the D1 and the D5, the $S^1$ collapses going near the horizon:

\bea
\lim_{r \rightarrow 0} \sqrt{g_{y_5y_5}} \sim  r
\eea

By letting units of momentum flow in the $S^1$ we can stabilize it and obtain a solution that has a correct interpretation at all scales greater than the compactification scale\footnote{From a purely gravitation perspective the momentum excitation can be seen as a gravitation wave moving in the $S^1$ direction}. As a last remark before writing the solution, we notice that the system has four unbroken supercharges ($1/8$ BPS) out of the starting 32 of IIB, since every atom is $1/2$ BPS and it's arranged in such a way to break different supercharges. Finally the solution \cite{Callan:1996dv,Maldacena:1996ky}:

 \bea
 ds^2_{D1,D5,P} &=& (H_1 H_5)^{-1/2} (-dt^2+dy^2_{5} + (H_P-1)(dt+dy_5)^2)+ \nn \\
 &+& (H_1 H_5)^{1/2} (dx^2_{1}+\dots  dx^2_{4})+H_1^{1/2} H_5^{-1/2} (dy^2_{6}+\dots  dy^2_{9})  \nn \\
  F_{0 5 r} &=&- \frac{1}{2}\partial_r H_1^{-1}  \qquad  F_{ijk}= \frac{1}{2} \epsilon_{ijkl}\partial_l H_5 \qquad i,j,k,l=1,2,3,4 \nn \\
  e^{\phi} &=& g_s H_1^{1/2}H_5^{-1/2} \qquad H_a = 1 + \frac{q_a^2}{r^2} \qquad a= 1,5,P
 \eea
 
 here $Q_a$ are non negative integers counting the number of branes and momentum charges, and indeed they will be the charges of the gauge fields obtained by dimensional reduction in 5d. They are related to the $q_a$ through:
 
 \bea
q_1^2 = \frac{g_s l_s^2}{V} Q_1 \qquad q_5^2 = g_s l_s^2 Q_5   \qquad q_P^2 = \frac{g_s^2l_s^2}{VR^2} Q_1 
 \eea
 
 where the physical volume of the $T^4$ is $(2\pi l_s)^4 V$ and the radius of the $S^1$ is $2\pi l_s R$, so $V$ and $R$ are dimensionless. The solution is reliable if the curvatures is small compared to the string scale, translating into:
 
 \bea
 g_s Q_1  >> 1 \qquad g_s Q_5  >> 1 \qquad g_s^2 Q_P  >> 1
 \eea
 
 for $V$ and $R$ of order 1. 
 
 Reducing to 5d, in the Einstein frame:
 
 \bea
ds^2_{D1,D5,P} &=& - \mathcal{H}^{-\frac{2}{3}}dt^2+\mathcal{H}^{\frac{1}{3}}(dr^2 + r^2 d\Omega_{3}^2) \qquad \mathcal{H} = H_1 H_5 H_P
 \eea
 
 leading to a non vanishing black hole area:
 
 \bea
 r_H = r \mathcal{H}^{\frac{1}{6}}|_{r=0} = (q_1q_5q_P)^{\frac{1}{3}} \qquad A = \Omega_3 r_H^3 = 2 \pi^2 q_1q_5q_P
 \eea
 
 To get the entropy we can use the Bekestein-Hawking formula, but it's better to use a system of units in which the gravitational constant in 5d is explicit. Using that $16 \pi G_{10} = (2\pi)^7 l_s^8 g_s^2$ we obtain:
 
 \bea
 \frac{1}{16 \pi G_5} = \frac{(2 \pi l_s R )(2 \pi l_s)^4 V}{16 \pi G_{10}} = \frac{RV}{(2\pi)^2 l_s^3 g_s^2}
 \eea
 
 resulting in a magical simplification:
 
 \bea
 S = \frac{1}{4} \frac{A}{G_5} = 2 \pi \sqrt{Q_1 Q_5 Q_P}
 \eea

\subsubsection{Microstate Counting}

The system of branes and momentum composing the black hole has a complementary description as bound state of branes with excitations traveling on the open strings. This description is valid in the regime \cite{Maldacena:1996ky}:

\bea
g_s Q_1  << 1 \qquad g_s Q_5  << 1
\eea 

that is the opposite of the SUGRA limit. Here the relevant dynamic can be given in term of the world volume theory, and since we are interested in the low energy regime we can neglect high energy excitations of the open strings and Kaluza-Klein excitations, apart from the $Q_p$ units of momentum on $S^1$. The entropically dominant configuration will be the one with the highest number of massless fields, and in such configuration will we count how many possible ways to distribute the momentum excitations are there. Since the quantity that we are computing it's BPS and moduli invariant, we can perform the calculation in the most convenient setting. In particular we take the volume of the $T^4$ very small, so that all the KK particles cannot contribute to the massless spectrum, de facto trivializing the $T^4$ contribution. Then we take the $S^1$ very large, so that the masses of the KK particles associated at $Q_P$ are much smaller than the string scale, and the direction $y_5$ can be seen as non compact, so we can neglect winding modes \cite{David:2002wn}. The relevant world volume field theory is  a strongly coupled $1+1$ dimensional $\mathcal{N} =(4,4)$ supersymmetric theory with gauge group $U(Q_1) \times U(Q_5)$, since the D1-D5 system leave 8 supercharges unbroken. 

After dimensional reduction the field content of the $1+1$ world volume theory is the following:

\begin{itemize}
\item $(D1,D1)$ : string starting and ending on the D1s, there are $U(Q_1)$ vector multiplets $\lbrace A_t^{(D1)},A_{y_5}^{(D1)},\Phi_i^{(D1)} \rbrace$ for the non compact directions $i=1,2,3,4$ and hypermultiplets $\Phi_a^{(D1)}$ with $a=6,7,8,9$.

\item $(D5,D5)$ : same as before, with $Q_1 \rightarrow Q_5$.

\item $(D1,D5)$ and  $(D5,D1)$: string stretched between different branes, respectively hypermultiplets in the bifundamental representation $(Q_1,\bar{Q}_5)$ or $(Q_5,\bar{Q}_1)$.
\end{itemize}  

We need to excite the system in such a way that the maximum number of particles remain massless, or equivalently to find the sector of the moduli space of vacua with more massless particles. The correct way to do it turns out to excite the hypers (Higgs branch, the one associated to a real bound state of branes), therefore giving mass to the scalars $\Phi_i^{(D1)}$ describing the transverse motion of the branes in the non compact space; the branes sit on top of each other, as needed. The total number of hypermultiplets is $4Q_1^2+4Q_5^2+4Q_1Q_5$, but we still need to impose the vanishing of the potential to have susy vacua (D-terms, imposing $3Q_1^2+3Q_5^2$ constraints) and to fix the gauge freedom ($Q_1^2+Q_5^2$), leaving only $4Q_1Q_5$ massless degrees of freedom.

Having taken the $S^1$ very large, the energy of each excitation is very small and we only need to know the infrared limit of the effective theory of the massless particles. The $\mathcal{N} =(4,4)$ susy implies that the IR fixed point is a superconformal sigma model with Hyper Kalher target space, so the central charge can be computed as if the fields were free. We must account for the fermions, contributing as $c=\frac{1}{2}$ to the central charge, so the total central charge is:

\bea
c= (1+\frac{1}{2})4Q_1Q_5 = 6 Q_1Q_5
\eea

since we are in $2d$ we can use the Cardy formula for the entropy:

\bea
S = 2 \pi \sqrt{\frac{c}{6}ER}
\eea

In our case the energy is due to the KK excitation, $E = \frac{Q_P}{R}$, therefore:

\bea
S =  2 \pi \sqrt{Q_1 Q_5 Q_P}
\eea

From a statistical entropy calculation of open string on branes we recover the same result obtained using Bekenstein-Hawking in gravity! This confirm the statistical interpretation of the quantum microstates and marks a compelling evidence in favor of string theory as quantum gravity theory.
This remarkable agreement can be attributed to the BPS nature of the system, meaning that susy has protected us from quantum corrections and allowed to extrapolate from low coupling to strong coupling. 

Since the agreement extends to not extremal (therefore not BPS) black holes, we might suspect that something bigger is going on here. And it is so. Starting from the low energy picture, we have a stack of branes sitting in flat space time at the point $r=0$. Connecting them with excited open strings we found agreement with a spacetime picture  of a bound state of branes. But in the spacetime picture the branes have distorted so much the transverse space near the horizon that $r=0$ is in fact a $3-$sphere and classically we can analitically continue past $r=0$, that is we can fall inside the horizon, heading to the singularity. In the open string picture the space stops at $r=0$, what is left are excitations of the open strings living on the D-branes. In a sense, the closed string dynamic inside the black hole is mapped to  open string physics on the horizon! This is a concrete realization of the holographic principle \cite{Susskind:1994vu}, ultimately leading to a precise mapping in the context of $AdS$ spaces, the $AdS/CFT$ correspondence \cite{Maldacena:1997re}. 

\subsection{The  Fuzzball Proposal}

The fuzzball proposal solves the information paradox by removing the black hole entirely \cite{Lunin:2001jy,Lunin:2002iz,Mathur:2003hj,Lunin:2004uu,Giusto:2004id,Giusto:2004ip,Mathur:2005zp, Skenderis:2008qn, Mathur:2008nj}. The strongest form of the proposal is the following: to every black hole microstate is associated a smooth and horizonless SUGRA solution \footnote{
The generic fuzzball is a solution of the full string-M theory and the reduction to SUGRA can hide the difference between different microstates; for instance two microstates cannot be distinguished in SUGRA if the difference is of the same order of  higher derivative corrections. It's likely that we will need to go beyond the SUGRA approximation to fully solve the information paradox.}. Having removed the singularity and the horizon, radiation processes from a fuzzball resemble more an ordinary burning encyclopedia that an evaporating black hole. The no-hair theorem it's violated, indeed different fuzzball are characterized by different multipole expansions. Every solution the paradox bring something difficult to digest and  the fuzzballs make no exception: contrary to basic expectations, quantum gravity effects would be relevant at horizon scale, that is even with a very weak gravitational fields for macroscopic black holes. 

Fuzzballs allow us to build the microstates directly in the gravity setting, even though it's often easier to study the corresponding world volume  theory of the branes composing the fuzzball or to use holography to have insights. The ultimate goal is to build a fuzzball associated to an astrophysical black holes, for instance Kerr or Schwarzschild. It's fair to say that we are a long way to that goal, since right now we mainly deal with supersymmetric charged black holes. The main results of the proposal were obtained studying the two charge system, which we will review next. Later we will talk about three and four charge black holes, the ones relevant to have non zero area in the SUGRA limit.

Previously, in \eqref{naive2},  we presented the D1-D5 system, and we found a naked singularity. We will from now on dub that system as "naive" D1-D5, since we argue that the correct description is given by the D1-D5 fuzzball, a smooth and horizonless solution:

\bea
ds^2 &=& \sqrt{\frac{H}{1+K}}[-(dt-A_i dx^i)^2+(dy_5+B_i dx^i)^2)]+\sqrt{\frac{1+K}{H}} dx_i^2+\sqrt{H(1+K)}dy_a^2 \nn \\
	H^{-1} &=& 1+\frac{ Q_1}{ L_T} \int_{0}^{ L_T}  \frac{dv}{|\vec{x}-\vec{F}(v)|^2}  \;\;\; K=\frac{ Q_1}{ L_T} \int_{0}^{ L_T}  \frac{dv ( \dot{F}(v))^2}{|\vec{x}- \vec{F}(v)|^2} \nn \\
	A_i &=& -\frac{ Q_1}{ L_T} \int_{0}^{ L_T}  \frac{dv  \dot{F}_i(v)}{|\vec{x}- \vec{F}(v)|^2} \;\;\; dB=-\star_4 dA\;\;\;  v=t-y_5  
\label{d1d5fu}
\eea

the complete solution comprise the dilaton, the B fields and RR fields, but here we limit ourselves to the metric, see \cite{Giusto:2009qq,Skenderis:2008qn} for more details. Here $\vec{F}(v)$ are the hairs of fuzzball:  an arbitrary closed, smooth and not intersecting curve discriminating between different fuzzballs. The constant $L_T$ is real and positive, its interpretation will be clear soon.
The simplest possible fuzzball has a circular profile singularity:

\bea
F_1=cos(\omega v), \;\; F_2=sin(\omega v), \;\; F_3=F_4=0
\eea

for some real parameter $\omega$. Of course this a pretty non generic fuzzball, being so symmetric, but it retains the main features of every other fuzzball, namely: from very far away ($r \rightarrow \infty $) has the same asymptotic charges of the naive solution, then for a sufficiently low values of the radius it approximate $AdS_3 \times S^3 \times T^4$ (similarly to the naive solution), but then starts to differ dramatically near the would be horizon, $r=0$. The singularity is replaced by a smooth cap which shape is parametrized by the function $F(v)$ living in the non compact $\mathbf{R}^4$. This is an important point that deserve to be treated in some detail, basically we want to prove that the singularity in the harmonics of   \eqref{d1d5fu} are harmless and the spacetime shrinks smoothly. Following \cite{Lunin:2002iz} we select a certain point $v_0$ on the curve, we zoom in near this point and  introduce a coordinate system with $z$ measuring the distance along the curve in the flat metric and spherical polar $(\rho, \theta, \phi)$ for the 3-plane perpendicular to the curve. Therefore near $v_0$:

\bea
z \sim |\dot{\vec{F}}(v)|(v-v_0)
\eea

\bea
H^{-1} &\sim & \frac{ Q_1}{ L_T} \int_{-\infty}^{+\infty}  \frac{dv}{\rho^2+z^2}  = \frac{Q_1\pi}{L_T  |\dot{\vec{F}}(v)| \rho} \nn \\
 K &\sim&  \frac{Q_1\pi  |\dot{\vec{F}}(v)|}{L_T  \rho} 
\qquad	A_i \sim - \frac{Q_1\pi }{L_T  \rho} 
\eea

Introducing $Q=\frac{Q_1 \pi}{L_T}$, the $(y_5,\rho,\theta,\phi)$ piece of the metric becomes:

\bea
ds^2_{y_5,\rho,\theta,\phi} = \frac{\rho}{Q} (dy_5 - Q(1-cos \theta) d\phi)^2 + \frac{Q}{\rho}(d \rho^2 + \rho^2(d \theta^2 + sin ^2 \theta d\phi^2))
\eea

This is a Kaluza-Klein monopole, which it's completely free of singularities and caps smoothly if the circle $y_5$ is compactified with the right periodicity, as it happens in this case. Apart from the trivial $T^4$, the other two directions are regular too:

\bea
ds^2_{t,z} = -\frac{\rho}{Q} dt^2 -2 dt dz
\eea

In summary, the spacetime is smooth, there are no singularities or horizon, and in fact the spacetime ends at the location of the would be horizon, with a shape characteristic of every single fuzzball solution. The missing interior is reminiscent of the old proposal of "bubbles of nothing", see \cite{Mathur:2014roa,Witten:1981gj}.

Some features of two charge fuzzballs, like the microstate counting or the spacetime size can be better appreciated in the duality frame $F1-P$, related by a chain of dualities  to $D1-D5$. Explicitly:

\bea
\begin{split}
& \begin{pmatrix} D1_{(5)} \\
D5_{(56789)} \\
 \end{pmatrix} \rightarrow^{S} 
\begin{pmatrix} F1_{(5)} \\
NS5_{(56789)} \\
\end{pmatrix}   \rightarrow^{T_{56}} 
\begin{pmatrix} P_{(5)} \\
N5S_{(56789)} \\
 \end{pmatrix} \rightarrow^{S} 
\begin{pmatrix} P_{(5)} \\
D5_{(56789)} \\
\end{pmatrix} \rightarrow^{T_{(6789)}}  \\
& \rightarrow \begin{pmatrix} P_{(5)} \\
D1_{(5)} \\
 \end{pmatrix}
  \rightarrow^{S}  \begin{pmatrix} P_{(5)} \\
F1_{(5)} \\
 \end{pmatrix}
\end{split}
\eea
Both entropy counting and size considerations are not affected by the duality frame, in particular the latter doesn't change in the Einstein frame, since S dualities act trivially on the metric, and T dualities are performed only in compact directions. Consider a fundamental string wrapped on an $S^1$ with $Q_w$ units of winding and $Q_p$ units of momentum, therefore with  waves traveling in only one direction around $S^1$, say only left moving. The many ways to partition the momentum excitations among different harmonics are the microstates we are going to count. From elementary string theory, the mass of a string excitation is given by:

\bea
m^2  = (2\pi R Q_w T - \frac{Q_P}{R})^2+8\pi T N_L= (2\pi R Q_w T + \frac{Q_P}{R})^2+8\pi T N_R
\eea

where $T$ is the string tension and $R$ the radius of $S^1$. Since we have a BPS string with only left moving excitations, $N_R=0$, therefore:

\bea
N_L = Q_wQ_P
\eea

Since $N_L$ is taken to be large (remember that $Q_w Q_P$ are dual to the D1 and D5 charges), we can use the Cardy formula for the entropy. We know, as apparent using  a light-cone gauge, that the dynamic degrees of freedom of a string are expressed by 8 bosonic oscillators in transverse directions and the corresponding 8 fermionic oscillators. The total central charge is therefore $c=8+8/2=12$ and the entropy is given by:

\bea
S \sim ln(e^{2\pi \sqrt{\frac{c}{6}N_L}})=  2 \sqrt{2} \pi \sqrt{Q_w Q_P}
\eea

Remarkably, the same result can be obtained in gravity in two completely different ways: firstly, by considering higher derivative terms in the lagrangian the naive D1-D5 system acquire a non zero, but still Planck sized, area. One should use the Wald entropy \cite{Wald:1993nt}, generalizing the Bekenstein entropy when there are higher derivatives corrections, and agreement has been found with the microscopic calculation \cite{Dabholkar:2004yr}. Secondly, we can quantize the supergravity space of fuzzballs, namely by imposing commutation relations to the string profiles characterizing the solution and quantizing the moduli space of regular solutions with fixed charge \cite{Rychkov:2005ji,Krishnan:2015vha}. It's highly non trivial that the latter computation has given the full microscopic entropy, since this is telling us that supergravity is powerful enough to contain all the microstate of the two charge system.

The complete SUGRA solution of the F1-P fuzzball is \cite{Mathur:2005zp}:

\bea 
ds^2 &=& H[-dudv + K dv^2 + 2 A_i dx_i dv] + dx_i^2+dy_a^2 \nn \\
B_{uv} &=& -\frac{1}{2}(H-1) \qquad B_{vi} = H A_i
\qquad e^{2 \phi} = H
\eea

with the same harmonics of \eqref{d1d5fu}, in which now we understand that $L_T$ it's the total length of the multiwound string. In this duality system the  function $F(v)$ has a manifest interpretation in terms of profile oscillation of the string in the non compact space. Indeed the traveling waves have no longitudinal component, thus all the momentum must be carried transversally to the string. This implies that the fuzzball has a non trivial size, and approximate calculations show that the region in which fuzzball start to differ substantially from the naive solutions has an area satisfying a Bekenstein-Hawking relation, namely:

\bea
\frac{A}{4G_{10}} \sim \sqrt{Q_w Q_P}
\eea

In summary, the two charge system is a full implementation of the fuzzball proposal: given a singular naive solution we are able to find the "true" SUGRA description, which turns out to be regular and horizonless. The region in which the fuzzball differ from the naive solution is horizon-sized and there are enough microstates to match the expectations from a microscopic counting. The main issue is that D1-D5 naive system it's not really a black hole, at least not a macroscopic one.

\begin{figure}
\begin{center}
\includegraphics[width=10cm]{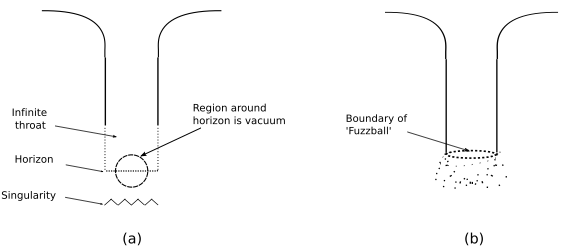} 
\end{center}
\caption{\cite{Mathur:2008nj} Comparison between a traditional black hole (a) and a fuzzball (b).}
\end{figure}

Macroscopic fuzzballs can be obtained from three and four charge black holes, the rationale is the same already seen for the two charge case, but there is no systematic way to construct these gravity solutions. The literature has mainly focused on the quest for three charge BPS regular solutions, in particular on the D1-D5-P system, in order to reuse much of the technology know from the D1-D5 system. Indeed building gravity solutions is hard, no escape from that (just remind that finding Kerr, in simple general relativity, took approximately 50 years!). Moreover we expect the typical SUGRA fuzzball to have all the possible fields turned on, since these fields work to regularize the black holes, similarly to regular black hole models that can be obtained in scalar tensor theories. Nevertheless, inspiration can come  from the microscopic side, which is sometimes easier to deal with. The main microscopic tools are: 

\begin{itemize}
\item worldvolume  theory: usually very difficult to deal with directly, but excellent for microstate counting.
\item string scattering amplitudes: computations are cumbersome, but it's relatively under control. It allows to get information on the SUGRA solution, for instance on which fields are turned on.
\item holography: exploiting the fact that extremal black holes contain AdS spaces as near horizon limit, one can use AdS/CFT to get insights on the SUGRA solution.
\end{itemize}  

The holographic approach is particularly popular for  $AdS_3/CFT_2$ and the D1-D5 CFT has been  thoroughly studied, explaining the huge literature on three charge black holes. 
To concretely find SUGRA solutions, two main techniques are used: supersymmetric classification and limits of known solutions. The former method consist in taking solution valid for arbitrary data, for instance an extremal solution given in terms of generic harmonic functions, and finding the data leading to regular solutions. The latter method has been used for non supersymmetric solutions, but has the disadvantage of finding rather non generic and symmetrical fuzzballs, witch are dual to atypical microstates. The general state of the art is that, even though many regular solution has been found, a fuzzball generic  enough to account for the expect behaviour of the entropy $S \sim \sqrt{Q_1Q_2Q_3}$ has not been discovered, see \cite{Bena:2007kg,Skenderis:2008qn,Bena:2016ypk} for details and recent results.

The  four charge case have been touched sporadically, since the $AdS_2/CFT_1$ is rather poorly understood and the possibility of lifting the problem to M-theory to use again $AdS_3/CFT_2$, with the so called MSW CFT \cite{Maldacena:1997de}, it's not an easy way out either, see for some recent progress \cite{Bena:2017geu}. In this thesis we will exploit the two alternatives to holography to get insight into four charge black hole physics.

We briefly touch two important  but still rather unexplored topics: the effective formation of the classical event horizon and the creation of astrophysical fuzzballs. The horizon can be thought as arising from a coarse-graining procedure of the horizonless geometries, even though no explicit calculation has ever been carried due to the lack of an explicit basis for macroscopic fuzzball geometries. This is analogous to thermodynamics, in which we assign some small set of macroscopic parameters, including entropy, to describe a bunch of particles as a gas. If we specifies position, velocity and charges of every particle we know exactly the state and the associated entropy is zero. In the same fashion we can choose between an exact description of a zero entropy (and horizonless) microstate or we can ignore the region in which the ensemble of microstates differs between each other, that is a space enclosed by the area of the would-be black hole with the same asymptotic charges, associating an entropy to our ignorance; this entropy it's  related via Bekenstein-Hawking to the area of the "ignored volume" and  enjoys a statistical description in terms of the compatible microstates. 

The dynamic formation of a fuzzball should be something pretty spectacular: a macroscopic collapsing  astrophysical soup suddenly obeys to the laws of quantum gravity and generates a fuzzball. Classical SUGRA fuzzballs can be thought as energy eigenstates of a (not well specified) quantum picture and in principle every object should have some probability to tunnel into one of such fuzzballs. Of course this probability is  exponentially small, otherwise we should see objects around us popping in and out of existence. The tunnel amplitude can be estimated as \cite{Mathur:2009hf}:

\bea
\mathcal{A} \sim e^{- \mathcal{S}_{tunnel}}
\eea

where from dimensional consideration for an object of mass $M$ spread over a region or size $GM$ we can estimate the action as:

\bea
\mathcal{S}_{tunnel} \sim \frac{1}{G} \int R d^4x  \sim \frac{1}{G} \frac{1}{(GM)^2} (GM)^4   \sim  GM^2
\eea

Even though tunneling is exponentially unlikely as the mass grows, black hole microstates have an huge exponential degeneracy 

\bea
\mathcal{N} \sim e^{- S_{bek-hawk}} \sim e^{- G M^2}
\eea

resulting in an overall $O(1)$ possibility of tunneling into a fuzzball. This heuristic picture can be backed up by a more concrete estimate in terms of spreading of the quantum gravity wavefunctional in the superspace of the 3-geometries \cite{Mathur:2017wxv}. It's fair to say that a definitive picture for the  formation of horizon-sized fuzzballs is still in his infancy and more work need to be done in this exciting direction. Regarding the dynamics of an infalling observer, a naive interpretation of the "non vacuum" nature of the fuzzball boundary would  suggest a "brick wall" at the would horizon, even though has been proposed (fuzzball complementarity) that an apparent harmful dynamic could be recovered for an infalling observer; we will not delve into this muddy territory limiting ourselves to point out relevant literature \cite{Mathur:2011wg,Mathur:2012jk,Avery:2012tf,Martinec:2014gka}.

\pagebreak

\section{From Open Strings  to Supergravity}

Consider a four charge BPS system in string theory, we will focus on these systems in the whole thesis. There are different u-duality frames from which to choose, for instance  D6-D2-D2-D2, D0-D4-D4-D4 in IIA superstring theory, M5-M5-M5-P in M-theory and the one that we will pick in this section, the D3-D3-D3-D3 system in IIB superstring theory. A possible BPS arrangement of the branes is given in the table, notice that each brane has exactly 4 ND directions with respect to each other brane. Here $x_i$ are non compact coordinates, while $y_I, \tilde{y}_I$ label the three internal $T^2s$ of the compact $T^6$, the latter being the simplest consistent  manifold that we can choose to compactify to four dimensions. The naive metric associated to this configuration can be constructed with the harmonic superposition rule:

 \begin{table}[]
\begin{center}
\begin{tabular}{|c|c|c|c|c|c|c|c|c|c|c|}
\hline
        Brane     &  t       & $x_1$   &  $x_2$  & $x_3$   & $ y_1 $ & $ \tilde y_1$    & $ y_2 $ & $ \tilde y_2$ & $ y_3 $ & $ \tilde y_3$        \\
\hline
 $ D3_0 $       &  $-$      &$.$    & $.$   &$.$    & $-$  &$.$   &$-$    &$.$     & $-$   & $.$     \\
$ D3_1$           &  $-$      &$.$    & $.$   &$.$    & $-$  &$.$   &$.$    &$-$     & $.$   & $-$     \\
$ D3_2$           &  $-$      &$.$    & $.$   &$.$    & $.$  &$-$   &$-$    &$.$     & $.$   & $-$     \\
 $ D3_3 $       &  $-$      &$.$    & $.$   &$.$    & $.$  &$-$   &$.$    &$-$     & $-$   & $.$     \\
\hline
\end{tabular}
\label{td3s}
\end{center}
\caption{D3-brane configuration: Neumann (N) and Dirichlet (D) directions are represented by lines and dots respectively.}
 \end{table}

\bea
ds^2 &=& (H_0 H_1 H_2 H_3)^{-\frac{1}{2}}dt^2+(H_0 H_1 H_2 H_3)^{\frac{1}{2}}\sum_{i=1}^3dx_i^2 +(\frac{H_2 H_3}{H_0 H_1})^{\frac{1}{2}}dy_1^2+ (\frac{H_0 H_1}{H_2 H_3})^{\frac{1}{2}}d\tilde{y}_1^2+\nn \\
&+& (\frac{H_0 H_2}{H_1 H_3})^{\frac{1}{2}}dy_2^2+ (\frac{H_1 H_3}{H_0 H_2})^{\frac{1}{2}}d\tilde{y}_2^2+(\frac{H_0 H_3}{H_1 H_2})^{\frac{1}{2}}dy_3^2+ (\frac{H_1 H_2}{H_0 H_3})^{\frac{1}{2}}d\tilde{y}_3^2
\eea

where $H_A=1+\frac{Q_A}{r}$. What is the fuzzball of the $D3^4$ black holes? As usual with extremal solutions, it's possible to express the solution in terms of arbitrary harmonic functions, of which the previous choice for $H_A$ it is the simplest possibility, leading to a black hole. Such a generic solution for a 10d four charge system in IIA  has been written in \cite{Dall'Agata:2010dy} and T-dualized, see the appendix for details, to the $D3^4$ system in \cite{Bianchi:2016bgx}, leading to:

      \bea
  ds^2  &=& -   e^{2U}( dt+w)^2 +e^{-2U} \,  \sum_{i=1}^3 dx_i^2 +   \sum_{I=1}^3   \left[  { d y_I^2 \over  V e^{2U} Z_I }  + V e^{2U} Z_I  \,  \tilde e_I^2  \right] \nn\\
  C_4 &=& \alpha_0  \wedge \tilde e_1\wedge \tilde e_2\wedge \tilde e_3+ \beta_0  \wedge dy_1\wedge  dy_2\wedge  dy_3 \nn\\ 
  && +\frac{1}{2} \epsilon_{IJK}\,   \left( \alpha_I  \wedge  dy_I \wedge \tilde e_J   \wedge \tilde e_K + \beta_I  \wedge  \tilde e_I \wedge dy_J   \wedge dy_K  \right)   \label{d34}
 \eea
 
   Parameterized in  terms of eight harmonic functions 
  ($a=1,\dots 8$, $I=1,2,3$) 
   \bea
   H_a=\{ V, L_I, K_I, M \} \label{harmh}
   \eea
  on (flat) $\mathbb{R}^3$ associated to the four electric and four magnetic charges in the type IIA description of the system. 
   These functions are conveniently combined into  
  \bea
 P_I &=& {K_I \over V} \nn\\
 Z_I &=& L_I +{ |\epsilon_{IJK}|\over 2} {K_J K_K\over V} \nn\\
 \mu &=& { M\over 2} +{L_I K_I \over 2\, V}+{ |\epsilon_{IJK}| \over 6} \, {K_I K_J K_K \over V^2}  \label{mudef}
 \eea
 Here
  $\epsilon_{IJK}$ characterise the triple  intersections  between two cycles on $T^6$ and:

  \bea
  e^{-4U} &=&  Z_1 Z_2 Z_3 V-\mu^2  V^2 \qquad ~~~~~~~~~~ \nn \\
   b_I &=&     P_I-\frac{\mu}{Z_I}    \qquad \qquad ~~~~~~~~~~~~~~~~~ \tilde e_I =   d \tilde y_I   -  b_I\, dy_I   \nn\\
 \alpha_0 &=& A-\mu\, V^2\, e^{4U}\, (dt+w) \qquad ~~~~~
\alpha_I =   -\frac{(dt+w)}{Z_I} + b_I\, A+w_I   \nn\\
\beta_0&=& -v_0 +{e^{-4U}\over V^2 Z_1 Z_2 Z_3}(dt+w)-b_I \,v_I+ b_1 \,b_2\, b_3 \, A +{|\epsilon_{IJK}|\over 2}\,b_I \,b_J\, w_K  \nn\\
\beta_I &=& -v_I + {|\epsilon_{IJK}|\over 2}\, \left(  {\mu\over Z_J Z_K}\, (dt+w)+b_J \, b_K\, A+2 b_J\, w_K  \right) 
 \eea
   and
  \bea
 {*_3}dA &=& d V     \qquad {*_3}dw_I = -d K_I  \qquad    {*_3}dv_0 =dM  \qquad    {*_3}dv_I =dL_I\nn\\
   \qquad  {*_3}dw  &=&  V d \mu-\mu dV-V Z_I dP_I  
  \eea

  For $K_I=M=0$,   the supergravity solution (\ref{d34})  describes a system of four stacks of intersecting D3-branes aligned according to  the table, but a suitable choice for $H_a$ can be the fuzzball that we are looking for. The probability of randomly guessing the correct harmonic between an infinity of possibilities is  zero, so to get some insights we can zoom into the microscopic world and compute the closed string emission of the branes. We will mainly follow the analysis of \cite{Bianchi:2016bgx,Pieri:2016pdt} in this section.

The microscopic picture is given by four stack of $D3$ branes connected by open strings, with Chan-Paton indices labeling branes on which the string endpoints lie. As every other object in a gravitational theory, this brane system must emit closed string from which we can reconstruct the leading orders of the SUGRA solution. 
\begin{figure}
\begin{center}
\includegraphics[width=8cm]{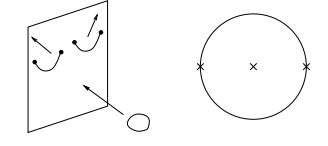} 
\end{center}
\caption{\cite{Hashimoto:1996bf} Depiction of the scattering of a closed string on a brane, followed by a split into two open strings living on the brane. The time reversal process can be seen as the emission from the brane. On the right the associated worldsheet diagram.}
\end{figure}
 The computation is performed at small string coupling $g_s$ and we can expect to match the SUGRA expansion in $g_s$ to scattering amplitudes involving a growing number of stacks; the scattering amplitudes are mixed open-closed disk amplitudes, since that is the leading contribution in the wordsheet $g_s$ expansion. The emission from a single stack will be associated to the naive SUGRA brane, but already for two stacks we can predict that higher multipole modes will be excited in SUGRA. To summarize the philosophy of the computation: to every microscopic open string configuration it  corresponds a precise multipole expansion, that is the microstates of the gravity picture identifying a particular fuzzball. Unfortunately with this analysis it's not possible to exactly reconstruct the harmonics, but only to get the leading behaviour. The distinctive features of four charge black hole microstates should be captured by scattering amplitudes involving a full circle of open strings touching every stack at least once. The number of different microscopic configurations will be proportional to $Q_1Q_2Q_3Q_4$, where $Q_a$ is the number of branes in the stack a: this is the same product appearing into the quartic Cremmer-Julia U-duality invariant of $E_{7(+7)}$ in $\mathcal{N}=8$, $d=4$ SUGRA, the invariant being of course the area (or the entropy) \cite{Kallosh:1996uy}:

\bea
S = 2\pi \sqrt{Q_1Q_2Q_3Q_4}
\eea

Coming back to the harmonics, since we noticed that the four functions $K_I$ and $M$ are associated to the non trivial configurations, we should demand that their behaviour at infinity doesn't modify the asymptotic charges, but only higher order multipoles. Indeed we expect the microstates to start differing from the coarse grained description only when approaching the black hole, whilst far away there shouldn't be any detectable difference. The expected falling off at infinity is therefore:

 \bea
 L_I &\approx& 1+\alpha_{D3} \, {N_I\over |x| } +\dots   \qquad     V \approx 1+ \alpha_{D3} \,{N_{0}\over |x| } +\dots     \nn\\
 K_I & \approx&  \alpha_{D3} \, c^{K_I}_i {x^i\over |x|^3} +\dots   \qquad  ~~  M \approx  \alpha_{D3}  \, c^M_i {x^i\over |x|^3}  +\dots
 \label{KLMasympt}
 \eea 
 with  $c_i$ some constants, $\alpha_{D3}=4\,\pi \,g_s \, {(\alpha')^2}/{V_{T_3}} $ the inverse D3-brane tension and $V_{T_3}$  the volume of the 3-torus wrapped by the stack of D3-branes, that  for simplicity we will take  to be equal for all the stacks.
The explicit string computations will show that the coefficients $c_i$'s entering in (\ref{KLMasympt}) 
 satisfy the relation
 \bea
c^M_i+\sum_{I=1}^3 c^{K_I}_i=0
 \eea
 or equivalently the function $\mu$ defined in (\ref{mudef}) starts at order $1/r^3$ \cite{Bianchi:2016bgx,Pieri:2016pdt}. 
    Since at linear order in $\alpha_{D3}$, the harmonic functions entering in the solution simply add, we can consider the contribution of each harmonic function separately, even though the complete gravity solution will  not simply be the "linear" superposition of this simple solutions.  In particular, we define three basic class of solutions: L solutions are associated to the harmonic functions $L_I,V$ with $K_I=M=0$. K solutions are generated by turning on a single
  $K_I$ harmonic function for a given $I$ and taking $M=-K_I$. Finally M solutions are associated to the orthogonal combination   $M+\sum_I K_I$ that is generated
  at order $1/r^3$, or equivalently  the monopole and dipole modes  are vanishing. More generally, the harmonic functions $H_a$  entering into the supergravity solutions   can be conveniently expanded in multi-pole modes.  The general 
expansion can be written  as
   \bea
H_a(x)=h_a+\sum_{n=0}^\infty  c^a_{i_1\dots i_n }  P_{i_1 \dots i_n} (x)       
\label{hhh}
 \eea
  with  $ P_{i_1 \dots i_n} (x)$ a totally symmetric and traceless rank $n$ tensor  providing a basis of harmonic functions on $\mathbb{R}^3$. The explicit
  form of  $ P_{i_1 \dots i_n} (x) $ can be found from the multi-pole expansion
 \bea
 {1\over  |x + a|} =\sum_{n=0}^\infty  a_{i_1} \dots a_{i_n }  P_{i_1 \dots i_n} (x)    
 \label{1multi}
 \eea
 For example for $n=0,1,2$ one finds
 \bea
 P(x)={1\over |x|} \qquad  \qquad  P_{i}(x)= - {x_i \over  |x|^3}   \qquad \qquad     P_{i j}(x)={3 x_i x_j - \delta_{ij} |x|^2 \over  |x|^5}   
 \eea
 and so on.
 We notice that the form of the Fourier transform is very simple:
  \bea
   P_{i_1 \dots i_n} (x)  = \int {d^3 k \over (2\pi)^3}   e^{i k x}  \tilde P_{i_1\dots i_n}(k)   
  \eea
  and using
  \bea
   {1\over   |x+a| }  =   \sum_{n=0}^\infty  a_{i_1} \dots a_{i_n }  P_{i_1 \dots i_n} (x) =4\pi   \int {d^3 k \over (2\pi)^3}  \frac{   e^{i k(x+a)}   }{k^2}  
  \eea
one finds that the Fourier transform of the harmonic functions   $P_{i_1 \dots i_n} (x)$ are simply polynomials of the momenta divided by $k^2$  
   \bea
  \tilde P_{i_1\dots i_n}(k)  =     { 4\pi \, i^n \over n! \, k^2}  k_{i_1} \dots  k_{i_n}  
  \eea
   Disk amplitudes will be conveniently written in this basis.  We notice that even if the single harmonic components $P_{i_1\dots i_n}(x)$ are singular at the origin, 
   in analogy with the 2-charge case \cite{Lunin:2002iz,Mathur:2005zp},   one may expect that for an appropriate choice of the coefficients $c_{i_1\dots i_n}$ the infinite sum (\ref{hhh}) produce a fuzzy and smooth geometry.
This was explicitly shown in \cite{Lunin:2015hma} for a class of solutions parametrized by one complex harmonic function ${\cal H}={\cal H}_1+ i {\cal H}_2$ obtained from our general solution by taking 

        \bea
 {\cal H}_1&=&L_1=L_2=L_3=V   \nn \\
 {\cal H}_2&=&K_1=K_2=K_3=-M  
  \eea
    
\subsection{String amplitudes and 1-charge microstates}
  
  To settle the notation and review the emission from branes we start with the trivial 1-charge microstates, that is the emission from a single brane. At weak coupling, the gravitational background generated by a D-brane  can be extracted from a disk amplitude  involving the insertion of a closed string states in the bulk and open string states specifying the D-brane micro-state on the boundary. The vev of the open string state, or condensate,  is specified by a boundary operator ${\cal O}$ given by a trace along the boundary of the disks as a product of constant open string  fields. The profile of a supergravity field $\Phi$ in the micro-state geometry can be reconstructed from the string 
 amplitude  $ {\cal A}_{\Phi,{\cal O}}(k)$  after Fourier transform in the momentum $k$ of the closed string state. Moreover, since we consider constant open string fields  we can consistently take all the open strings momenta to be vanishing. 
 The closed string momentum satisfies  $k^2=0$ and $k^{M}E_{MN}=0$, moreover we concentrate on a purely spatial momentum $k^M=\lbrace k^0,k^i,k^{int} \rbrace=\lbrace 0,k^i,0\rbrace$,  analytically continuing the momentum to complex values in order to be consistent with the mass-shell condition.  
   Explicitly, the deviation  $\delta\tilde \Phi(k)$ from flat space of a field $\tilde \Phi(k)$ is extracted from the string amplitude via the bulk-to-boundary  formula
      \be
 \delta\tilde \Phi(k)
  =  \left( - {i  \over k^2} \right) {\delta {\cal A}_{\Phi,{\cal O}}(k) \over \delta \Phi} 
  \ee
  with  $(- i  /k^2)$ denoting the massless closed string propagator for the (transverse) physical modes.

  We notice that the limit of zero momenta for open string states 
  is well defined only in the case the disk diagram cannot factorize into two or more disks via the exchange of open string states. Indeed, in presence of factorization channels 
  the string amplitude has poles in the open string momenta, the underlying world-sheet integral diverges and there is no corresponding SUGRA emission. We will always carefully choose the open string  polarizations  in such a way that no factorization channels be allowed. As a result, one finds that, up to $1/k^2$, $\delta \widetilde \Phi (k)$ is a polynomial in $k_i$ and therefore can be expressed as a linear combination
  of harmonic functions  $\tilde P_{i_1\dots i_n}(k) $ in momentum space
   \be
   \delta \widetilde \Phi (k)=  \sum_n  c^{\Phi}_{i_1\dots i_n}  \tilde P_{i_1\dots i_n}(k) 
   \ee
In this section we illustrate  the stringy description of the  micro-state geometry  for the very familiar case of parallel (wrapped) D3-branes. The vacua of this system are parametrised by the  expectation values of the (three) scalar fields $\phi_i$, along the non-compact directions. 
  The general open string condensate involves  an arbitrary number of insertions of the untwisted field $\phi_i$ on the boundary of the disk, where on the other hand twisted strings are the one connecting different branes. The associated micro-state
  geometry is described by a D3-brane solution characterized by a smooth multi-pole harmonic function determined  by the open string condensate  
  \cite{Klebanov:1999tb}.  The string computation will allow us to fix the relative normalization between the NS and RR vertices that will be used in the analysis of 2- and 4-charge intersections in the next sections.

   For concreteness we consider  the supergravity solution generated by a single stack of D3-branes labeled as $D3_0$ in the previous table. We dub them L solutions, defined as:
  
   \be
  V=L(x) \qquad\qquad M=K_I=0 \qquad\qquad L_I=1 
  \ee
   in (\ref{d34}). This results into
      \bea
   e^{-2U} &=& L^{1/ 2} \qquad\qquad  \tilde e_I=d\tilde y_I  \qquad\qquad \mu=0       \nn\\
   d A &=& {*_3}dL   \qquad    \alpha_0= A \qquad  \beta_0=L^{-1}\, dt  \qquad w=\beta_I=0   \qquad \alpha_I=-dt
   \eea
   The Type IIB supergravity solution reduces to  (discarding d-exact terms) \footnote{For a spherical symmetric   harmonic function
  $
   L(r)={Q/r}     
   $
   one finds  $A=Q\, \cos\theta \, d\varphi$    in spherical coordinates where $dx_i^2=dr^2+r^2( d\theta^2+\sin^2\theta\, d\varphi^2) $. }  
  \bea
 ds^2 &=&  L^{-{1/2}} \left(-dt^2+ \sum_{i=1}^3 dy_i^2 \right )  +  L^{{1/2}}  \sum_{i=1}^3 \left(   dx_i^2+  d\tilde y_i^2 \right)   \nn\\
 C_4 &=&  L^{-1}  \wedge dt\wedge dy_1 \wedge  dy_2 \wedge  dy_3  +A \wedge d\tilde y_1\wedge d\tilde y_2\wedge d\tilde y_3   \label{d34l}
 \eea

   At linear order in $\alpha_{D3}$ one finds 
     \bea
 \delta g_{MN} dx^M dx^N  &=&   { \delta L \over 2}  \left[  dt^2- \sum_{i=1}^3 (dy_i^2 -dx_i^2-d \tilde y_i^2)    \right] +\dots  \nn\\
  \delta C_4 &=&  -\delta L  \wedge dt\wedge dy_1 \wedge  dy_2 \wedge  dy_3  + A \wedge d\tilde y_1\wedge d\tilde y_2\wedge d\tilde y_3 +\dots
   \label{metricv}
 \eea
     with $\delta L=L-1$ and $A$ both or order $\alpha_{D3}$.  In particular one can take
   \be
  L=1+ {\alpha_{D3} N_0\over  |x| }   + \dots  \qquad ~~~~~~  {*_3}dL=d A
  \label{simpleL}
    \ee
     corresponding to $N_0$ D3 branes  on $\mathbf{R}^3$. In the next section, we will try to recover this SUGRA structure from the closed string emission of the brane.

   \subsubsection{NS{-}NS amplitude}

 Let's  consider a single D3-brane.
 The moduli space is parametrized in terms of  vacuum expectation value of open string scalar fields  $\phi_i$  describing the position of the brane in the transverse space. The associated supergravity solution can be extracted from disk amplitudes involving the  insertion of one closed string field and a bunch of open scalar fields $\phi_i$ at zero momenta.The open string background can be conveniently encoded in a generating function   
\be
\xi(\phi)=\sum_{n=0}^\infty \xi_{i_1\dots i_n } \phi^{i_1} \dots \phi^{i_n}
\ee
    
The relevant vertex operators are\footnote{ Here the correlator is evaluated for generic $n\neq 0$ and the $n=0$ term is obtained by extrapolation. A direct evaluation of this term  is subtler since there are not enough insertions to completely fix the $SL(2,\mathbf{R})$ world-sheet invariance.}  
\bea
W_{NS{-}NS}(z, \bar z) &=& c_{\rm NS} \,  (E R)_{MN}  \,e^{-\varphi} \psi^{M} e^{ik X} (z)  \, e^{-\varphi} \psi^{N} e^{ik  R X} (\bar z) \nn\\
V_{\xi(\phi)}({x}_a) &=& 
 \sum_{n=0}^\infty  \xi_{i_1\dots i_n}  \,   \partial X^{i_1}  (x_1) \, 
\prod_{a=2}^n\, \int_{-\infty}^\infty \,    {d{x}_a\over 2\pi  }  \,   \partial X^{i_a}  ({x}_a)
\eea
with $E=h + b$  the polarization tensor containing the fluctuation $h$ of the metric and $b$ of the B-field and $c_{NS}$ a normalisation constant; moreover $\varphi$ is a scalar field that bosonize the superghost, and $\psi$ and $X$ are the worldsheet fermions and bosons respectively.   The momentum of the closed string state will be labeled by $k$.  Henceforth we denote by $X(z)$ the holomorphic part of the closed string field $X(z,\bar z)=X(z)+R X(\bar z)$ with $R$ the reflection matrix implementing the boundary conditions on the disk. More precisely, $R$ is a diagonal matrix  with plus one ({+}1) and minus one ({-}1) along the Neumann 
  and Dirichelet directions, respectively. In particular for a disk of type $0$, we have plus along $t,y_I$ and minus along $x_i,\tilde y_I$, consequently $kR=-k$.

Left and right string sectors are related by:

\begin{equation}
X_{right}^{M}=R^{M}_{N} X^{N}_{left} \;\;\;\; \psi_{right}^{M}=R^{M}_{N} \psi_{left}^{N} \;\;\;\;  \varphi_{right}=\varphi_{left}
\end{equation}
 
   We choose length units whereby  $\alpha'=2$. 
  Moreover we exploit $SL(2,\mathbb{R})$ invariance of the disk to fix the positions of the closed string and the first of the open string insertions.
  Other open string vertices are integrated along the real line to take into account all possible orderings of the insertions.
  
  The resulting string amplitude can be written as 
\begin{equation}
\mathcal{A}_{NSNS,\xi(\phi)}=  \langle c(z) c(\bar z) c(z_1) \rangle  \left\langle  W_{NSNS}(z,\bar z)      V_{\xi(\phi)}   \right\rangle
\end{equation}
    The basic contributions to the correlators are 
  \bea
  \langle c(z) c(\bar z) c(z_1) \rangle &=&(z-\bar z)(\bar z-z_1)(z_1-z)\nn\\
  \left\langle   e^{i k X}(z)  e^{-i k  X}  ( \bar z)   \, \partial X^{i_a}  ({x}_a)  \right\rangle   &=&    i  k^{i_a}  \,    \left( {1\over z-{x}_a}-{1\over \bar z-{x}_a  } \right) =
   i k^{i_a}   \,   { \bar z-z\over |z-{x}_a|^2 }   \nn\\
  \left\langle   e^{-\varphi} \psi^{M} (z) e^{-\varphi} \psi^{N}  ( \bar z)     \right\rangle   &=&  { \delta^{MN} \over (z-\bar z)^2}    \label{nsns1}
  \eea
 Using 
  \bea
 \int_{-\infty}^\infty   {dx_a}  {   (\bar z-z) \over   | z-x_a|^2 } =  - 2 \pi i
  \eea
  one finds 
  \be
\begin{boxed}{\mathcal{A}_{NS{-}NS, \xi(\phi) } = i \, c_{ NS}\, \, { tr} ({E}R)  {  \xi(  k)  } }\end{boxed}    
 \ee
  The asymptotic deviation from the flat metric can be extracted from   
  \be
  \delta \tilde g_{MN} (k) =  \left( -{  i \over k^2} \right) \sum_{n=0}^\infty  \,  \,  {\delta \mathcal{A}_{NS{-}NS, \phi^n } \over \delta h_{MN} } =  c_{\rm NS}
  { \xi(  k) \over  k^2 } \, (\eta R)_{MN}  
   \ee
    After Fourier transform one finds 
    \be
     \delta  g_{MN}= \int {d^3 k \over (2\pi)^3}  \delta\tilde g_{MN}=    -\frac{1}{2} (\eta R)_{MN}  \, \delta L (x)
    \ee
  with the identification 
    \bea
     \delta L (x) = -2 \,c_{\rm NS} \, \int {d^3 k \over (2\pi)^3}   { \xi(  k) \over  k^2 }   e^{ikx}    
  \label{lfourier}
    \eea
  We conclude that the function $\xi(\phi)$ codifies the complete multipole expansion of $L(x)$ in terms of  vacuum
   expectation values for untwisted scalar fields 
   \be
  \xi_{i_1\dots   i_n}= \langle {\rm tr} \, \phi_{i_1} \dots \phi_{i_n} \rangle 
  \ee
   In particular,  the harmonic function for a single D3-brane at position $a$, is  realized by taking
   \be
    \xi (\phi) \sim e^{i \, a\,  \phi}
   \ee
     that  reproduces the multipole expansion (\ref{1multi}) of $L(x)$ according to (\ref{lfourier}).  In summary, the one boundary computation found a SUGRA emission is compatible with the monopole   \eqref{simpleL}, as expected for a single stack of branes.

   \subsubsection{R{-}R amplitude}

Next we consider the R{-}R string amplitude
\begin{equation}
{\cal A}_{R{-}R, \xi(\phi)}=  \langle c(z) c(\bar z) c(z_1) \rangle  \left\langle  W_{R{-}R}(z,\bar z) 
 V_{\xi(\phi)}   \right\rangle
\end{equation}
with
\bea
W_{R{-}R}(z, \bar z) &=& c_{\rm R}\,   \mathcal{(FR)}_{\Lambda \Sigma} \, e^{-{\varphi\over 2} }  \, S^{\Lambda} e^{ik X}(z)\, e^{-{\varphi \over 2} } \, S^{\Sigma} \, e^{ik R X}(\bar z)  \nn\\
V_{\xi(\phi)}   &=&  \sum_{n=0}^\infty \, \xi_{i_1\dots i_n}  e^{-\varphi} \psi^{i_1}  (z_1)    \prod_{a=2}^n  \int_{-\infty}^{\infty}   \, {d{x}_a \over 2\pi }  \, \partial X^{i_a}  ({x}_a)
\eea
Here 
\be
S^\Lambda=e^{{i \over 2} (\pm \varphi_1 \pm \varphi_2 \dots \pm \varphi_5)}   \qquad {\rm with~even}~\#~{\rm of}~-'s 
\ee
represents a ten-dimensional spin field of positive chirality with $\Lambda = 1, \dots 16$ and 
\be
{\cal F}=\sum_{n} \frac{1}{n!} F_{M_1\dots M_n} \Gamma^{M_1\dots M_n} \qquad ~~~~~~ {\cal R} =  \Gamma^{t y_1 y_2 y_3}
\ee

where $F$ are the R-R field strengths and ${\cal R} $ is the reflection matrix in the spinorial representation.

The basic contributions to the correlators are given by the first two lines of (\ref{nsns1}) and
  \bea
&&  \left\langle   e^{-{\varphi \over 2}}  \, S^{\Lambda} (z)  e^{-{\varphi \over 2}}  \, S^{\Sigma} \,   ( \bar z)     e^{-\varphi }\, \psi^M(z_1)   \right\rangle   =  {1\over \sqrt{2} } { (\Gamma^M)^{\Lambda \Sigma} \over (z-\bar z) |z-z_1|^2} 
  \eea
 Altogether  one finds
 \be
\begin{boxed}{\mathcal{A}_{R{-}R, \xi(\phi) } = i \,{c_{\rm R} \over \sqrt{2}}   \, {\rm tr}_{16} (  {\cal C  R} ) {   \xi( k)} ={16 \, i \, c_{\rm R} \over \sqrt{2}} \, { \xi( k)} 
C_{t y_1 y_2 y_3}}\end{boxed}
 \ee
  where we wrote ${\cal F}=i k_i \Gamma^i {\cal C}$. Here  and below we will  always restrict ourselves to the `electric' components $C_{t M_1 M_2 M_3 }$ since the remaining `magnetic' components are determined
  by the self-duality of $F_5$. The   R-R fields  at the disk level is then determined from the corresponding variation of the string amplitudes
  \be
  \delta \tilde C_{t y_1 y_2 y_3}=\left( -{  i \over k^2} \right) \sum_{n=0}^\infty \,  {\delta \mathcal{A}_{R{-}R, \xi(\phi)  } \over \delta C_{t y_1 y_2 y_3}  } = 8 \,\sqrt{2}     \,c_{\rm R} \, {\xi(k)  
   \over k^2 } 
   \ee
   Choosing 
   \be
   c_{\rm R} ={ c_{\rm NS} \over 4 \sqrt{2}}      \label{crcns}
   \ee
   and using (\ref{lfourier}) one finds
   \be
   \delta C_{t y_1 y_2 y_3}=- \delta L(x)
   \ee
    in agreement with (\ref{metricv})  for an arbitrary harmonic function $L(x)$ specified by $\xi(k) $ via   (\ref{lfourier}).

 \subsection{ 2-charge microstates  }
 
 Next we consider the K solutions:
  \be
  K_3 =-M=K(x)  \qquad\quad  \mu=0 \qquad\quad L_I=V=1  \qquad\quad K_1=K_2 =0 
\ee
   For this choice one finds
 \bea
   e^{-2U}&=&1\qquad {*_3}dw  = - d K \qquad \tilde e_{1}=d\tilde y_1 \qquad \tilde e_{2}=d\tilde y_2 \qquad \tilde e_3 =d\tilde y_3 -K \,dy_3 \nn   \\
    \alpha_0 &=& \beta_I=0  \qquad\quad  \alpha_1=\alpha_2=-(dt+w)  \qquad\quad \alpha_3=-dt  \qquad\quad  \beta_0=dt  
   \eea
  and the resulting   supergravity solution reduces to (discarding d-exact terms) 
\bea
 ds^2 &=& -   ( dt+w)^2 +  \sum_{i=1}^3 \left(  dx_i^2 +dy_i^2 + d\tilde y_i^2 \right)   - 2 \,K \,dy_3 \,d\tilde y_3 +K^2\, dy_3^2  \nn \\
 C_4 &=&  - (dt+w)   \wedge  \tilde e_3\wedge  (dy_1 \wedge d\tilde y_2+ d\tilde y_1 \wedge d y_2  )   
  \label{d34k}
 \eea
  Similar solutions are found by turning on $K_1$ or $K_2$.   
 At  leading order in $\alpha_{D3}$, K solutions  reduce to
\bea
 \delta g_{MN} dx^M dx^N & = & - 2\, dt \, w     - 2  \,  K \,  dy_3 d\tilde y_3 +\dots  \label{deltag2} \\
  \delta C_4 &=&  (K\,  dt \wedge  dy_3    -w   \wedge  d\tilde y_3)\wedge  (dy_1 \wedge d\tilde y_2+ d\tilde y_1 \wedge d y_2  )   
\eea
  with 
  \be
  {*_3}dw=-dK
  \ee
  and $K$ a harmonic function starting at order $|x|^{-2}$.  For example one can take $K$  to be
\be
 K=    { v_i x_i \over  |x|^3}     + \dots  \label{k3}
\ee
 while
\be
w= \epsilon_{ijk} \,v_i  { x_j \, dx_k \over  |x|^3} + \dots  
\ee

This solution is interesting since it contains higher order multipoles, therefore corrections to the naive brane solution.

\subsubsection{NS{-}NS amplitude}

 In this section we compute the NS-NS disk amplitude generating the metric of the K solutions. The relevant string amplitude has been computed in \cite{Giusto:2009qq} for the case of D1-D5-branes and in \cite{Bianchi:2016bgx,Pieri:2016pdt} for the D3-D3' setup.   The system considered is a non threshold bound state of D-branes, meaning that the solution is not in a naive BPS superposition and individual branes cannot be freely separated (Higgs branch). From the microscopical point of view, this is signaled by the presence of a non zero open string condensate, or in other words of a non zero  v.e.v. for the fields associated to open strings stretched between different branes. 
 
  Consider a non-trivial open string condensate  
  \be
  {\cal O}^{AB}={ \rm tr } \, \bar \mu^{(A}  \mu^{B)} \, \xi(\phi)  
  \ee
    with $\bar \mu^A$ 
   a fermionic excitation of the open string  starting from D3$_0$ and ending on D3$_3$ and  $\mu^B$ a fermion excitation of the open string with opposite orientation. Here and below we denote with trace the sum over Chan-Paton indices along the boundary of the disk. The open string condensate microscopically encodes the different microstates of the black hole geometries. 
   The  condensate can be linked to the profile function $f(v)$ appearing as a displacement of the center of the $D1D5$ fuzzball harmonic functions or alternatively as the profile of the oscillation of the string in the U-dual F1-P (fundamental string with momentum) system \cite{Lunin:2001fv,Lunin:2001jy,Lunin:2002bj,Lunin:2002iz,Kanitscheider:2007wq}. More precisely the condensate is linked to the derivative of $f(v)$ \cite{Giusto:2009qq}, therefore it gives the dynamic part of the displacement, whether the fixed part can be encoded in the untwisted scalars, as we have already said. While here the condensate is an arbitrary real number, in a proper treatment based on the microscopic theory on the world-volume of the D-branes it will be a definite number, once all the underlying normalizations are fixed.

\begin{figure}
\begin{center}
\includegraphics[width=8cm]{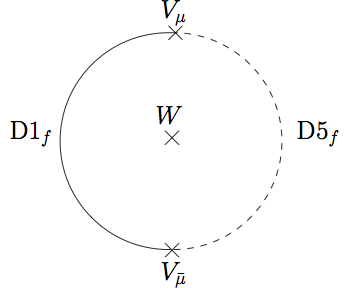} 
\end{center}
\caption{\cite{Giusto:2011fy} Worldsheet diagram for two stack of branes connected by twisted fermionic strings. The closed string vertex operator is inserted in the bulk. Here D1-D5 and D3-D3' are interchangeable. }
\end{figure}
   
   We organize the states in representations of the $SO(1,5)$ Lorentz symmetry rotating the six-dimensional hyper-plane along which the two stacks of branes are NN or DD, {\it i.e.}
    the space-time and the $y_3,\tilde y_3$ directions.  
 Upper (lower)  indices $A=1,\dots 4$ and $\hat  M=1,\dots 6$  run over the right (left) spinor and vector representations of this group, respectively. 
    The projection onto the symmetric part is required by the irreducibility of the string diagram since the anti-symmetric part of the fermionic bilinear 
   produces a   scalar field $\phi_{[AB]}$ and consequently a  factorization channel\footnote{More in detail: the fermionic condensate $ \mu^A \bar{\mu}^B$, where $A,B=1, \dots ,4$, factorizes as $\bf{4} \times 4 = 6 + 10$. The $\bf{6}$, the anti-symmetric part of the product, is simply the vectorial representation, corresponding to the $\Gamma^{\hat{M}}_{[AB]}$  ($\hat{M}=1,\dots ,6$) appearing in the fermionic kinetic term in the effective lagrangian, therefore giving a factorization channel for the $NSNS$-fermion-fermion diagram into two diagrams linked by a massless scalar field $\phi_{[AB]}$. To get a non factorizable diagram, one must restrict to the symmetric part of the product for which no fields with suitable indices are present. }.  
 
The relevant disk amplitude can be written as
 \bea
\mathcal{A}^{NS{-}NS}_{\mu^2, \xi(\phi)} &=&   \int     dz_4  \langle c(z_1)\,c(z_2)\,c(z_3) \rangle  \left\langle V_{\bar \mu}(z_1) \,V_{\mu}(z_2)  \,   W(z_3,z_4)  \,   V_{ \xi(\phi)}  \right \rangle
\eea
with 
    \bea
 V_{\bar\mu}(z_1) &=& \bar \mu^{A} \,e^{-\varphi /2}\,  S_{A}   \sigma_2 \sigma_3  \nn  \\
 V_{\mu}(z_2) &=&  \mu^{B} \,e^{-\varphi /2}\,  S_{B}    \sigma_2 \sigma_3     \\
 V_{\xi(\phi)}    &=&     \sum_{n=0}^\infty \xi_{i_1\dots i_n}  \prod_{a=1}^{n}\, \int_{-\infty}^\infty \, {dx_a \over 2\pi }  \,   \partial X^{i_a}  (x_a)      \nn\\
W(z_3,z_4) &=& c_{\rm NS}\, ({E}R)_{MN}  e^{-\varphi} \psi^{M} \, e^{i k X}(z_3)
(\partial  X^{N}+i \, k  \psi \, \psi^{N}) e^{-i k  X}(z_4) \nn
 \eea
Here $\sigma_I$ denotes the $\mathbb{Z}_2$-twist field along the $I^{\rm th}$ $T^2$ inside $T^6$ with conformal dimension 1/8 and
 \bea
 S_{A}=e^{ \pm \frac{1}{2}(i \varphi_3 \pm i \varphi_4 \pm i \varphi_5) }        \qquad { even~number~-'s} 
 \eea  
the spin field on the six-dimensional plane along which D3-branes are NN or DD.
 
 To evaluate the correlator one can consider a specific component and then use $SO(6)$ invariance to reconstruct its covariant form.  
For instance, if we take  $A=B=\frac{1}{2}(+++)$, the open string condensate contributes a net charge $+2$ along the first three complex directions, so that only the $\psi{:}\psi \psi{:}$ term can contribute to the correlator. The relevant correlators are  
 \bea
 &&  dz_4\langle c(z_1)\,c(z_2)\,c(z_3)\rangle = dz_4\, z_{12}\, z_{23}\, z_{31}
 = dw\, (z_{14} z_{23})^2   \nn\\
 &&     \left\langle e^{ik X}(z_3) e^{ -ik  X}(z_4)  \,  V_{\xi(\phi)}  \right \rangle  =\xi(k) \nn\\
&& \left\langle   e^{-\varphi/2}(z_1)   e^{-\varphi/2}(z_2)   e^{-\varphi}(z_3)     \right\rangle =  z_{12}^{-1/4} (z_{13} z_{23}  )^{-1/2}  \nn\\  
&& \left\langle    \sigma_2 \sigma_3 (z_1)   \sigma_2 \sigma_3 (z_2)   \right\rangle =   z_{12}^{ -1/2  } \nn \\
&& \left \langle S_{A}(z_1)  S_{B}(z_2)  \psi^{\hat M} (z_3)  \psi^{\hat N}\psi^{\hat P}(z_4)  \right\rangle  =  
   {  \Gamma^{\hat M \hat N \hat P}_{AB} \,  z_{12}^{3/4}        \over {2\sqrt{2}} (z_{13} z_{23} )^{1/2}   z_{14} z_{24}   }   
   \label{corr2ns}
 \eea
 with $w={z_{13} z_{24} \over z_{14} z_{23} }$. We notice that the scalar fields  factorize from the rest, since they are always oriented along the non compact space directions, while twist fields are along $T^6$.
 
  The $z_i$ dependence  boils down to the worldsheet integral
\be
{\cal I}=\int_\gamma  {dw \over w} =2\pi i   \label{ii2pi}
\ee
where $\gamma$ is a small contour around the origin.  On the other hand the open string condensate can be written as 
  \be
 \left\langle{\rm tr} \, \bar \mu^{(A}  \mu^{B)}  \right \rangle  = \frac{c_{\cal O} }{3!}\, v_{\hat M\hat N\hat P}\, (\Gamma^{\hat M \hat N \hat P})^{AB}  \label{mumu}
 \ee
 with $v_{MNP}$ a self-dual tensor  in six-dimensions   and $c_{\cal O}$ a normalisation constant. 
 Combining (\ref{mumu}) with (\ref{corr2ns}) and taking
 \be
 c_{\cal O} ={1\over \sqrt{2}\, {\cal I} \, c_{ NS}}    \label{co}
 \ee
  one finds
 \begin{equation}
\begin{boxed}{
\mathcal{A}^{NS{-}NS}_{\mu^2, \xi(\phi)}  =  \frac{1}{3!}\, (E R)_{\hat M\hat N} k_{\hat P}  \, v^{\hat M\hat N\hat P}\,  \xi(k) 
} 
\end{boxed}
  \label{nsamp2}
\end{equation}
 As before, the factor $\xi(k)$ comes from untwisted insertions and is associated  to higher multi-pole 
 modes of the underlying harmonic function, so  it is enough to match the leading term, {\it i.e.} we set $\xi(k)=1$. 
 Assuming  a $v_{\hat I\hat J\hat K}$ of the form 
  \be
 v_{y_3 \tilde{y}_3  3} = - v_{12t}= 4 \pi \, v   \label{v3}
  \ee
 and using (\ref{nsamp2}) one finds
\be
\delta \tilde g_{2t}=- 4\pi v\,  k_1     \qquad  \delta \tilde g_{1t}= 4\pi v \, k_2     \qquad     \delta \tilde g_{y_3 \tilde y_3}= -4\pi v\,  k_3     
\ee
  After Fourier transform one finds
 \bea
 \delta g_{2t} = -v\,  { x_1 \over  |x|^3}      \qquad  \delta g_{1t}= v { x_2 \over  |x|^3}     \qquad     \delta  g_{y_3 \tilde y_3}= -v\,  { x_3 \over   |x|^3}     
 \eea
   in agreement with (\ref{deltag2}) for the choice (\ref{k3}) with $v_i=\delta_{i3} \,v$. We conclude that the harmonic function $K_3$ describes  a particular component of the  fermion bilinear  condensates  connecting two branes parallel along the $(y_3,\tilde y_3)$-plane. Similarly $K_{1,2}$ represents  fermion bilinear condensates connecting branes parallel along
the 1 and 2 planes (tori). 

 Other choices for the 2-charge condensate give rise to solutions that do not belong to the $H_a$ harmonic family (\ref{d34}). In particular, taking non-zero values of $v_{t y_3 \tilde{y}_3}$, $v_{t y_3 i}$ or $v_{t \tilde{y}_3 i}$ give rise to solutions with non-trivial  $b_{ij}$, $g_{{y}_3 t}$ or $b_{\tilde{y}_3 t}$ components: 
 
 \bea
 v_{t y_3 \tilde{y}_3} \rightarrow None \qquad v_{ i j k} \rightarrow b_{ij}
 \eea
 
 \bea
 v_{t y_3 i} \rightarrow b_{{y}_3 t} \qquad v_{\tilde{y}_3 j k} \rightarrow b_{\tilde{y}_3 k}
 \eea
 
 \bea
v_{t \tilde{y}_3 i} \rightarrow g_{t \tilde{y}_3} \qquad v_{{y}_3 j k} \rightarrow g_{{y}_3 k}
\eea

\subsection{R{-}R amplitude}

Next we consider the disk amplitude with  the insertion of a R{-}R vertex operator
 \bea
\mathcal{A}^{R{-}R}_{\mu^2,\xi(\phi)} &=&    \int     dz_4  \langle c(z_1)\,c(z_2)\,c(z_3)\rangle   \left\langle V_{\bar\mu}(z_1) V_{\mu}(z_2)     W_{R{-}R}(z_3,z_4)     V_{\xi(\phi)} \right \rangle
\eea
with
\bea
W_{R{-}R}(z_3,z_4)&=&  c_{\rm R} \, \mathcal{(FR)}_{CD}   (e^{-{\varphi \over 2} }C^{C}C^{\dot{\alpha}} e^{i k X})(z_3)(e^{-{\varphi \over 2}} C^{D}C_{\dot{\alpha}}e^{- i k  X})(z_4)\nn
\eea

\bea
 C^{A} &=&e^{ \pm  \frac{1}{2}(i \varphi_3 \pm i \varphi_4 \pm i \varphi_5) }        \qquad {\rm odd~number~-'s}  \\
  C_{\dot \alpha}&=&e^{ \pm {i\over 2}(\varphi_4-\varphi_5) }  
   \eea  
The  ghost and untwisted bosonic correlators are given again by the first two lines in (\ref{corr2ns}) while the fermionic and twist-field correlators are
\bea
&& \langle S_{(A}(z_1) S_{B)}(z_2)  C^{C}(z_3) C^{D}(z_4)  \rangle = \delta_{(A}^{C} \delta_{B)}^{D} \left( \frac{z_{12} z_{34}}{z_{13} z_{14}z_{23} z_{24}} \right)^{3/4}   \nn\\
&&  \langle \sigma_2 \sigma_3(z_1) \sigma_2 \sigma_3(z_2) \rangle=z_{12}^{-1/2} \nn\\
 &&  \langle \prod_{i=1}^4 e^{-\varphi/2}(z_i)  \rangle=\prod_{i<j}^4  z_{ij}^{-1/4}   \nn\\
 && \langle C^{\dot{\alpha}}(z_3) C_{\dot{\alpha}}(z_4) \rangle=2 \, z_{34}^{-1/2}
\eea
 Again one finds that the amplitude is proportional to ${\cal I}$ given in (\ref{ii2pi}) and  can be written in the form  
\begin{equation}
\begin{boxed}{
\mathcal{A}^{R{-}R}_{\mu^2,\xi(\phi)} ={ 1 \over \, 3! \, 4}  \, v_{\hat M\hat N\hat P} \, {\rm tr}_4 \left( \mathcal{ FR}  \Gamma^{\hat M\hat N\hat P} \right)  \xi(k)
} \end{boxed}
\label{rrd3d3}
\end{equation}
  where   we  used  (\ref{mumu}), (\ref{co}) and (\ref{crcns}).  
  Taking $\xi(k)=1$, specializing to the condensate in (\ref{v3})  and focussing on the $C_{t \hat M_1\hat M_2 \hat M_3}$ components one finds
 \be
\mathcal{A}^{R{-}R}_{\mu^2,\xi(\phi)} =    4\pi v  \left(   F_{ t 3  y_1 \tilde y_2  y_3 } +  F_{ t 3  \tilde y_1  y_2  y_3 }  \right) 
\ee
Varying with respect to the four-form potential one finds
   \be
   \delta \tilde C_{t y_1 \tilde y_2 y_3 } =   \delta \tilde C_{t \tilde y_1 y_2 y_3 }=   k_3 \,  4\pi v   
   \ee
 reproducing the Fourier transform of  (\ref{deltag2})  for the choice (\ref{k3})  with $v_i=\delta_{i3} \,v$.

 \subsection{ 4-charge microstates  }
 Finally we consider solutions with 
  \be
  K_2=M=M(x)  \qquad  \mu=M \qquad L_I=V=1  \qquad K_1=K_3 =0 
\ee
   Now one finds
 \bea
   e^{-2U}&=&\sqrt{1-M^2} \quad \tilde e_{1}=d\tilde y_1+M \, dy_1  \quad \tilde e_{2}=d\tilde y_2 \quad \tilde e_3 =d\tilde y_3 +M \,dy_3 \quad w =0    \nn\\
   \alpha_0 &=& {-M dt\over 1-M^2}\qquad\alpha_1=\alpha_3= -dt\qquad\alpha_2=-dt+w_2\qquad{dw}_2=-{*_3}dM    \\
   \beta_0 &=& (1-M^2)\,dt+(1+M^2)\, w_2 \qquad \beta_1=\beta_3=M(dt-w_2)  \qquad \beta_2=M \,dt\nn
   \eea
 The Type IIB supergravity solution reads (discarding d-exact terms) 
 \bea
 && ds^2 = -   e^{2U} dt^2 +e^{-2U} \,  \sum_{i=1}^3 (dx_i^2+dy_i^2)    +e^{2U}   \,  \left[ \sum_{i=1,3}(d \tilde y_{i}   + M \, dy_{i} )^2      + d\tilde y_2^2 \right] \label{d34mu} \\
&& C_4 =- {1\over 1-M^2} \, dt  \wedge  \tilde e_3   \wedge  ( dy_1\wedge d \tilde y_2+ M\, d\tilde y_1\wedge d  \tilde y_2 )  \nn\\
&& ~~~~~~~~~~~+ w_2 \wedge ( dy_1   \wedge dy_2 \wedge dy_3 +     d\tilde y_1 \wedge dy_2\wedge  d\tilde y_3   )  
  \eea

    At  leading order in $\alpha_{D3}$, M solutions (\ref{d34mu}) reduce to
\bea
&&  \delta g_{MN} dx^M dx^N = 2 M \,  \left( dy_1\, \tilde dy_1+dy_3\, \tilde dy_3  \right)  +\dots \nn  \\
&& \delta C_4 =-  M\, dt  \wedge\, (     dy_1\wedge d \tilde y_2  \wedge d y_3  +   d\tilde y_1\wedge d  \tilde y_2\wedge d\tilde y_3 )  \nn\\
&& ~~~~~~~~ + w_2 \wedge ( dy_1   \wedge dy_2 \wedge dy_3 +     d\tilde y_1 \wedge dy_2\wedge  d\tilde y_3   )  
   +\dots\label{solm}
\eea
  with  $M$ a harmonic function starting at order $|x|^{-3}$.   For example one can take $M$  to be of the form
\be
 M=    v_{ij} {   3 \,x_i\, x_j -\delta_{ij} |x|^2 \over   |x|^5}     + \dots
\ee

\subsubsection{NS{-}NS amplitude}

We consider now string amplitudes on a disk with boundary on all four types of D3-branes. In particular we consider the insertions of four fermions $\mu_a$  
starting on a D3-brane of type $(a)$ and ending on a D3-branes of type $(a+1)$ with $a=0,1,2, 3$ (mod 4),  in a cyclic order.  
We notice that unlike the case of two boundaries now the condensate is complex. Indeed, even if each intersection preserves ${\cal N}=2$ SUSY (1/4 BPS), so that each fermion $\mu_a$ comes together with its charge conjugate $\bar \mu_a$, the overall configuration preserves only ${\cal N}=1$ SUSY (1/8 BPS) and fermions connecting all four type of D3 brane form two pairs of opposite chirality. The charge conjugate condensate can be defined by replacing each $\mu_a$ with its charge conjugate field $\bar \mu_a$  and running along the boundary of the disk  with the same cyclic order but in  reversed sense.  The real and imaginary parts of the string amplitude can be selected by turning on the
real and imaginary parts of the condensate respectively. 

We consider the following NS{-}NS amplitude 
\bea
\mathcal{A}^{NS{-}NS}_{\mu^4, \xi(\phi)}&=&  \langle c(z_1)\,c(z_2)\,c(z_4)\rangle \int dz_3 dz_5 dz_6  \\
&& \left \langle V_{\mu_1}(z_1) V_{\mu_2}(z_2) V_{\mu_3}(z_3) V_{\mu_4}(z_4)    W_{NSNS}(z_5,z_6)  V_{\xi(\phi)} \right\rangle\nn
\label{4p}
\eea
with 
\bea
 V_{\mu_1}(z_1) &=& \mu_1^{\alpha} \,e^{-\varphi /2}\,  S_{\alpha}  S_{1}  \,\sigma_{2} \sigma_3 (z_1)   \\
 V_{\mu_2}(z_2) &=& \mu_2^{\beta} \,e^{-\varphi /2}\,  S_{\beta}  S_{3} \sigma_{1} \sigma_2 (z_2)  \nn\\
 V_{\mu_3}(z_3) &=& \mu_3^{\dot\alpha} \,e^{-\varphi /2}\,  C_{\dot\alpha} \bar S_{1} \sigma_{2}\sigma_3   (z_3)  \nn\\
 V_{\mu_4}(z_4) &=& \mu_4^{\dot\beta} \,e^{-\varphi /2}\,  C_{\dot\beta} \bar S_{3} \sigma_{1} \sigma_2 (z_4)   \nn\\
  V_{\xi(\phi)}  &=&     \sum_{n=0}^\infty \xi_{i_1\dots i_n}  \prod_{a=1}^{n}\, \int_{-\infty}^\infty \, {dx_a \over 2\pi }  \,   \partial X^{i_a}  (x_a)     \nn\\
W_{NSNS}(z_5,z_6) &=& c_{\rm NS} (E R)_{MN}  (\partial X^{M}-i k \cdot \psi \psi^{M})e^{i kX}(z_5)(\partial  X^{N}+i k  \cdot  \psi \psi^{N}) e^{-i k  X}(z_6) \nn
 \eea
 We use the notation
 \bea
 S_I=e^{i \varphi_I \over 2} \qquad \bar S_I=e^{-{i \varphi_I \over 2} } \qquad   S_{\alpha}=e^{ \pm {i\over 2}(\varphi_4 +\varphi_5 ) }   \qquad   C_{\dot \alpha}=e^{ \pm {i\over 2}(\varphi_4-\varphi_5) }    
 \eea  
for internal and spacetime spin fields. The condensate is now the tensor
  \be
 {\cal O}^{\alpha\beta\dot\alpha\dot\beta} =  {\rm tr}  \, \mu_1^{(\alpha} \, \mu_2^{\beta)}\,  \bar\mu_3^{(\dot\alpha}\,  \bar\mu_4^{\dot\beta)} \, \xi(\phi)\label{omcond}
  \ee
where  the sum over all the Chan-Paton indices $\mu^1_{i_1}{}^{i_2} \mu^2_{i_2}{}^{i_3} \bar\mu^3_{i_3}{}^{i_4} \bar\mu^4_{i_4}{}^{i_1}$ is understood. As before  we discard components proportional to $\epsilon^{\alpha\beta}$ and 
  $\epsilon^{\dot\alpha\dot\beta}$ to ensure the irreducibility of the string diagram. As a result $ {\cal O}^{\alpha\beta\dot\alpha\dot\beta}$ transforms in the $({\bf 3},{\bf 3})$ of the
  $SU(2)_L\times SU(2)_R$ Lorentz group and can be better viewed as   a symmetric and traceless tensor $v^{\mu\nu}$.

\begin{figure}
\begin{center}
\includegraphics[width=8cm]{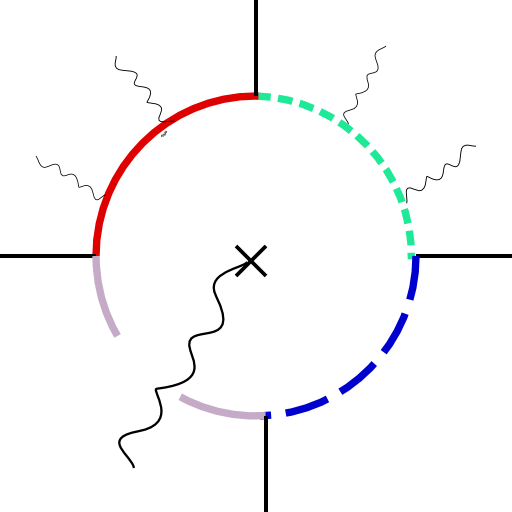} 
\end{center}
\caption{ Worldsheet diagram for four stack of branes connected by twisted fermionic strings. The closed string vertex operator is inserted in the bulk and multiple untwisted scalars can be inserted on the same stack.}
\end{figure}

To evaluate the correlator we specialize to $\alpha=\beta=\frac{1}{2}(++)$ and  $\dot{\alpha}=\dot{\beta}=\frac{1}{2} (-+)$. Only the four fermion piece of the closed string vertex can 
    compensate for the net $+2$ charge  of the open string condensate and two of these fermions have to be taken along  the fifth plane.    
        In addition since the net internal $U(1)^3$ charge of the condensate is zero, the only choices for the   $\psi^M \psi^N$ fermions in the closed string vertex 
  are $(M,N)=(I,\bar I)$ with $I=1,2,3,4$  

For example, taking  $(M,N)=(1,\bar 1)$ the relevant correlators are   
 \bea
 && \langle c(z_1)\,c(z_2)\,c(z_4)\rangle =  z_{12}\, z_{24}\, z_{41}  \nn \\
  &&     \left\langle e^{ik X}(z_3) e^{ -ik  X}(z_4)  \,  V_{\xi(\phi)}  \right \rangle  =\xi(k) \nn\\
&& 
\left\langle   \sigma_{3} (z_1)   \sigma_{1} (z_2)   \sigma_{3} (z_3)   \sigma_{1}  (z_4)  \right\rangle 
= (z_{13} z_{24})^{-1/4 }    \nn\\
&& \left\langle \prod_{j=1}^4 e^{-\varphi/2}(z_j) \right\rangle =
\prod_{i<j}  z_{ij}^{-1/4 } \nn\\ 
&& 
\left\langle  S_1   (z_1)   \bar S_1   (z_3)  \psi^1 (z_5)     \psi^{\bar 1} (z_6) \right\rangle \left\langle  S_3  (z_2)   \bar S_3  (z_4)      \right\rangle  =\frac{1}{2}  z_{13}^{-{1\over 4} }  z_{56}^{-1}   \left( { z_{15} z_{36} \over z_{16} z_{35}   } \right)^{1/2}   ( z_{24})^{-1/4 }   \nn\\
 && \left\langle    \sigma_{2} (z_1) \sigma_{2} (z_2)   \sigma_{2} (z_3)  \sigma_{2}  (z_4)  \right\rangle =   f\left(\frac{z_{14} z_{23} }{z_{13} z_{24} } \right)  
 \left({ z_{13} z_{24} \over  z_{12} z_{23} z_{34} z_{41} } \right)^{1/4 }  \nn\\
&& \left \langle S_{(\alpha}(z_1) S_{\beta)}(z_2) C_{(\dot \alpha}(z_3) C_{\dot\beta)}(z_4) \psi^{ \mu} (z_5)     \psi^{ \nu} (z_6)  \right\rangle   =  
  { (z_{12} z_{34} )^{1/2} z_{56} \over  
2 \prod_{i=1}^4 (z_{i5} z_{i6})^{1/2}  }  \sigma^{(\mu}_{\alpha \dot{\alpha}  }  \sigma^{\nu)}_{\beta \dot \beta}   \label{corr4ns}
 \eea
 with $f(x)$ the four twist correlator \cite{Cvetic:2003ch}: 
 \be
 f(x)= { \Lambda(x)   \over   (F(x)F(1-x))^{1/2} } 
 \ee
 where $F(x) ={}_2F_1(1/2,1/2; 1; x)$ is a hypergeometric function and 
 \be
 \Lambda (x) =  \sum_{n_1,n_2}  e^{- {2\pi \over \alpha'}   \left[   { F(1-x) \over F(x) } \, n_1^2 R_1^2  +    { F(x) \over F(1-x) } \, n_2^2 R_2^2   \right]}    
 \ee
 accounts for the classical contribution  associated to world-sheet instantons\footnote{ 
 In the supergravity limit $\sqrt{\alpha'} << R_{1,2}$, worldsheet instantons are exponentially suppressed  and one can simply take $\Lambda(x)\to 1$}, where $R_x$ is the radius of the $x$ torus, $x = 1,2$. 
 
 Assembling all the pieces together and taking
  \be
  z_1=-\infty \qquad z_2=0\qquad z_3=x \qquad z_4=1 \qquad z_5=z \qquad z_6=\bar{z}
  \ee
  one finds that the string amplitude is proportional to the integral 
      \bea
 I_1 &=&  \int_0^1    \,     dx  \,  f(x)\,  {\cal I}_1  (x)
   \label{ii0}
  \eea
  with
  \be
   {\cal I}_1 (x) = \int_{\mathbb{C}^+}  { d^2z \over |z(1-\bar z)|(x- z)} 
 \label{i1}
  \ee
  Similarly for other choices of $(M,N)=(I,\bar I)$, $I=2,3,4$,  one finds integrals of the form (\ref{ii0}) with ${\cal I}_1$ replaced by 
  \bea
  {\cal I}_2 (x)&=&   \int_{\mathbb{C}^+}  { d^2z \over |z(1-\bar z)(x- z)|} \nn\\
  {\cal I}_3 (x) &=& \int_{\mathbb{C}^+}  { d^2z \over \bar z (1- z) |x- z|}  \nn\\ 
  {\cal I}_4 (x)  &=&   \int_{\mathbb{C}^+}  { d^2z \over \bar z(1-z)(x- z)} 
  \label{i234}
 \eea
 The integrals (\ref{ii0}),  (\ref{i1}), (\ref{i234}) can be computed numerically for arbitrary values of the radii and one always finds  a finite result,  with ${\cal I}_{2}, {\cal I}_4$ real and   ${\cal I}_{1}, {\cal I}_3$
 purely imaginary\footnote{Similar integrals appear in \cite{Stieberger:2009hq}.} and $ {\cal I}_{1}= {\cal I}_3 $ 
   
   Noticing that real and imaginary parts of the amplitude contribute to symmetric $({E}R)_{(MN)}$ and antisymmetric parts $({E}R)_{[MN]}$ 
 we conclude that a purely imaginary  $ v^{ij} $ generates the string amplitude 
 \bea
\begin{boxed}{
\mathcal{A}^{NS{-}NS}_{\mu^4, \xi(\phi)}=   \left[ ({E}R)_{[1\bar 1]} +  ({E}R)_{[3\bar 3]}  \right] k_{i} k_{j}  \,    v^{ij} \, \xi(k)
}\end{boxed}  
\label{a4nsns}
\eea
with
 \be
 \left\langle {\rm tr} \, \mu_1^{(\alpha} \, \mu_2^{\beta)}\,  \bar\mu_3^{(\dot\alpha}\,  \bar\mu_4^{\dot\beta)}  
     \right\rangle = {2 \pi v^{ij}  \over c_{\rm NS} \, {\cal I}_1} \,  \sigma_i^{\alpha\dot\alpha } 
   \bar\sigma_j^{\beta \dot\beta}   
  \ee
   We notice that $({E}R)_{[1\bar 1]}$ and $({E}R)_{[3\bar 3]} $ always involve a Neumann and a Dirichlet direction, so only the metric contributes
   to the antisymmetric part of the matrix.  Varying with respect to the symmetric part $h_{ij}$  of  the polarization tensor $E=h+b$ one finds
\bea
\delta \tilde g_{1\bar 1 } &=&\delta \tilde g_{3\bar 3 }  = -2\pi   i \,  v^{ij} \, \frac{\, k_{i} k_{j} }{k^2}    \, \xi(k) 
\eea
 that reproduces (\ref{solm}) after Fourier transform. 
 
We conclude this section by noticing that there  other choices for the 4-charge condensate besides (\ref{omcond}) lead to solutions that go beyond the eight harmonic function family 
(\ref{d34}). On the one hand one can consider  different orderings of the branes along the disk. These condensates lead again to solutions of type M but with internal coordinates that are permuted among one another. Notice that some of these solutions can look very different from each other since $L_I$ and $V$ do not enter in a symmetric fashion.
On the other hand, turning on the real part of the condensate (\ref{omcond}) will produce solutions with a non-trivial B-field given in terms of the real worldsheet integrals ${\cal I}_2$ and ${\cal I}_4$.  Finally turning one can consider condensates of type
 \bea
 {\cal \widetilde O}^{\alpha\dot\alpha\beta\dot\beta}&=&{\rm tr}  \, \mu_1^{(\alpha} \,  \bar\mu_2^{(\dot\alpha}\, \mu_3^{\beta)}\,  \bar\mu_4^{\dot\beta)} \,  \nn \\
\hat{\cal O}^{(\alpha\beta\gamma)\dot\beta} &=&{\rm tr}  \, \mu_1^{(\alpha} \,  \mu_2^{\beta}\, \mu_3^{\gamma)}\,  \bar\mu_4^{\dot\beta} 
 \eea
These condensates generate a new class of solutions with off-diagonal terms mixing either two different $T^2$'s inside $T^6$ or space-time and internal components.

\subsubsection{R{-}R amplitude}

Finally we consider the amplitude with the insertion of a R{-}R vertex operator
\bea
\mathcal{A}^{R{-}R}_{\mu^4, \xi(\phi)}&=&  \langle c(z_1)\,c(z_2)\,c(z_4)\rangle \int dz_3 dz_5 dz_6  \\
&& \left \langle V_{\mu_1}(z_1) V_{\mu_2}(z_2) V_{\mu_3}(z_3) V_{\mu_4}(z_4)    W_{R{-}R}(z_5,z_6)  V_{\xi(\phi)} \right\rangle\nn
\label{ampl4D3R{-}R}
\eea
with 
\bea
 W_{R{-}R}^{(-1/2,+1/2)}(z_5,z_6)&=& c_{\rm R}\,  (\mathcal{FR} \Gamma_M)_\Lambda{}^\Sigma \, e^{-{\varphi \over 2} } S^{\Lambda}\, e^{ik X} (z_5) e^{\varphi \over 2}   (\partial X^{M}+ i  k  \psi \psi^{M} )  C_{\Sigma} \, e^{ -ik  X}  (z_6)\nn
\eea
the R-R vertex operator in the asymmetric   (-1/2,+1/2) super-ghost picture.
  Notice that $ \Psi^M_\Sigma (k) = {:}k\psi \psi^{M}C_{\Sigma}{:}$ is a world-sheet primary field of dimension 13/8, belonging to the {\bf 128} of $SO(1,9)$, with $C_{\Sigma}$ the spin field of negative chirality.
  Again taking $\alpha=\beta=\frac{1}{2}(++)$, $\dot\alpha=\dot\beta=\frac{1}{2}(-+)$, it is easy to see that the net charge $-2$ of the condensate can be only compensated by the  
   $ k  \psi \psi^{M}$ term in the closed string vertex   with   $S^{\Lambda}$ and $C_\Sigma$ both carrying charge $-\frac{1}{2}$ along the fifth 
   direction. This leads to eight choices for  $S^{\Lambda}$. 
   On the other hand, once $S^{\Lambda}$ is chosen, the charges of $ \Psi^M_\Sigma (k) = {:}k\psi \psi^{M}C_{\Sigma}{:}$ are completely determined by charge conservation.  
   For example, for the choice
  \bea
  S^\Lambda &=&  e^{{i\over 2} (\varphi_1+\varphi_2-\varphi_3+\varphi_4-\varphi_5)}  \nn\\
   k \psi \, \psi^M\, C_\Sigma &=&k_{\bar 5} \psi^{\bar 5} \psi^{\bar 1} e^{{i\over 2} (\varphi_1-\varphi_2+\varphi_3-\varphi_4-\varphi_5)} 
  \eea
  one finds again the first four lines in (\ref{corr4ns}), while the last  four lines are replaced by
   \bea
&& \langle  \prod_{i=1}^5 e^{-{\varphi \over 2}}(z_i) e^{\varphi \over 2}(z_6)  \rangle=\prod_{j=1}^5  z_{j6}^{1\over 4} \prod_{i<j}^5  z_{ij}^{-{1\over 4} }  \\ 
 && \left\langle  S_1   (z_1) S_3 (z_2) \bar S_1 (z_3) \bar S_3   (z_4)   S_1 S_2  \bar S_3 (z_5)\, \bar S_1 \bar S_2  S_3     (z_6) \right\rangle =   \left( { z_{15} z_{45} z_{26} z_{36}  \over z_{13} z_{24} z_{25} z_{35} z_{16} z_{46} z_{56}^3  }\right)^{1\over 4}  \nn\\
 && \left \langle S_{(\alpha}(z_1) S_{\beta)}(z_2)    C_{(\dot \alpha}(z_3) C_{\dot\beta)}(z_4)   C_{\dot \gamma} (z_5)   \psi^\mu S_{ \gamma} (z_6)  \right\rangle \nn\\
 &&~~~~~~~= \frac{ 1 }{\sqrt{2}} 
     \left( {z_{12} z_{34}  z_{56} \over z_{35} z_{45}   z_{36} z_{46}  z_{16}^2 z_{26}^2  } \right)^{1\over 2}  \epsilon_{\gamma(\alpha} \sigma_{\beta)(\dot\beta}^\mu   \epsilon_{\dot \alpha)\dot\gamma }\nn
   \eea
Combining the various correlators and taking
  \be
  z_1=-\infty \qquad z_2=0\qquad z_3=x \qquad z_4=1 \qquad z_5=z \qquad z_6=\bar{z}
  \ee
 one finds that the string amplitude is again given by the integral (\ref{ii0}) involving ${\cal I}_1$.  A similar analysis can be performed for other choices of spin-field components (labelled by
 $\Lambda, \Sigma$) leading to similar answers in terms of the four characteristic integrals ${\cal I}_1, \dots {\cal I}_4$ on $\mathbf{C}$. 
 The results are summarised in table 2.   Since ${\cal I}_2$, ${\cal I}_4$ 
 are real and ${\cal I}_1, {\cal I}_3$ purely imaginary,  a purely imaginary condensate selects the ${\cal I}_1$, ${\cal I}_3$ and ${\cal \bar I}_1$, ${\cal \bar I}_3$ components.   
 \begin{table}[h]
\begin{center}
\begin{tabular}{|c|c|c|c|c|}
\hline
 & ${\cal I}_1 $ & ${\cal I}_2$ & ${\cal I}_3$ & ${\cal I}_4$ \\ 
 \hline
 $\Lambda$ &  $\frac{1}{2}(++-+-) $  & $ \frac{1}{2}(+++--)$ & $ \frac{1}{2}(-+++-)$ & $\frac{1}{2}(+-++-)$ \\
 \hline
 & ${ \cal \overline{I} }_1 $ & ${\cal \bar I}_2$ & ${\cal \bar I}_3$ & ${\cal \bar I}_4$ \\ 
 \hline
  $\Lambda$    &  $\frac{1}{2}(--+--)$  & $ \frac{1}{2}(---+-)$ & $ \frac{1}{2}(+----)$ & $\frac{1}{2}(-+---)$ \\
 \hline
\end{tabular}
\end{center}
\label{tspinor}
\caption{Contributions to the string correlator of the various spinor components.  }
\end{table}%
  The resulting string amplitude can then be written as
 \be
\begin{boxed}
{ \mathcal{A}^{RR}_{\mu^4, \xi(\phi)}=\frac{ 1}{4 } {\rm tr}_{16} (\mathcal{F\, R} \, {\cal P}\, \Gamma^i    )  2 \pi  v_{ij}  k_j \, \xi(k)
}\end{boxed}
 \ee
 with 
 \be
 {\cal P}   = \frac{1}{2}(1-\Gamma^{y_1\tilde y_1 y_3\tilde y_3} )  \Gamma^{y_2 \tilde y_2} 
 \ee
 a projector on  ${\cal I}_1$, ${\cal I}_3$ and ${\cal \bar I}_1$, ${\cal \bar I}_3$ components   
  with $+i$ and $-i$ eigenvalues respectively. These are precisely the eigenvalues of the matrix ${\cal P}$ justifying our claim. 
 Using ${\cal R}=\Gamma^{t y_1 y_2 y_3}$ one   finds
 \be
\begin{boxed} { \mathcal{A}^{RR}_{\mu^4, \xi(\phi)}=2 \pi i  \,  k_i v_{ij}  k_j \, \xi(k)  \left(  C_{t  y_1  \tilde y_2  y_3} + C_{ t   \tilde y_1 \tilde y_2 \tilde y_3}          \right) }\end{boxed}
 \ee
 Taking derivatives with respect to $C_4$ one finds agreement with (\ref{solm}) after Fourier transform.

 \pagebreak

\section{Smooth Horizonless geometries in Supergravity}

Even though in the previous chapter we found a dictionary between open string condensates and sugra solutions, to find an explicit fuzzball with a string interpretation we need to make a consistent ansatz in SUGRA and verify that the solution is indeed smooth and horizonless. We reviewed how for five dimensional black holes the mechanism of regularization involves the presence of a Kaluza-Klein monopole (also known as Euclidean Taub-NUT): this space is free of curvature singularities and the throats that open up near the centers of the monopoles are not infinitely extended, but instead cap smoothly at a finite distance due to the shrinking of a compact dimension. One may wonder if a regular four charge fuzzball solution can be found regular directly in four dimensions. Unfortunately the topology  is rather limited in four dimension \cite{Breitenlohner:1987dg,Gibbons:2013tqa}, in fact we lose the extra 5d angular variable that was shrinking going near the horizon, so some new regularization mechanism is required. One possible loophole to get a regular solitonic solution is to employ more than  one asymptotic region. Even though some asymptotically $AdS_2$ solutions  have been found \cite{Lunin:2015hma}, an extension to asymptotically flat seems problematic and it's nevertheless unclear whether a wormhole like solution could fit as a description for a regular fuzzball while mimicking a black hole \cite{Pieri:2016cqz}ì, therefore in this section we will focus on finding a four charge solutions that are regular in higher dimension, but still requiring only 3+1 dimension to be non compact and only one asymptotic region. 

We consider again the BPS $D3^4$ system, this time compactified to four dimensions (see the appendix for details), mainly following the work \cite{Bianchi:2017bxl}. 
The four-dimensional geometries can be viewed as solutions of an ${\cal N}=2$ truncation of  ${\cal N}=8$ supergravity involving the gravity multiplet and three vector multiplets. The  scalars $U_I$ in the vector multiplets,  usually referred as STU, parametrise the complex structures of the three internal $T^2$'s and span the moduli 
space  ${\cal M}_{STU}= [SL(2,R)/U(1)]^3 \subset  E_{7(+7)}/SU(8)= {\cal M}_{{\cal N} =8}$. Setting $16\pi G=1$, the lagrangian can be written as  
\eq
\lagr= \sqrt{g_4}\left( R_4-\sum_{I=1}^3 \frac{\partial_\m U_I\partial^\m \bar{U}_I}{2\pt{\Imm U_{I}}^2}-\frac{1}{4}F_a \mathcal{I}^{ab}F_{b}-\frac{1}{4}F_{a}\mathcal{R}^{ab} \widetilde{F}_{b} \right)
\feq
where $F_{a}=dA_a$, $\widetilde F_{a}=*_4 F_a$ with $a=0,1,2,3$, including the graviphoton, and 
\begin{align}
U_I=(\si+is,\tau+it,\nu+i u )
\end{align}
 are the three complex scalars in the vector multiplets. In these variables the gauge kinetic functions read  
\begin{align*}
\mathcal{I}^{ab}&= stu \pmat 1+\frac{\si^2}{s^2}+\frac{\tau^2}{t^2}+\frac{\nu^2}{u^2} & \:\:{-}\frac{\si}{s^2}&\:\:{-}\frac{\tau}{t^2}&\:\:{-}\frac{\nu}{ u^2} \\
-\frac{\si}{s^2}  & \frac{1}{s^2} &0 &0 \\
-\frac{\tau}{t^2}  & 0 &\frac{1}{t^2}&0\\
-\frac{\nu}{ u^2}  & 0 & 0 &\frac{1}{u^2}\fpmat	 \quad	 
\mathcal{R}^{ab}=\pmat 2\si\tau\nu &\:\:{-}\tau\nu&\:\:{-}\nu \sigma& \:\:{-}\sigma\tau \\
-\tau\nu & 0 & \nu &\tau \\
-\nu \sigma  &\nu &0&\sigma\\
-\sigma\tau   &\tau &\sigma &0 \fpmat \num
\end{align*}
  The solutions  will be written in  terms of eight harmonic functions 
    \eq
  \{ V, L_I, K^I, M \} \label{harmh}
   \feq   
  on $\mathbb{R}^3$. It is convenient to introduce  the  combinations
\begin{align}\label{pzm}
Z_I&=L_I+\frac{|\ve_{IJK}|}{2}\frac{K^JK^K}{V},\nn\\
\m&=\frac{M}{2}+\frac{L_IK^I}{2V}+\frac{|\ve_{IJK}|}{6}\frac{K^IK^JK^K}{V^2}. 
\end{align}
 Here
  $\epsilon_{IJK}$ characterise the triple  intersections  among the three $T^2_I$ 2-cycles in $T^6$.
 
 The solutions can then be written as
\begin{align} \label{4dsolution}
ds^2&=-e^{2U}\pt{dt+w}^2+e^{-2U} d|\mbf{x}|^2  \nn\\
A_a& =(A_0,A_I)=w_a +a_a\pt{dt+w}  \nn\\    
U_I &= -b^I+ i \pt{V e^{2U}Z_I}^{-1}
   \end{align}  
with
\begin{align}
b^I  &={K^I\over V} -\frac{\m}{Z_I}   \quad  \quad a_0=-\m V^2 e^{4U}  \quad  \quad a_I= V e^{4U} \, \left(  -{Z_1 Z_2 Z_3\over Z_I}  +K^I \, \m    \right)  \nn\\
   *_3dw_a&=-dK_a   \quad \quad    K_a=\left(-V, K^I\right) \quad \quad
*_3dw=\half\pt{VdM-MdV+K^IdL_I-L_IdK^I}  \label{4dsolution0}
\end{align}
and
\bea
&&e^{-4U} =  {\cal I}_4(V,L_I,K^I,M)\equiv Z_1Z_2Z_3V-\m^2V^2 = \\
&&L_1\, L_2 \, L_3 \, V- K_1\, K_2\, K_3\, M +\sum_{I>J}^3   {K^I K^J  L_I L_J \over 2}- {M V\over 2}  \sum_{I=1}^3 K^I L_I    -\frac{M^2 V^2}{4}
- \sum_{I=1}^3 {(K^I)^2 L_I^2 \over 4}  \nn
\eea
which has the same structure as the quartic U-duality invariant.

For general choices of the eight harmonic functions, the solution (\ref{4dsolution}) is singular. Both naked and `horizon-dressed' curvature singularities can be present. 
The generic solution is characterised by a mass $\mathfrak M$,   associated to the Killing vector $\xi^{(t)M} \partial_M= \partial_t$,  four  electric charges $Q^a$ and four magnetic charges $P_a$. Introducing the symplectic vector 
\bea
{\cal F}= \left(
\begin{array}{c}
  F_a  \\
 {\delta {\cal L}\over \delta F_a}   \\
\end{array}
\right) = \left(
\begin{array}{c}
  F_a  \\
  \star_4 \, \mathcal{I}^{ab} \, F_b-\mathcal{R}^{ab}\, F_b   \\
\end{array}
\right)
\eea
    one finds for the charges
    \bea
\mathfrak M &=& -\frac{1}{8\pi G}   \int_{  S^2_{\infty}  } \star_4 \,  d \xi^{(t)}    
     \nn\\
  \left(
\begin{array}{c}
  P_a  \\
  Q^a   \\
\end{array}
\right)    &=&      -\frac{1}{4 \pi}   \int_{  S^2_{\infty}  }  {\cal F}
\eea
 with  $\xi^{(t)} =\xi^{(t)}_M\, dx^M   $ and  $S^2_{\infty}$   the two-sphere at  infinity.   Solutions with extra symmetries   arise for special choices of the harmonic functions. Axially symmetric solutions are characterised by the existence of an additional Killing vector $\xi^{(t)M} \partial_M= \partial_\phi$ associated to rotations around an axis in $\mathbb{R}^3$, and carry an extra quantum number, the angular momentum $J$ given by\footnote{Being a surface integral, the expression for $J$ holds true even when $\xi^{(\phi)}$ is only an asymptotic Killing vector. $J$ is the $z$ component of the vector $\mbf{J}$, defined in (\ref{jjj}).}
   \be
J = -\frac{1}{16 \pi G}   \int_{  S^2_{\infty}  } \star_4 \,  d \xi^{(\phi)}  \label{jj}
     \ee
  with  $\xi^{(\phi)} =\xi^{(\phi)}_M\, dx^M$. Spherical symmetric solutions are invariant under rotations around the origin and are characterized by  zero angular momentum. 

 In this paper we consider fuzzballs of spherically symmetric (single center) black holes, that is solutions that have the same asymptotic charges as single center black holes, but differ from them when approaching the would-be horizon.  The harmonic functions specifying the general spherically symmetric  solution can be written in the single-center form
   \begin{align}
 V=v_0+{v\over r}    \qquad     L_I={\ell}_{0I}+{{\ell}_I\over r}  \qquad  K^I=k_{0}^I+ {k^I\over r} \qquad M=m_0+{m\over r}  \label{bcH}
 \end{align}
  and describe a general system of intersecting D3-branes wrapping three cycles on $T^2\times T^2\times T^2$ with one leg on each of the three $T^2$.  
  The absence of Dirac-Misner strings requires that $w$ vanishes at infinity or, equivalently, that  $*_3dw \sim r^{-3} $ at infinity leading to the constraint 
  \be
  v_0 \, m-m_0 \, v+ k_{0}^I \, {\ell}_I-{\ell}_{0I}\, k^I=0 \label{dmstring}
  \ee
   For simplicity  we take $m_0=m=0$. For this choice one finds
  \be
  e^{-4U}= V \, L_1\, L_2\, L_3-\frac{1}{4} \left( \sum_{I=1}^3 K^I L_I \right)^2 
  \ee
   Poles and zeros of this function are associated to horizons and curvature singularities respectively. If  $e^{-4U}>0$ for all $r>0$  the solution describes a black 
   hole with near horizon geometry $AdS_2\times S^2$ and entropy proportional to
 \be
 \lim_{r\to 0} r^4\, e^{-4U}  = {\cal I}_4(v, {\ell}_I, k^I, m=0)  =  v \, {\ell}_1\, {\ell}_2\, {\ell}_3 -\frac{1}{4} \left( \sum_{I=1}^3 k^I {\ell}_I \right)^2>0
 \ee
    If $e^{-4U}$ has zeros for some positive $r$, the solution exposes a naked singularity.   

The charges of the solution (or its fuzzball) are computed by the integrals (\ref{charges}) evaluated in the  asymptotic geometries (\ref{bcH}). Writing the 
three-dimensional metric   in spherical coordinates
\be
ds^2=-e^{2U} (dt+w)^2+e^{-2U}\,  (dr^2+r^2 \, d\theta^2+r^2\,\sin^2\theta\, d\phi^2)
\ee
 and setting $G=(16\pi)^{-1}$ one finds for the charges\footnote{In our conventions $\star_4 \, dr\wedge dt=e^{-2U}\, r^2\, \sin\theta \, d\theta\wedge d\phi$ and 
 $\int \sin\theta d\theta \wedge d\phi  =4\pi$.}
    \bea
\mathfrak M &=&  8\pi  \, {r^2 \, \partial_r \, e^{2U}}  
     \nn\\   
\left(
\begin{array}{c}
  P_a  \\
  Q^a   \\
\end{array}
\right)    &=&        
\left(
\begin{array}{c}
  (v,-k^I)^T \\
    -r^2 \, \partial_r\left( {\cal I}^{ab}\,  a_b + \mathcal{R}^{ab}\, K_b\right)    \\
\end{array}
\right)   
   \label{charges}
\eea
   where we used the fact that at infinity $w=0$ and $F_a=dw_a+d a_a \, dt$.  
   On the other hand the angular momentum of the  fuzzball is computed by the integral (\ref{jj}). 
   The evaluation of this integral requires a more detailed knowledge of the asymptotic geometry since angular momentum arises from the first dipole mode in the expansion of the harmonic function. Indeed, denoting by
   \be
 H =h_0+  {h_1\over r} +{ \mbf{h}_2 \cdot \mbf x \over r^3}      
   \ee 
    one finds for the angular momentum
    \be
{\mbf J} =4\pi\,    \left[ m_0 \,\mbf{v}_2- v_0 \,{\mbf m}_2 + {\ell}_{0I} \, \mbf{k}^I_{2} - k_{0}^I \,\boldsymbol{\ell}_{2I}  \right]   \label{jjj}
   \ee    
    We anticipate here that apart from the scaling solutions, all fuzzball solutions we will find here carry a non-trivial angular momentum.  We observe that for orthogonal branes  angular momentum is carried by K-components (see Appendix \ref{appN1basic} for details)  corresponding to open string condensates on disks with boundary on two different D3-brane stacks.  Indeed, an explicit  microscopic  description  of the general supergravity solution  exists if the harmonic functions satisfy the boundary conditions  \cite{Bianchi:2016bgx}
    \be
     m_2+\sum k_{2}^I=0   \label{micro}
    \ee

Let us focus on the following harmonics: 
  \begin{align}
 V=1+{v\over r}    \qquad     L_I= 1+{{\ell}_I\over r}  \qquad  K^I= M= 0  \label{bcH1}
 \end{align} 
 describing a system of four stacks of D3-branes intersecting orthogonally on $T^6$.  At large distances one finds
 \bea
 e^{-2U} &=& \sqrt{VL_1L_2L_3}=1+{( v+{\ell}_1+{\ell}_2+{\ell}_3)\over 2\,r}  +\ldots   \nn\\
 a_I &=& -L_I^{-1}=-1+{{\ell}_I\over r}+\ldots    \qquad   a_0=0 \nn\\ 
 U_I &=&  i \pt{V e^{2U} L_I}^{-1} = i+\ldots 
 \eea
  leading to 
    \bea\label{massort}
\mathfrak M &=& 4\pi\, ( v+{\ell}_1+{\ell}_2+{\ell}_3) 
     \\
    \label{chargeort} \left(
\begin{array}{c}
  P_a  \\
  Q^a   \\
\end{array}
\right)    &=&        
\left(
\begin{array}{c}
  (v, 0,0,0)^T \\
   (0,{\ell}_1,{\ell}_2,{\ell}_3)^T  \\
\end{array}
\right)   
\eea
  The extremal Reissner Nordstrom solution corresponds to the choice ${\ell}_I=v=Q/2$, or equivalently 
  \be
  L_I=V=1+{\ell \over  r}   \qquad M=K^I=0
  \ee 
   after the identification  $ F_{\rm RN}=\frac{1}{2} \left(   * F_0+\sum_{I=1}^3 F_I  \right)$.

\subsection{Regularity Analysis in Higher Dimensions}

 In this section we review the Bena-Warner  multi-Taub NUT ansatz \cite{Bena:2007kg}
 for fuzzball geometries of four- and five-dimensional black holes and in the next section we will present explicit horizon-free solutions with three centers.

The four-dimensional solution (\ref{4dsolution})  lifts to an eleven dimensional solution representing a systems of intersecting M5-branes with four electric and four magnetic charges. The eleven dimensional metric is given by:
\bea
ds^2=ds^2_5+ds^2_{T^6}  \label{11d}
\eea
where
\bea
ds_5^2 &=& -(Z_1\, Z_2\, Z_3)^{-{2\over 3}} \left[dt+\mu(d\Psi+w_0) +w\right]^2 + (Z_1\, Z_2\, Z_3)^{1\over 3}\left[ V^{-1}  (d\Psi+w_0)^2+ V d|\mbf{x}|^2 \right] \nn\\
\vec{\nabla} \times \vec{\omega_0} &=& \vec{\nabla} V\nn \\
ds_{T^6} &=& \sum_{I=1}^3 \left( { Z_1\, Z_2\, Z_3 \over Z_I^3 } \right)^{ 1\over 3} (dy_I^2+d\tilde y_I^2)  
\label{11h}
\eea
in which $\lbrace t, \, \mbf x ,\, \Psi, \, y_I, \,\tilde y_I \rbrace$ with $I=1,2,3$ are the coordinates of $\mathbb{R}\times \mathbb{R}^3\times S^1 \times T^6$, respectively.
The 11d BPS solutions just written comes from a more general solution in which the 5d metric is given by

\bea
ds_5^2 &=& -(Z_1\, Z_2\, Z_3)^{-{2\over 3}} \left[dt+\mu(d\Psi+w_0) +w\right]^2 + (Z_1\, Z_2\, Z_3)^{1\over 3} h_{\mu \nu }dx^{\mu}dx^{\nu} \ \nn
\eea

with $h_{\mu \nu }$ a 4d hyper-kalher metric. The choice for  $h_{\mu \nu }$  performed in \eqref{11h} is a Gibbons-Hawking metric, that is an $U(1)$ fibration over a flat base $\mathbf{R}^3$. The behaviour of $V$ is crucial, taking 

\bea
 V &=& v_0+\sum_{i=1}^N  {q_i\over r_i }  \qquad  ,  \qquad    
 \eea
 
  there are orbifold singularities at the location of the centers. Indeed if we zoom in near the location of the centers $\vec{x}^{(j)}$ and define a local frame centered there, with radial coordinates $\rho = 2 \sqrt{\vec{x}-\vec{x}^{(j)}}$, the metric is locally:
  
  \bea
  ds_4^2 = d\rho^2 + \rho^2 d\Omega^2_3(q_j)
  \eea
  
  with  $d\Omega^2_3(q_j)$ being the metric on $S^3/\mathbb{Z}_{|q_j|}$. In the following we will consider $|q_j|=1$ so that the space is locally $\mathbb{R}^4$, even though orbifold singularities arising from $|q_j| \in \mathbb{Z}$ are  under control in string theory. If $v_0 = 0$ the whole 5d solution is asymptotically $\mathbb{R}^{1,4}$, suitable for a 5d black hole, but since we are interested in 4d black holes we will consider $v_0 \neq 0$, leading to a KK-monopole (or equivalently, an euclidean Taub-NUT) near the centers

\bea
ds^2 &=& \left[ V^{-1}  (d\Psi+w_0)^2+ V d|\mbf{x}|^2 \right] \nn\\
\vec{\nabla} \times \vec{\omega_0} &=& \vec{\nabla} V
\eea

 and with the final 5d space that is asymptotically $\mathbb{R}^{1,3}\times S^1$.

  Micro{-}states of the four dimensional black holes can be generically defined as smooth geometries with no horizons or curvature singularities in eleven dimensions carrying the same mass and charges as the corresponding black hole; regular solutions can be constructed in terms of multi-center harmonic functions $(V,L_I,K^I,M)$ with the positions of the centers and the charges chosen such that  $Z_I$ are finite and $\mu =0$ near the centers. Under these assumptions one finds that the eleven dimensional metric (\ref{11d})  near the centers is $\mathbb{R}\times T^6\times \mathbb{R}^4/\mathbb{Z}_{|q_i|}$.  
  Moreover, the  absence of horizons and closed time{-}like curves  requires that
    \bea
  && Z_I V >0 \qquad {\rm and} \qquad   e^{2U}>0
  \eea
   Let us remark that the  condition $Z_I  V>0$  near the centers requires 
  \be
\left.Z_I\,V\right|_{r_i=0}=q_i\,\left({\ell}_{0I}+\sum_{j\ne i}\frac{{\ell}_{I,j}}{r_{ij}}\right)+{\ell}_{I,i}\left(v_0+\sum_{j\ne i}\frac{q_j}{r_{ij}}\right)+|\varepsilon_{IJK}| k^J_i\left( k_{0}^K+\sum_{j\ne i}\frac{k_j^K }{r_{ij}}\right) >0
  \ee
It turns out that these necessary conditions are often sufficient to ensure the positivity of both $Z_I V$ and $e^{2U}$ on the whole $\mathbb{R}^3$. 
  In the next section we look for explicit  solutions of these requirements satisfying the boundary conditions (\ref{bcH}). We stress that the resulting solutions are regular everywhere in five dimensions and fall off at infinity to $\mathbb{R}^{1,3}\times S^1$. The four-dimensional fuzzball solution  follows from reduction of this five-dimensional geometry down to four dimensions where the singularity in the geometry is balanced by a blow up of the scalar fields.

  Let us analyze in more detail the regularity conditions near the centers. Consider N-center harmonic functions following the Bena-Warner ansatz \cite{Bates:2003vx,Bena:2007kg,Gibbons:2013tqa}
 \bea
 V &=& v_0+\sum_{i=1}^N  {q_i\over r_i }  \qquad  ,  \qquad      L_I = {\ell}_{0I}+ \sum_{i=1}^N   { {\ell}_{I,i} \over  r_i }  \nn\\ 
 K^I &=& k^I_{0}+\sum_{i=1}^N  {k^I_{i}\over r_i }  \qquad  ,  \qquad M =  m_0+\sum_{i=1}^N   { m_i \over  r_i }  \label{ansatz0}
 \eea
 with $ r_i =|{\bf  x}_i -{\bf x}|$  and  ${\bf x}_i$ the position of the $i^{\rm th}$ center. We notice that $({\ell}_{Ii} , m_i)$ and $(q_i,k^I_i)$ describe the electric and 
 magnetic  fluxes of the four dimensional gauge fields through the sphere encircling  the $i^{\rm th}$- centers, so Dirac quantisation requires that  
 they be quantised.  Here we adopt units such that they are all integers. 
 Alternatively, one can think of the eight charges as parametrising the number of D3-branes wrapping one of the eight three{-}cycles with a leg on each of the three tori $T^2_I$ in the factorisation $T^6 = T^2_1 \times T^2_2\times T^2_3$. In other words, in our units  each charge describes the number of D3-branes of a certain kind.
 
 We look for regular five dimensional geometries behaving as $\mathbb{R}\times {\rm Taub{-}NUT}$ near the centers. It is easy to see that $w$ vanishes near the centers, so the Taub-NUT geometry factorises if $Z_I$ are finite and $\mu$ vanishes near the centers, i.e. 
  \bea
 Z_I\big|_{r_i\approx 0}  \approx  \zeta^I_i    \qquad , \qquad
      \mu  \big|_{r_i \approx 0} \approx  0
 \eea
  with $\zeta^I_i$ some finite constants.  The conditions that $Z_I$ is finite near the centers  can be solved by taking
 \bea
  {\ell}_{I,i} = -   {|\epsilon_{IJK}|\over 2}  {k^J_{i} \, k_i^K\over q_i  }   \qquad , \qquad
    m_i  =    {k^1_i \, k^2_{i} \, k_i^3\over q_i^2 } \label{halfbps} 
 \eea
The vanishing of $\mu$ near the centers boils down to the so called {\it bubble equations}
 \bea \label{eq: bubble equations}
 \sum_{j=1}^N \, {\Pi_{ij}  \over  r_{ij} }+v_0 \,{k^1_i\, k^2_i\, k^3_i \over q_i^2 } - {\ell}_{0I}\, k^I_i  -  
|\epsilon_{IJK}| \, {  k_{0}^I \, k_i^J \, k_i^K \over 2\,q_i} -m_0 q_i=0
 \eea
 where $r_{ij}=| {\bf x}_i-{\bf x}_j|$ and
\be
  \Pi_{ij}=   q_i \,q_j \prod_{I=1}^3 \left( {k^I_i\over q_i}-{k^I_j\over q_j}  \right)  \qquad\qquad     \ee
represents the symplectic form $P^a_i Q_{aj} - P^a_j Q_{ai}$ that counts the number of D3-brane intersections.  The bubble equations  are also known as "integrability conditions" in the the context of 4d black hole physics \cite{Denef:2000nb,Denef:2002ru,Bates:2003vx} and ensure the absence of Dirac-Misner strings. To see this, we notice that using 
  the bubble equations, the $w$ function  defined by (\ref{4dsolution0}) can be written in the form
   \bea
   *_3 dw &=&  \sum_{i,j=1}^N ( q_{[i} \, m_{j]}   +k^I_{[i} \, {\ell}_{j],I }   )  {1\over r_i}d  {1\over r_j } +\half\sum_{i=1}^N  \left( v_0\, m_i-m_0 \, q_i- {\ell}_{0I}\, k_i^I +
    k_{0}^I \, {\ell}_{i,I} \right)\, d {1\over r_i }  \nn \\  
     &=&  \half\sum_{i,j=1}^N\, \Pi_{ij}\, \,\left(   {1\over r_j} -{1\over r_{ij} }  \right)  d{1\over r_i}   \label{dwdm}
   \eea
    where in the second line we used equations (\ref{halfbps}), (\ref{eq: bubble equations}) and $A_{[BC]}$ means $\frac{1}{2}(A_{BC}-A_{CB})$. The solution can be written in the form
      \bea
   w  &=& \quar\sum_{i,j=1}^N\, \Pi_{ij} \,\omega_{ij}   \label{wiwij}
   \eea 
    in terms of the one forms $\omega_{ij}$ defined via the relation
\eq 
*_3 d\om_{ij}=\pt{\frac{1}{r_j}-\frac{1}{r_{ij}}}d\frac{1}{r_i}-\pt{\frac{1}{r_i}-\frac{1}{r_{ij}}}d\frac{1}{r_j},
\feq
that is
     \be
   \omega_{ij} =\frac{\pt{\mbf{n}_i+\mbf{n}_{ij}}\cdot\pt{\mbf{n}_j-\mbf{n}_{ij}}}{ r_{ij} }d\phi_{ij}   \label{wij}
   \ee
with
 \be
       {\bf n}_i ={{\bf x}- {\bf x}_i   \over r_i}   \qquad      {\bf n}_{ij} ={ {\bf x}_i -{\bf x}_j \over r_{ij} } \nn\\
   \qquad 
 d{\phi}_{ij} =  \frac{ {\bf n}_{ij}\times {\bf n}_i \cdot d{\bf x}}  {    r_i  \left[ 1- ({\bf n}_{ij} \cdot {\bf n}_i)^2 \right]   }.  
\ee
  It is easy to see that $\omega_{ij}$ is free of Dirac-Misner singularities. Indeed along the dangerous lines connecting any two centers the numerator of  (\ref{wij}) always vanish so no string{-}like singularity arises. One can also see that near the centers $w$ goes to a constant exact form. 
 
  Finally we notice that  if the coefficients $k^I_i$ satisfy the  relation
   \be
   v_0 \, m_i - m_0 q_i - {\ell}_{0I}\, k^I_i  +  
 k_{0}^I \,  {\ell}_{Ii} =0
    \label{scaling}
   \ee
   the system of equations is invariant under  overall rescalings of the center positions ${\bf x}_i \to \lambda {\bf x}_i$. These solutions are known as ``scaling solutions".  Multiplying equation (\ref{scaling}) by the positions of the centers ${\mbf x}_i$ and summing one finds that  the scaling solutions satisfy 
   \be
    m_0 \,\mbf{v}_2- v_0 \,{\mbf m}_2 + {\ell}_{0I} \, \mbf{k}_{2}^I - k_{0}^I \,\boldsymbol{\ell}_{2I} =0
   \ee    
   and therefore according to (\ref{jjj}) they carry zero angular momentum.

\subsection{Regular Solutions}

  We look for regular geometries with the asymptotics (\ref{bcH1}), i.e.   
  \be
  {\ell}_{0I}=v_0=1  \qquad  m_0=m=k_{0}^I=k^I=0
  \ee
  For concreteness we take $q_i=1$. 
   The  charges  of the fuzzball solutions are then
  \bea
P_0 = N \qquad , \qquad
Q_I =    -    \sum_{i=1}^N \, {   |\epsilon_{IJK}|   k^J_{i} \, k_i^K \over 2}  \label{chargesortho}
 \eea 
    The  solution is specified by the  the positions ${\bf x}_i$ of the centers and the fluxes  $k_i^I$. The positions of the centers are constrained by the bubble equations
  \bea
       &&  \sum_{j \neq i}^N  { \Pi_{ij}  \over  r_{ij} } + \Lambda_i  -\Gamma_i  =0   
                       \label{eqsf}
   \eea
   with
\be
\Pi_{ij}=\prod_{I=1}^3 (k^I_i -k^I_j)  \qquad\qquad \Gamma_i=\sum_{I=1}^3 k^I_{\ i}\qquad\qquad \Lambda_i=k^1_{\ i}k^2_{\ i}k^3_{\ i}
\ee
   while the consistency with the boundary conditions, namely that $K$ and $M$ fall off faster than $r^{-1}$, require
    \bea
         \sum_{i=1}^N \,k_i^I=   \sum_{i=1}^N \,   k^1_i \, k^2_{i} \, k_i^3  =0 \label{bcasymp}
    \eea
  Configurations with one or two centers fail to meet the requirement $Q_I>0$, so  we will consider solutions with three centers.

The bubble equations  (\ref{eqsf})  for three centers can be solved in general by taking
\be
r_{12}=\frac{\Pi_{12}\,r_{23}}{\Pi_{23}-r_{23}\left(\Gamma_2 -\Lambda_2\right)}\qquad\qquad r_{13}= \frac{\Pi_{13}\,r_{23}}{-\Pi_{23}+ r_{23}\left(\Gamma_1+\Gamma_2-\Lambda_1-\Lambda_2 \right)}     \label{rij3}
\ee
  A solution given by (\ref{rij3}) makes sense if the distances $r_{ij}$ between the three centers are positive and they satisfy the triangle inequalities.  This  restricts significantly the choices for the $k_i^I$. A quick scan over the integers shows that boundary conditions are solved only if at least one of the fluxes $k_i^I$ vanishes.
  Without loss of generality the general solution can then be parametrised in the form (up to permutations of rows and columns)
   \bea
 k^I {}_i &=&
\left(
\begin{array}{ccc}
-\kappa_1\,\kappa_2 & -\kappa_1\,\kappa_3   & \kappa_1\,(\kappa_2+\kappa_3)  \\
\kappa_3 & \kappa_2  & -\kappa_2-\kappa_3  \\
 -\kappa_4&  \kappa_4 &  0
\end{array}
\right)  
\eea
 Consequently the harmonic functions takes the general form
 \bea
V &=& 1+\sum_{i=1}^3 {1\over r_i}   \qquad   M= \kappa_1 \kappa_2\kappa_3\kappa_4 \, \left( { 1\over r_1}-{1\over r_2}\right) 
\nn\\
 L_1 &=&1+  \kappa_4 \left( {\kappa_3 \over r_1}-{\kappa_2\over r_2} \right)  \qquad  L_2 =1+  \kappa_1 \kappa_4 \left(  -{\kappa_2\over r_1}+{\kappa_3\over r_2} \right) \nn \\
 L_3 &=&1+  \kappa_1 \left( {\kappa_2\kappa_3 \over r_1}+{\kappa_2 \kappa_3\over r_2} +{(\kappa_2+\kappa_3)^2 \over r_3} \right) \qquad
K_1 =  \kappa_1 \left( -{\kappa_2 \over r_1}-{\kappa_3\over r_2}+{\kappa_2+\kappa_3\over r_3} \right) \nn\\
  K_2 &=&  {\kappa_3 \over r_1}+{\kappa_2 \over r_2}-{\kappa_2+\kappa_3 \over r_3}  \qquad 
  K_3 =  \kappa_4 \left( -{1\over r_1}+{1\over r_2} \right)  \nn
 \eea
 The charges and distances between the centers reduce to
 \bea
Q_1&=& \kappa_4(\kappa_3-\kappa_2)\qquad  Q_2= \kappa_1\kappa_4(\kappa_3-\kappa_2)\qquad  Q_3= \kappa_1(\kappa_2^2+4\kappa_2\kappa_3+\kappa_3^2) \nn\\
r_{12}&=& \frac{2\kappa_1\kappa_4(\kappa_2-\kappa_3)^2r_{23}}{\kappa_1\kappa_4(2\kappa_2^2+5\kappa_2\kappa_3+2\kappa_3^2)+(\kappa_2+\kappa_4-\kappa_1\kappa_3+\kappa_1\kappa_2\kappa_3\kappa_4)r_{23}}\nn\\
r_{13}&=& \frac{\kappa_1\kappa_4(2\kappa_2+\kappa_3)(\kappa_2+2\kappa_3)r_{23}}{\kappa_1\kappa_4(2\kappa_2^2+5\kappa_2\kappa_3+2\kappa_3^2)-(\kappa_1-1)(\kappa_2+\kappa_3)r_{23}}.
\eea

The scaling solution corresponds to the choice 
\be
\kappa_2=0  \qquad    \kappa_1=1  \qquad \kappa_3=\kappa_4=\kappa
\ee
One finds
 \bea
 k^I {}_i &=&
\left(
\begin{array}{ccc}
 0 & -\kappa   & \kappa  \\
 \kappa & 0  & -\kappa  \\
  -\kappa & \kappa  &   0
\end{array}
\right)   \qquad   r_{12}=r_{23}=r_{13}=\ell  \nn\\
P_0 &=& 3  \qquad ~~~~~~~~~ Q_1= Q_2=Q_3=\kappa^2    
 \label{kk}
 \eea
for any given $\ell$.   The regularity conditions  become  
  \begin{align}
& e^{-4U} =1+\frac{r_1 r_2+r_1
   r_3+r_2 r_3}{r_1 r_2 r_3}+\frac{\kappa^2 \left(r_1 r_2+r_1 r_3+r_2r_3+3 r_1+3 r_2+3 r_3\right)}{r_1 r_2 r_3}\nn\\
&+\frac{\kappa^4 \left(r_1+r_2+r_3+9\right)}{r_1 r_2
   r_3}  +\frac{\kappa^6 \left(2 r_1 r_2+2 r_1 r_3+2 r_2r_3+ r_1 r_2 r_3-r_1^2 -r_2^2-r_3^2\right)}{r_1^2 r_2^2
   r_3^2}>0\nn\\
   Z_1V &=1+\frac{ r_1 \,r_2 +r_1\, r_3+ r_2\, r_3 + \kappa^2 (2 \, r_2 +2 r_3-r_1+r_2\, r_3  ) }{r_1r_2r_3}>0
\end{align}
 The conditions  $Z_2 V>0$ and  $Z_3 V>0$ follow from  $Z_1V >0$ and the permutation symmetry of the system. 
The two conditions can be shown to be satisfied using  the  property
 \be
r_1+r_2-r_3 \geq 0
 \ee
 satisfied by the distances $r_i$ from any  point $x\in \mathbb{R}^3$ to the three vertices of an equilateral triangle. This inequality can be proved using triangle inequalities.  
  
   We conclude that the five-dimensional geometry defined by the multi-center solution is regular everywhere. We notice that the  fluxes satisfy the scaling condition
   (\ref{scaling}) and consequently a rigid rescaling of the positions of the centers generate a new solution; similar scaling solution has been already identified as interesting microstate candidates in \cite{deBoer:2008zn,Bena:2012hf}.  More precisely, the moduli space of solutions with this charge is spanned by a single continuous parameter $\ell$ and permutations of the rows or columns of the matrix (\ref{kk}). 
    There are 12  inequivalent choices corresponding to the $3!$ permutations of the entries in the first line in (\ref{kk}) times the two choices for the position of the 0 in the second line. The remaining entries are determined by the conditions that the sum along rows and columns of the matrix $k_i^I$ should vanish.  The solutions satisfy $\sum_{I=1}^3 \mbf{k}^I_{2}=\mbf{m}_2=0$ and   
therefore according to (\ref{jjj}) and (\ref{micro}) they carry zero angular momentum and admit a microscopic description in terms of orthogonal intersecting D3-branes along the lines of \cite{Bianchi:2016bgx}.

The analysis  above can be repeated for more general choices of the fluxes but  regularity conditions in general can only be verified numerically.  For instance other class of solutions are:

\begin{itemize}

\item{ $\kappa_2=0$,  $\kappa_1=\kappa_3=1$, $\kappa_4=\kappa:$
 \bea
 k^I {}_i &=&
\left(
\begin{array}{ccc}
0 & -1   & 1  \\
 1 & 0  & -1  \\
-\kappa&  \kappa &  0
\end{array}
\right)   \qquad   r_{13}=r_{23}\qquad r_{12}=\frac{2\, \kappa\, r_{23}}{2\,\kappa +(\kappa-1)\,r_{23}} \nn\\
P_0&=& 3 \qquad Q_1=Q_2=\kappa\qquad Q_3=1  
\eea
 Interestingly, triangle inequalities in this case do not constrain $r_{23}$ that can take arbitrarily large value. 
 
 }
 
  \item {$\kappa_2=\kappa_4=\kappa$, $\kappa_1=1$,\  $\kappa_3=2\,\kappa$ 
  \bea
 k^I {}_i &=&
\left(
\begin{array}{ccc}
-\kappa & -2\,\kappa   & 3\,\kappa \\
2\kappa & \kappa  & -3\,\kappa \\
 -\kappa&  \kappa &  0
\end{array}
\right)   \qquad   r_{12}=\frac{r_{23}}{10 + r_{23}}\qquad r_{13}=r_{23}.\nn\\
P_0&=& 3 \qquad Q_1=Q_2=\kappa^2\qquad Q_3=13\,\kappa^2  
\eea
 As before $r_{23}$  can take arbitrarily large value. 
}

 \item{ $\kappa_2=0$, $\kappa_1=3\,\kappa$,\  $\kappa_3=2\,\kappa$,   $\kappa_4=\kappa$
  \bea
 k^I {}_i &=&
\left(
\begin{array}{ccc}
0 & -3\,\kappa   & 3\,\kappa  \\
 \kappa & 0  & -\kappa  \\
-2 \kappa&  2\kappa &  0
\end{array}
\right)   \qquad   r_{12}=\frac{12\, \kappa^2\, r_{23}}{12\,\kappa^2 -r_{23}} \qquad r_{13}=\frac{6\, \kappa^2\, r_{23}}{6\,\kappa^2 -r_{23}}  \nn\\
P_0&=& 3 \qquad Q_1=2\,\kappa^2\qquad Q_2=6\,\kappa^2\qquad Q_3=3\,\kappa^2  \nn\\
r_{23} &<& 6\,(2-\sqrt{2})\,\kappa^2 
\eea
 We notice that triangle inequality in this case imposes an upper bound on $r_{23}$ leading to a moduli space of finite volume.
 
}

\end{itemize}

\pagebreak

\section{From Supersymmetric quantum mechanics to Supergravity}

In the string theory scattering computation we came across the open fields vevs signaling the bindings between different branes. In the calculations present in literature one usually works in a simplified settings with trivial closed string moduli, in which the open string theory has flat directions traslating into arbitrary open string vevs. In a generic point of the moduli space the flat directions should lift and only a discrete number of points associated to the allowed open string configurations remain. The main goal of this final section is to sketch how the open string vevs can be computed in a generic point of the closed string moduli space.

If we are only interested in counting the number of microstates it's enough to limit ourselves to the classical level of the SQM arising on the worldvolume of the branes and count the number of supersymmetric ground states. These microstates can  be directly related to the one found in SUGRA and in string theory, even though in general the relation can be subtle and it is an active area of research. A nice  overview has been given in  \cite{Denef:2002ru}\footnote{ See also \cite{Balasubramanian:2006gi} in which the correspondence has been extended to M-theory } particularly in the setup of two center solutions: performing the limit $g_s \rightarrow 0$ we first go  from a  SUGRA configuration in the large scale regime to a quiver QM (QQM) description. Indeed after quantum corrections, the Coulomb branch ($x_i^{(a)} \neq 0$, where $x_i$ are the scalars in the vector multiplet) of the QQM is actually identical to the SUGRA “solution space,” as both are subject to the Bubble equations as a necessary condition for being BPS. If we keep on lowering $g_s $, the  system  decay into a configuration with nonzero vevs for the chiral fields: we go from the Coulomb to the Higgs Branch. This Coulomb-Higgs map  for two centers  has been found to be one-to-one \cite{Denef:2002ru}, while with three centers the relation is more involved   \cite{Bena:2012hf}. In particular there are some states called "pure-Higgs" that do not map to Coulomb-SUGRA, but can be mapped to SUGRA if they are scaling solutions \footnote{This states live in the middle cohomology of the moduli space of classical solution ( D and F constraints) and  they grow exponentially with the charges, much faster than the Coulomb states.}.  Finally, at  $g_s = 0$, the chiral fields can be interpreted as the moduli of a suitable geometric object, namely open stings of the intersecting  susy D-Branes.

%

In the next subsection we will discuss about the classical counting of the microstates for the purely D-brane system, while  later we will worry about quantum corrections.

\subsection{Microstate counting}

In this subsection we follow~\cite{Chowdhury:2014yca,Chowdhury:2015gbk} and review the derivation of the QQM describing the BPS states for the purely $1/8 \, BPS$ D-branes brane system in the D6-D2-D2-D2 frame. We consider four stacks of D-branes in type IIA string theory compactified on $T^6$  as summarized in table~\ref{tab1}.

\begin{table}[h]
\begin{center}
\begin{tabular}{|c|c|c|c|c|c|c|c|c|c|c|}
\hline
        Brane     &  $t$       &$x_1$   & $x_2$  &$x_3$  & $y_1$ &$y_2 $  &$y_3$  &$y_4$   & $y_5$ &$y_6$          \\
\hline
 $ D2_1 $      & $-$      &$.$     & $.$   &$.$    & $-$   &$-$    &$.$    &$.$    & $.$   &$.$        \\
  $ D2_2 $      & $-$      &$.$     & $.$   &$.$    & $.$   &$.$    &$-$    &$-$    & $.$   &$.$        \\
   $ D2_3 $      & $-$      &$.$     & $.$   &$.$    & $.$   &$.$    &$.$    &$.$    & $-$   &$-$        \\
$ \overline{D6}_4$           &  $-$      &$.$    & $.$   &$.$    & $-$  &$-$    &$-$   &$-$    & $-$   &$-$  \\
\hline
\end{tabular}
\end{center}
\caption{The $y$-coordinates parametrize a $T^6$, while $t,\;x_i$ are Cartesian coordinates of $\mathbf{R}^{1,3}$. We consider three stacks of orthogonal D2-branes and a stack of anti D6-branes wrapped on $T^6$: Neumann and Dirichlet directions of the D-branes are indicated by lines and dots respectively.}
\label{tab1}
\end{table}

Globally this D-brane configuration preserves four supercharges, which are characterised by a ten-dimensional Majorana-Weyl spinor $\eta_{10}$ in ten dimensions satisfying the chirality  conditions
\begin{equation}
  \label{eq:supc-proj}
  \Gamma^4 \Gamma^5 \Gamma^6 \Gamma^7 \eta_{10} = 
  \Gamma^6 \Gamma^7 \Gamma^8 \Gamma^9 \eta_{10} = -\eta_{10}\;,
\end{equation}
where the $\Gamma$'s are the 10D Gamma matrices. In the explicit representation of appendix \eqref{notaz}, the solutions to~\eqref{eq:supc-proj} yield the string supercharges
\begin{equation}
    Q^\alpha =  e^{-\frac{\varphi}{2}}  \, S^{\alpha}  \, e^{-\frac{i}{2}\sum_{\mathbf{j}=1}^3 h_\mathbf{j}}\;,~~~ 
  \bar{Q}^{\dot \alpha} =  e^{-\frac{\varphi}{2}}  \, S^{\dot{\alpha}} \, e^{\frac{i}{2}\sum_{\mathbf{j}=1}^3 h_\mathbf{j}}\;,
  \label{qcrea}
\end{equation}
where $h_{\mathbf{j}}$ are the bosonised fields related to the Ramond-Neveu-Schwarz (RNS) fermions in the plane $(y_{2\mathbf{j}-1},\;y_{2\mathbf{j}})$, $\varphi$ is the superghost and $S^\alpha$ ($S^{\dot{\alpha}}$) are $\mathbf{R}^{1,3}$ spin fields of positive (negative) 4D chirality. 

The theory describing the dynamics of the massless excitations of open strings connecting the various D-branes is a ${\cal N}=4$ SQM with Gauge group $U(N_1)\times U(N_2)\times U(N_3)\times U(N_4)$; the field depend only on time since the overall world volume is zero plus one dimensional. Here $N_a$ are the number of branes in every stack and in the following we will focus on configurations will small $N_a$ for which calculations can be done explicitly.  The field content and the QM action can be obtained via dimensional reduction from the field content and action  of the four-dimensional system describing a four-stack bound state of D5-D9 branes; we will use the  $D=4$, ${\cal N}=1$ conventions in the following. As already stated, we will work at a generic point in the closed moduli space.

  We have a vector superfield $V^{(a)}$ and three adjoint chiral multiplets $\Phi^{(a)}_{\mathbf{j}}$  for each stack $a=1,\ldots,4$ of D-branes. These are the fields associated to open string starting and ending on the same stack of branes.  
   The bosonic sector of the vector multiplet collects the NS states of the open strings starting and ending on the same stack with the polarizations along $\mathbf{R}^{1,3}$. The bosonic sector in the adjoint chiral multiplet consists of NS states with the polarizations along $T^6$ labelled by $\mathbf{j}=1,2,3$ as in~\eqref{qcrea}. Finally for each pair $(a,b)$ of D-brane stacks we have  a chiral and antichiral multiplet that we will denoted by $Z^{(ab)}$ and $Z^{(ba)}$ respectively, associated to open strings binding the branes. The actual field content is therefore: 

\subsubsection*{Vector superfields}

The adjoint vector superfields contain a gauge field, three scalars, four fermionic variables and an auxiliary field:
\begin{equation}
  \label{eq:Vsf}
   V^{(a)} = \lbrace A^{(a)}_0, \, x^{(a)}_{i}  , \, \lambda_\alpha^{(a)}  , \, \bar\lambda_{\dot \alpha}^{(a)} , \, D^{(a)}  \rbrace ~,
\end{equation}

The scalars $x^{(a)}_{i}$ comes from the dimensional reduction of the 10d vectors and represent the displacement of the brane $a=1,2,3,4$ in the non compact directions $i=1,2,3$.  If we consider more than one D6, then $x^{(4)}_{i}$ is a matrix and diagonal components represent the displacement of the brane in the non compact directions while off-diagonal component arises from open strings stretched between different D6. 

\subsubsection*{Adjoint chiral superfields}

The adjoint chiral superfields contain a complex scalar,  four fermionic variables and a complex auxiliary field:
\bea
  \label{eq:Phisf}
  \Phi^{(a)}_{\mathbf{i}} &=& \lbrace \phi^{(a)}_{\mathbf{i}}  , \, \chi^{(a)}_{\alpha,\mathbf{i}} ,  \, F^{(a)}_{\mathbf{i}}  \rbrace     \qquad\qquad 
  \bar\Phi^{(a)}_{\mathbf{i}} = \lbrace \bar\phi^{(a)}_{\mathbf{i}}  , \, \bar\chi^{(a)}_{\dot\alpha,\mathbf{i}} ,  \, \bar F^{(a)}_{\mathbf{i}}  \rbrace    
  \eea
  
  This fields can be seen as the displacement of the a-brane in the internal directions transverse to the D2 (for $ \mathbf{i}=1,2,3$) and the Wilson lines along parallel directions. Only Wilson lines for the D6.

\subsubsection*{Bifundamental chiral superfields}

Also the bifundamental chiral superfields contain a complex scalar, four fermionic variables  and a complex auxiliary field:
\bea
  \label{eq:Zsf}
  Z^{(ab)} &=& \lbrace z^{(ab)}  , \, \mu_\alpha^{(ab)} ,  \, F^{(ab)}  \rbrace ~ \qquad\qquad 
  \bar Z^{(ab)} = \lbrace \bar z^{(ab)}  , \, \bar \mu_{\dot \alpha}^{(ab)} ,  \, \bar F^{(ab)}  \rbrace  
\eea

\subsubsection{The action}

 The action of the quantum mechanics can be written in the superfield formalism as
 
\bea
\mathcal{L}&=& \int\!dx_0 \,d^2\theta\,d^2\bar{\theta} \, \mathrm{tr}  \left[  Z^{\dagger \, ba}
\,e^{2gV}\,Z^{ab}  + 2 \,  e^{-2gV}\,\Phi^{a \dagger}_{\bf i}
\,e^{2gV}\,\Phi^a_{\bf i}    + \xi \, V  \right]\nn \\
&+&\!\int\!dx_0\,d^2\theta\, \mathrm{tr} \left( \frac{1 }{8 g^2}  W^\alpha W_\alpha  +  {\cal W}(\Phi,Z) \right)  
+h.c.
\eea 
The superpotential for a generic vev of the closed string fields can be written as
\bea
 \mathcal{W}  &=&   \sum_{a,b=1}^4  \, Tr(Z^{(ab)} \Phi^{(a)}_{\textbf{i}} Z^{(ba)}\eta^{\textbf{i}}_{ab}) +
 \sum_{ a,b,c=1}^4  Tr( Z^{(ab)} Z^{(bc)} Z^{(ca)}) - \nn\\ 
 &-&\sum_{ a,b=1}^4  \, Tr(c^{(ab)} \Phi^{(a)}_{\textbf{i}} \eta^{\textbf{i}}_{ab}) + c^{(a)}\,  D_a 
\eea
with $Z^{(aa)}=0$, $c_{ab}$ and  $c^{(a)}$ parameterizing the closed string vevs and $\eta^i_{ab}$ the t'Hooft symbols 
\be
\eta^i_{jk}=\epsilon_{ijk} \qquad  \eta^i_{j4}=-\eta^i_{4j}=\delta^i_j
\ee

The superpotential  contains other terms in principle, for instance  quartic terms in $Z^{(ab)}$, but
 if we take small closed string moduli, and therefore small $c_{ab}$ and  $c^{(a)}$, then
 the bifundamental and chiral fields solving the eq. of motion can be shown to be small, and then it's consistent to ignore this higher order terms in the 
￼superpotential.

The moduli-dependence of the quantum mechanics action can be derived from string scattering amplitudes involving open and closed strings (in \cite{Chowdhury:2014yca,Chowdhury:2015gbk} they arrived to the same result with a much longer direct computation).  At linear order in the closed string fields two types of terms are generated
\bea
\mathcal{L}_D &=& \int\! d^2\theta\,d^2\bar{\theta} \, \mathrm{tr}  \, c^{(a)} \, V_a  =c^{(a)}\,  D_a \nn \\
  \mathcal{L}_F &=& -\int\!  d^2\theta\,  \sum_{ a,b=1}^4   \eta^{\textbf{i}}_{ab}\, c^{(ab)} \, \mathrm{tr}   \Phi^{(a)}_{\textbf{i}}+{\rm h.c}  =
   - \sum_{ a,b=1}^4   \eta^{\textbf{i}}_{ab}\, c^{(ab)} \, \mathrm{tr}  F^{(a)}_{\textbf{i}}+{\rm h.c }
\eea 
  The functions $c^{(a)}$ and $c^{(ab)}$ depends on the closed string moduli and characterize the generated  
  Fayet-Iliopoulos and F-terms respectively.  At linear order the coupling is computed by the two point  string amplitudes on the disk:
\begin{equation}
\mathcal{A}_{D,F}=   \left\langle  W_{g,b}   V_{D,F}   \right\rangle
\end{equation}
where $V_{D,F}$ are the open string vertices for the D aor F fields and $W_{g,b}$ the closed string vertex. Explicitly, the vertex operators are \cite{Bertolini:2005qh}:
\bea
W(z_1,z_2) &=& (ER_a)_{MN}e^{-\varphi} \,c\, \psi^{M} e^{ik X}(z_1) c\, e^{-\varphi} \psi^{N} e^{ikR X}(z_2) \nn\\
V_D(z_3) &=& \xi  \, \Omega_{MN} : c\, \psi^{M}\, \psi^{N}(z_3) :\nn\\
V_{F_\mathbf{i}}(z_3) &=& \zeta^{\mathbf{i}}_{MN} : c\, \psi^{M}\, \psi^{N}(z_3): \label{closedw}
\eea
with $R_a$ the reflection matrix, 
  $E_{MN}=g_{MN}+b_{MN}$  describing the graviton $g_{MN}$ and the B-field $b_{MN}$ polarizations  and
$\zeta^{\mathbf{i}}_{MN}$ a tensor with the only non-trivial components
\be
 \zeta^{\mathbf{i}}_{\bar{\mathbf{j}} \bar{\mathbf{k}} } ={1\over 2} \zeta^{\mathbf{i}}\, \delta^{ \mathbf{i} \bar{\mathbf{j}} }\, \epsilon_{\bar{\mathbf{j}} \bar{\mathbf{k}}  \bar{\mathbf{l}} } 
 \ee
 Finally $\Omega_{MN}$ is an off-diagonal anti-symmetric matrix with non-tivial components $\Omega_{45}=1$,  $\Omega_{67}=\Omega_{89}=-1$.

The string amplitude becomes 
\begin{equation}
\mathcal{A}_{D,F}\sim     \langle c\, e^{-\varphi}\, \psi^M(z_1) c\, e^{-\varphi}\,  \psi^N( z_2)  c\, :\psi^P \psi^Q (z_3): \rangle  = 
 (\eta^{PM}\eta^{QN}-\eta^{QM}\eta^{PN})
\end{equation}
leading to
  \bea
\mathcal{A}_{D}  &=&  \xi \, {\rm tr}  ( \Omega ER_a)  \nn\\
\mathcal{A}_{F_\mathbf{i}}  &=&  \zeta^{(\mathbf{i})MN}  (ER_a)_{[MN]} =  {1\over 2} \zeta^{\mathbf{i}}\, \epsilon^{\mathbf{i}\mathbf{j}\mathbf{k}   }  (ER_a)_{\mathbf{j}\mathbf{k}}
\eea
where by $[MN]$  we denote the antisymmetric part of the tensor.  We note that $(R_0)_{MN}=\epsilon_M \delta_{MN}$ with $\epsilon_M=\pm$, with plus and minus signs  for Neumann and Dirichlet directions respectively.  Specifying to the various types of branes and denoting by 45, 67 and 89 the Neumann directions for the $D2_1$, 
$D2_2$ and $D2_3$ branes respectively one then finds

\begin{itemize}

\item{D-terms: 
\bea
{\rm tr}  ( \Omega E \,D2_1)      &=&  ~~b_{45}-b_{67}-b_{89} =c^{(1)}\nn\\
{\rm tr}  ( \Omega E \,D2_2)   &=&  -b_{45}+b_{67}-b_{89} =c^{(2)}\nn\\
{\rm tr}  ( \Omega E \,D2_3)   &=&  -b_{45}-b_{67}+b_{89} =c^{(3)}\nn\\
{\rm tr}  ( \Omega E \,D6)    &=&  ~~b_{45}+b_{67}+b_{89} =c^{(4)}\nn\\
  \eea
 }

 \item{F-terms: 
\bea
\zeta^{\mathbf{1}}\,  (ER_{D6} )_{23}&=& -\zeta^{\mathbf{1}}\,  (ER_{D2_1} )_{23} = b_{68}-i \, b_{69}-i \, b_{78}-  b_{79}=c^{(14)}\nn\\
 \zeta^{\mathbf{2}}\,  (ER_{D6} )_{31}&=& -\zeta^{\mathbf{2}}\,  (ER_{D2_2} )_{31} =  b_{48}-i \, b_{49}-i \, b_{58}-  b_{59}=c^{(24)}\nn\\
 \zeta^{\mathbf{3}}\,  (ER_{D6} )_{12}&=& -\zeta^{\mathbf{3}}\,  (ER_{D2_3} )_{12} =  b_{46}-i \, b_{47}-i \, b_{56}-  b_{57}=c^{(34)}  \nn\\
 \zeta^{\mathbf{1}}\,  (ER_{D2_2} )_{23}&=& -\zeta^{\mathbf{1}}\,  (ER_{D2_3} )_{23} = g_{69}-i \, g_{68}-i \, g_{79}-  g_{78}=c^{(23)} \nn\\
  \zeta^{\mathbf{2}}\,  (ER_{D2_3} )_{31}&=& -\zeta^{\mathbf{2}}\,  (ER_{D2_1} )_{31} = g_{49}-i \, g_{48}-i \, g_{59}-  g_{58}=c^{(13)} \nn\\
   \zeta^{\mathbf{3}}\,  (ER_{D2_1} )_{12}&=& -\zeta^{\mathbf{3}}\,  (ER_{D2_2} )_{31} = g_{47}-i \, g_{46}-i \, g_{57}-  g_{56}=c^{(12)}  
  \eea
 }

 \end{itemize}

\subsubsection{The Ground States}

The classical ground states are  defined by zeros of the scalar potential, which is the sum of F terms, D terms and Gauge contributions:

\bea
 V_F=\sum_{a=1}^4 \sum_{\mathbf{i}=1}^3 \left| \frac{\partial \cal W}{\partial \Phi^{(a)}_\mathbf{i}}\right|^2+\sum_{a=1}^4 \sum_{b=1}^4 \left| \frac{\partial \cal W}{\partial Z^{(ab)}}\right|^2
\eea

\bea
   V_D=  \frac{1}{2}\sum_{a=1}^4  \left(\sum_{b=1}^4  (\bar{Z}^{(ab)}Z^{(ab)}-\bar{Z}^{(ba)}Z^{(ba)}) -c^{(a)}\right)^2
\eea

\bea
   V_{Gauge}=\sum_{i=1}^3 \sum_{a=1}^4 \sum_{b=1}^4 (x^{(a)}_i-x^{(b)}_i)(x^{(a)}_i-x^{(b)}_i)(\bar{Z}^{(ab)}Z^{(ab)}+\bar{Z}^{(ba)}Z^{(ba)})
   \eea

being positive, each of these terms must be separately zero. One can show that this lead to the equations of motion:

\bea
&&  \sum_{b=1}^4  \, Tr(Z^{(ab)}  Z^{(ba)}\eta^{\textbf{i}}_{ab}) -  \sum_{b=1}^4  \, Tr(c^{(ab)} \eta^{\textbf{i}}_{ab}) =0 \qquad \forall ~ \textbf{i},a \nn\\
&&   \sum_{\textbf{i}=1}^3  Tr\left[ (\Phi^{(a)}_{\textbf{i}} - \Phi^{(b)}_{\textbf{i}} ) Z^{(ba)}\eta^{\textbf{i}}_{ab} \right]   + 
     \sum_{c=1 }^4 Tr( Z^{(bc)} Z^{(ca)}) = 0        \qquad  \forall ~ a,b 
\eea

Is it possible to numerically solve this set of equations for low number of branes. For instance fixing the closed moduli and for $N_a = 1$ with $a = 1,2,3,4$ one finds exactly 12 distinct solutions \cite{Chowdhury:2014yca,Chowdhury:2015gbk}  characterized by different value of $z^{(ab)}$ and $\phi^{(a)}$, but with $x^{(a)}_i = 0$. This explicit SQM counting matches with the exact index computation in the D1-D5-KK-P frame \cite{Shih:2005qf}. Notice that the D1-D5-KK-P index for  $N_a = 1$ is exactly 12, therefore each microstate has zero angular momentum. More generally, since the moduli space is composed by disconnected zero dimensional pieces, every microstate will have zero angular momentum and will be a "pure Higgs" state (pure Higgs states are the ones arising from the middle cohomology of the moduli space \cite{Bena:2012hf}, but here the middle cohomology is the whole space, since the space is zero dimensional).

As we increase the number of branes in each stack the computations become quickly intractable, since the fields are now matrices, therefore there are new terms due to the commutators and more unknowns to be solved; nevertheless for the few $N_a >1 $ that have been checked one finds perfect agreement between index and SQM calculations.  

\subsection{Open String vevs}

The microstate counting was performed at the classical level, with all the fermions vanishing; for every branes configurations we found different vacua, that can be seen as delta function wave function, while the real quantum solutions will be  wave functions with finite spread around the classical solutions and we expect them to be relevant for the exact values assumed by the open string vevs in the string scattering computations. More precisely the fermionic-bosonic polarizations in the open string vertex will be promoted to operators in the SQM and the vev over a particular ground states will have a definite value depending on the ground state itself.

To be concrete, let's focus on the string theory 4 fermions correlator \eqref{4p}, that we rewrite here schematically:

\bea
 \left \langle V_{\mu_1}(z_1) V_{\mu_2}(z_2) V_{\mu_3}(z_3) V_{\mu_4}(z_4)    W_{NSNS}(z_5,z_6)  \right\rangle
\eea

Previously this correlator was computed only over the worldsheet CFT (integration over $z_p$), while now given the string vertex operator

\bea
V_{\mu_\alpha}=\mu_{\alpha}^{(ab)} \,  e^{-\frac{\varphi}{2}}  \, S^{\alpha} \, e^{\frac{i}{2} h_{\mathbf{j}_{ab}  }} \,\prod_{\mathbf{j}\not=\mathbf{j}_{ab} } \Delta_{\mathbf{j}}
\eea

we promote the open string polarizations $\mu_{\alpha}^{(ab)}$ to operators, impose the adequate anticommutation relationships and identify them with the fields of the quantum mechanics. The explicit form of the other bifundamental vertex operators consistent with the supercharges preserved by the D6-D2-D2-D2 system can be found in appendix \eqref{bifVer}. 
Finally, for the polarizations we have an additional vev over the SQM ground state $ |f \rangle $:

\bea
 \langle  f |\, \mu_{A_1}^{(a_1  a_2)} \,\mu_{A_2}^{(a_2 a_3)} \mu_{A_3}^{(a_4  a_1)} \mu_{A_4}^{(a_4  a_1)}  |f \rangle 
 \eea

To found the explicit form of $|f \rangle$, that will include even fermions creation operators, one must go beyond the classical approximation and solve the full Schrodinger equation for every ground states.

\subsubsection{Quantum Ground States: Toy Model}

To illustrate how to compute the ground state in a generic SQM we will use a toy model, while we will only give some partial result on the computation with the real D6-D2-D2-D2 system, as this is still work in progress \cite{wrk}. See  \cite{Hori:2003ic} for a comprehensive overview of SQM.

Consider a SQM with only one scalar field $x$, a  fermionic superpartner $\psi$ and with superpotential $W$. The lagrangian of such a theory is given by:

\bea
L = \frac{1}{2} \dot{x}^2 - \frac{1}{2} (\partial_xW(x))^2 +\frac{i}{2}(\bar{\psi} \dot{\psi}-\dot{\bar{\psi}} \psi) - \partial^2_xW(x)\bar{\psi} \psi
\eea

where $\bar \psi $ is the complex conjugate of $\psi$, satisfying $\lbrace \psi, \psi\rbrace = 0$ and $\lbrace \bar \psi, \bar{\psi}\rbrace = 0 $ . The lagrangian is invariant under supersymmetric transformations

\bea
 \delta x = \epsilon \bar{\psi} - \bar{\epsilon} \psi  \qquad \delta \psi = \epsilon (i \dot{x} + \partial_xW) \qquad \delta \bar{\psi} = \bar{\epsilon} (-i \dot{x} + \partial_xW)
 \eea

where $\epsilon$ is the complex fermionic parameter and  $\bar \epsilon$ is the conjugate. The associated conserved charges are two real supercharges:

\bea
\delta \int L dt = \int dt \left(  -i \dot{\epsilon} Q -i \dot{\bar{\epsilon}} \bar{Q}  \right)  
\eea

\bea
 Q =  \bar{\psi }  (i \dot{x} + \partial_xW) = \bar{\psi }  (i p + \partial_xW) \qquad \bar{Q}=\psi    (-i \dot{x} + \partial_xW) = \psi    (-i p + \partial_xW)
 \eea
 
 where $Q^{\dagger} = \bar{Q}$ and $p = \frac{\partial L }{\partial \dot{x}} = \dot{x}$. Using that $p_{\psi} = \frac{\partial L }{\partial \dot{\psi}}= i \bar \psi$ we can impose the quantization relations:

\bea
 [x,p] = i \qquad \lbrace \psi, \bar{\psi}\rbrace = 1 
 \eea
 
 where the second relation follows from $\lbrace \psi, p_{\psi}\rbrace = i$. The bosonic Hilbert space  is given by the space of square-normalizable wave functions, while the fermionic space is built from a vacuum state $|0 \rangle $ defined by the action of the annihilation operator:

\bea
  \psi \ket{0} = 0 
 \eea
 
 In this simple theory the fermionic Hilbert space is two dimensional:

\bea
   \ket{0} \qquad \bar{\psi} \ket{0} 
 \eea
 
 since $\bar{\psi}^2 = 0$ and there no other fermions. Therefore the most general state is given by:
 
\bea
  \Psi = f_1(x) \ket{0} + f_2(x) \bar{\psi} \ket{0} 
 \eea

One can show that the Hamiltonian satisfies 

\bea
H = \frac{1}{2} \lbrace Q \, ,  \, \bar Q\rbrace > 0
\eea

therefore zero energy states (ground states) are annihilated by the supercharges and are supersymmetric, and vice versa\footnote{Moreover it can be shown the space of supersymmetric ground states coincide with the cohomology of the supercharges.}. Since $p$ acts as a derivative in the space of the $x$, the supersymmetric conditions

\bea
 Q \Psi = \bar{Q} \Psi = 0 
\eea

translates into differential equations for the bosonic piece:

\bea
 \left( \frac{d}{dx}+ W'(x) \right)f_1(x) = 0
\eea
\bea
 \left( -\frac{d}{dx}+ W'(x) \right)f_2(x) = 0
\eea

which are solved by

\bea
f_1(x) = c_1 \, e^{-W(x)} \qquad f_2(x) = c_2 \, e^{W(x)}
\eea

Since we are looking for normalizable solutions, the final answer depend on the behaviour of the superpotential at infinity. Beside the case with no solutions, the two possibilities are:

\bea
 \Psi = e^{-W}  \ket{0} \qquad \Psi = e^{W} \bar{\psi} \ket{0} 
\eea

In a more realistic scenario with $n$ boson-fermion pairs $x^A, \, \psi^A$ with $A = 1, \dots, n$, the superpotential can be expanded around its critical points, which coincide with the ground states \footnote{More precisely, this is true at the perturbative level in the parameter $\lambda$ used to rescale the superpotential $W \rightarrow \lambda W$, to take advantage of the scale invariance of the index. Non perturbative effect can lift some zero energy states. } :  

\bea
W(x) = W(x_i) +\frac{1}{2} \left[\frac{\partial^2 W}{\partial x^A x^B}\right]_{x=x_i} (x^A-x^A_0)(x^B-x_0^B)+ \dots \nn \\
 = W(x_i)  + \sum_A c_A^{(0)} (\tau^A_{(0)})^2 
\eea

where $\tau$ are some adapted coordinates. The supersymmetric ground states in this theory are given by 

\bea
\Psi_i =e^{- \sum_A  \, |c_A^{(i)}| (\tau^A_{(i)})^2 }\prod_{B: c^{(i)}_B <0 } \bar{\psi}^B \ket{0} 
\eea

where $i$ labels the particular critical point $x_i$ around which we are expanding the superpotential, $c^{(i)}_B$ are the eigenvalues of the Hessian matrix $\partial_I \partial_J W $ at $x_i$. The number of fermionic creation operators is the Morse index, that is the number of negative eigenvalues of the Hessian matrix at $x_i$.

\subsubsection{Quantum Ground States: D-brane system}

 Coming back to the D-brane system, the kinetic part of the action can be written as
\bea
S_{\rm kin} 
 &=&    \sum_{a=1}^4   \left[\frac{1}{2}  \sum_{i=1}^3 \left(\partial_t x_i^{(a)}\right)^2 + i \bar{\lambda}^{(a)}\, \bar{\sigma}_0 \,   \partial_t \, \lambda^{(a)}  \right]
+ \sum_{(ab)} \bigg[ \partial_t z^{(ab)} \partial_t \bar{z}^{(ba)}   + i \bar{\mu}^{(ba)} \bar{\sigma}^{0}  \partial_t \mu^{(ab)} \bigg]  \nn\\
&&+ \sum_{a=1}^4 \sum_{\mathbf{i} = 1}^3  \left(  \partial_t \phi_{\mathbf{i}}^{(a)} \partial_t \bar{\phi}_{\mathbf{i}}^{(a)}     + i \bar{\chi}_\mathbf{i}^{(a)} \bar{\sigma}_0   \partial_t\chi_{\alpha,\mathbf{i}}^{(a)} \right)   
\eea
  After quantization, time derivative of the scalars are momenta conjugate to the original fields and can be represented by derivatives
\begin{equation}
  \label{pz2d}
  \frac{\delta {\cal L}}{\delta \dot{z}^{(ab)}} = \dot{\bar{z}}^{(ba)} ~ \qquad \dot{z}^{(ab)} \to - i \frac{\partial}{\partial {\bar{z}^{(ba)}}}~.
\end{equation}

Similarly for the fermions we have
\begin{equation}
  \label{ppsi2c}
  \frac{\delta {\cal L}}{\delta \dot{\mu}_\alpha^{(ab)}} =- i \bar{\sigma}_0^{ \dot\alpha \alpha} \, \bar{\mu}_{\dot \alpha}^{(ba)} 
  \end{equation}
which, using that $\lbrace\mu,\pi_{\mu} \rbrace = i$, implies the anticommutation relations
\begin{equation}
  \label{eq:psicr2}
  \{ \bar{\mu}_{\dot \alpha}^{(ba)}     , \mu_\beta^{(ab)} \} =- {\sigma}^0_{  \beta \dot\alpha}  =  \delta_{  \beta \dot\alpha}
  \end{equation}
%
%
%
 
Finally we define the vacuum of the fermionic Fock space as follows
\begin{equation}
  \label{eq:psivac}
   \mu_\alpha^{(ab)} |0\rangle =0 
\end{equation}
  for all values of $(ab)$ and for $\alpha=1,2$. In an analogous fashion one finds:
  
  \begin{equation}
  \{ \bar{\lambda}_{\dot \alpha }^{(a)}     , \lambda_{\beta}^{(a)} \} = \delta_{\dot\alpha  \beta }     \qquad , \qquad 
   \{ \bar{\chi}_{\dot \alpha \mathbf{i}}^{(a)}     , \chi_{\beta\textbf{j}}^{(a)} \} =\delta_{  \dot\alpha \beta} \, \delta_{   \mathbf{i} \textbf{j}  } 
    \quad , \quad  \mathbf{i},\mathbf{j}=1,2,3
  \end{equation}
  and
  \begin{equation}
   \chi_{\alpha \mathbf{i}}^{(a)} |0\rangle =0 \qquad , \qquad 
   \lambda_\alpha^{(a)} |0\rangle =0 
\end{equation}

Given the lagrangian, we can extract the supercharges by making the $\epsilon$ depend on time and looking at:
\bea
\delta \int L dt = \int dt \left(  \dot{\epsilon}^{\alpha} Q_{\alpha} + \dot{\bar \epsilon}_{\dot\alpha} \overline{Q}^{\dot\alpha}   \right)
\eea
with $\delta= \sqrt{2} (\epsilon^{\alpha}\delta_\alpha+ \bar\epsilon_{\dot\alpha} \delta^{\dot \alpha})$. 
The terms with time derivatives of $\epsilon$ come only from the susy variation of the kinetic terms, so we find
\bea
Q_\alpha=\sum_s {\partial  L \over \partial \dot{\Phi}_s } \delta_\alpha \Phi_s 
\eea
with the sum running over all fields  $\Phi_s$ of the quantum mechanics. The relevant variations are

\bea
\delta_\alpha z^{(ab)}&=& \mu_\alpha^{(ab)}         \qquad \qquad\qquad~~~~~~~~~~~    \delta_{\dot \alpha} \bar z^{(ab)}= \bar\mu_{\dot \alpha}^{(ab)} \nn\\
\delta_\alpha \mu_{\beta}^{(ab)} &=&- {\partial \overline{\mathcal W} \over \partial \bar{z}^{(ba)} }  \epsilon_{\alpha\beta}\qquad \qquad  ~~~~~~~~~  \delta_{\dot \alpha} \bar\mu_{\dot \beta}^{(ba)}
= - {\partial {\mathcal W} \over \partial z^{(ab)} } \epsilon_{\dot \alpha\dot \beta} \nn\\
\delta_\alpha \bar\mu_{\dot\alpha}^{(ba)} &=& - i \sigma^0_{\alpha\dot\alpha} \dot{\bar z}^{(ba)} - \sigma^i_{\alpha\dot\alpha}\,  x_i^{(ba)}\, \bar z^{(ab)}  \qquad
\delta_{\dot\alpha} \mu_{\alpha}^{(ab)} =i \sigma^0_{\alpha\dot\alpha} \dot{z}^{(ab)} - \sigma^i_{\alpha\dot\alpha}\,  x_i^{(ab)}\,  z^{(ab)}  
\eea

\bea 
\delta_\alpha \phi^{(a)}_{\bf{i}}&=& \chi_{\alpha\bf{i}}^{(a)}         \qquad \qquad\qquad~~~~~~~~~~~    \delta_{\dot \alpha} \bar \phi^{(a)}_{\bf{i}}= \bar\chi_{\dot \alpha \bf{i}}^{(a)} \nn\\
\delta_\alpha \chi_{\beta \bf{i}}^{(a)} &=&- {\partial \overline{\mathcal W} \over \partial \bar \phi^{(a)}_{\bf{i}} }  \epsilon_{\alpha\beta}\qquad \qquad  ~~~~~~~~~  \delta_{\dot \alpha} \bar\chi_{\dot \beta \bf{i}}^{(a)}
= - {\partial {\mathcal W} \over \partial \phi^{(a)}_{\bf{i}} } \epsilon_{\dot \alpha\dot \beta } \nn\\
\delta_\alpha \bar\chi_{\dot\alpha \bf{i}}^{(a)} &=& - i \sigma^0_{\alpha\dot\alpha} \dot{\bar \phi}_{\bf{i}}^{(a)} - \sigma^i_{\alpha\dot\alpha}\,  x_i^{(a)}\, \bar \phi^{(a)}_{\bf{i}}  \qquad
\delta_{\dot\alpha} \chi_{\alpha \bf{i}}^{(a)} =i \sigma^0_{\alpha\dot\alpha} \dot{\phi}_{\bf{i}}^{(a)} - \sigma^i_{\alpha\dot\alpha}\,  x_i^{(a)}\,  \phi^{(a)}_{\bf{i}}
\eea

\bea
\delta_\alpha x^{(a)}_{\mu}&=& -i \bar \lambda^{(a)}_{\dot \alpha}    (\bar{\sigma}^{\mu})^{\dot \alpha \beta}  \epsilon_{\alpha \beta}   \qquad \qquad\qquad~~~~~~~~~~~    \delta_{\dot \alpha} x^{(a)}_{\mu}= i    \epsilon_{\dot \alpha\dot \beta}   (\bar{\sigma}^{\mu})^{\dot  \beta \alpha}  \lambda^{(a)}_{\alpha} \nn\\
\delta_\alpha \lambda_{\beta}^{(a)} &=& (\sigma^{\mu \nu} F_{\mu \nu}^{(a)}  + i D^{(a)})_{\alpha}^{\, \, \,\gamma}\epsilon_{\gamma\beta}\qquad \qquad   \delta_{\dot \alpha} \bar\lambda_{\dot \beta}^{(a)}
= \epsilon_{\dot \alpha \dot \gamma}  (\bar \sigma^{\mu \nu} F_{\mu \nu}^{(a)}  - i D^{(a)})_{\, \, \, \dot \beta}^{\dot \gamma}   
\eea

where $x_i^{(ab)} = x_{i}^{(a)}-x_{i}^{(b)}$.
 Additional details can be found in the appendix \eqref{sqmA}.

Using the supersymmetry variations the supersymmetry charges can be written as\footnote{To obtain this  symmetric form we rewrite the fermionic kinetic terms in a symmetric fashion (for instance, $\bar{\mu}  \dot{\mu}  \rightarrow \frac{1}{2}(\bar{\mu}  \dot{\mu}-\dot{ \bar{\mu}} \mu )$.}
\bea
  \label{eq:Qm}
  Q_\alpha &=&   \sum_{a\not=b}  \left(  2  \dot{ \bar z}^{(ba)}\mu_\alpha^{(ab)}  + i \sigma^i_{\alpha \dot \beta}  (\bar{\sigma}^0)^{\dot \beta \beta} x_i^{(ab) }\bar{z}^{(ba)}\mu_\beta^{(ab)} - i  \bar{\mu}_{\dot \alpha}^{(ba)} (\bar{\sigma}^0)^{\dot \alpha \beta} \epsilon_{ \alpha \beta}\frac{\partial  \overline {\cal W} }{\partial \bar z^{(ba)}} \right)  + \nn\\
  &+&   \sum_{a,\bf{i}}  \left(  2  \dot{ \bar \phi}_{\bf{i}}^{(a)}\chi_{\alpha\bf{i}}^{(a)}  + i \sigma^i_{\alpha \dot \beta}  (\bar{\sigma}^0)^{\dot \beta \beta} x_i^{(a) }\bar{\phi_{\bf{i}}}^{(a)}\chi_{\beta\bf{i}}^{(a)} - i  \bar{\chi}_{\dot \alpha \bf{i}}^{(a)} (\bar{\sigma}^0)^{\dot \alpha \beta} \epsilon_{ \alpha \beta}\frac{\partial  \overline {\cal W} }{\partial \bar \phi_{\bf{i}}^{(a)}} \right)+ \nn \\
  &+& \sum_{a,i} \left( -2\,{\rm i}\, \dot{ x}_i^{(a)}\bar \lambda^{(a)}_{\dot \alpha}    (\bar{\sigma}^{i})^{\dot \alpha \beta}  \epsilon_{\alpha \beta} - \bar \lambda^{(a)}_{\dot \alpha}      (\bar{\sigma}^0)^{\dot \alpha \beta}    D^{(a)} \, \epsilon_{\alpha\beta}\right)
  \eea
  
  \bea
   \bar Q_{\dot \alpha} &=&  \sum_{a\not=b}    \left(  2   \dot{ z}^{(ab)} \bar\mu_{\dot \alpha}^{(ba)}  -  i \sigma^i_{\alpha \dot \alpha} (\bar{\sigma}^0)^{\dot \gamma  \alpha} x_i^{(ab) }z^{(ab)}\bar\mu_{\dot \gamma}^{(ba)}  + i \mu^{(ab)}_{\alpha}  (\bar{\sigma}^0)^{\dot \beta \alpha} \epsilon_{\dot \alpha \dot \beta } \frac{\partial  {\cal W}  }{\partial z^{(ab)}}  \right)    + \nn \\
   &+&  \sum_{a,\bf{i}}    \left(  2   \dot{ \phi_{\bf{i}}}^{(a)} \bar\chi_{\dot \alpha \bf{i}}^{(a)}  -  i \sigma^i_{\alpha \dot \alpha} (\bar{\sigma}^0)^{\dot \gamma  \alpha} x_i^{(a) }\phi_{\bf{i}}^{(a)}\bar\chi_{\dot \gamma \bf{i}}^{(a)}  + i \chi^{(a)}_{\alpha \bf{i}}  (\bar{\sigma}^0)^{\dot \beta \alpha} \epsilon_{\dot \alpha \dot \beta } \frac{\partial  {\cal W}  }{\partial \phi_{\bf{i}}^{(a)}}  \right) \nn \\
  &+&  \sum_{a,i} \left( 2\, {\rm i}\,  \dot{ x}_i^{(a)} \epsilon_{\dot \alpha\dot \beta}   (\bar{\sigma}^{i})^{\dot  \beta \alpha}  \lambda^{(a)}_{\alpha} - \epsilon_{\dot \alpha \dot \beta}  \,  D^{(a)}  (\bar{\sigma}^0)^{\dot \beta \beta}  \lambda^{(a)}_{\beta} \right)
\eea
 where we used the fact that the only non-trivial components of $F_{\mu\nu}$ is  $F_{0 i} = \dot x_i$ since we choose the gauge $x_0=0$. The auxiliary field D is fixed by the equations of motions as:
 
\bea
c^{(a)} + \sum_{b}   \, (\bar{ z}^{(ba)}z^{(ab)}-\bar{ z}^{(ab)}z^{(ba)}) = D^{(a)}
\eea

We define the  ground state $|f\rangle$ as the solution to the equations
  \begin{equation}
    \label{eq:Schreq_g}
    Q_{\alpha} \,| f\rangle=\bar Q_{\dot\alpha} \,| f\rangle= 0~,
  \end{equation}
  
  Using \eqref{pz2d} and analogous relations, these conditions are differential equations for the open string fields. When solved \cite{wrk}, one can compute the open string vevs  as

\bea
 \langle  f |\, \mu_{A_1}^{(a_1  a_2)} \,\mu_{A_2}^{(a_2 a_3)} \mu_{A_3}^{(a_4  a_1)} \mu_{A_4}^{(a_4  a_1)}  |f \rangle 
 \eea
 
 Since the string scattering processes  have an associated  SUGRA solution, it's in principle possible to identify these microstates from the subleading terms in the asymptotic expansion of the SUGRA solutions.
   
\pagebreak

\section{Conclusions}

After nearly 50 years from his introduction \cite{Hawking:1974rv} the black hole information paradox is still alive and unsolved. The experimental detection of gravitational waves from a black hole merger may be the experimental hint that will guide the community towards the solution, discriminating among the countless theoretical proposals or letting us finding a new one. In this thesis we made the initial assumption that string theory is the correct theory of quantum gravity and performed calculations in this setting; this belief is supported by the matching of many calculations from the macroscopic and microscopic side. We mainly focused  on the Fuzzball proposal for four charges black holes in string theory, which conjecture that every microstates is associated to a regular solution in supergravity, even though many of the results are in fact more general.

In the first part of the thesis we explored the black hole problem from the point of view of probes strings scattering on a configuration of intersecting branes. We compared some known supergravity solutions with the emission from a system of four D-branes stacks. The goal was to find the subclass of supergravity solutions able to carry the string theory microstates impersonated by the open strings configurations. Here the underlying idea is that while every open string configuration is a valid microstate in the full string theory, is it not obvious what appearance this microstate will have in low energy gravity and if it will be regular. The outcome of this analysis is the general expression of the supergravity subclass compatible with the string theory microstates, even though a loot of freedom remain since the original family was constructed in term of generic harmonic functions. 

The next step is to study these harmonic functions; in light of the fuzzball proposal there should be regular solutions associated to these microstates therefore we worked purely in supergravity to find regular solutions. The study of the supergravity solution can be essentially reduced to five dimensions, where many class of solutions are already known, even though since we ultimately interested to regular solutions associated to four dimensional black holes we still demand the correct amount of charges and the correct asymptotic in five dimensions to be $\mathbb{R}^{1,3}\times S^1$, so that the number of large space-time dimensions is four. There are two kind of constraints on the solution: firstly the regularity constraints coming from classical gravity like absence of closed time-like curves, Taub-NUT singularities and divergences in the curvature scalars; secondly there constraints coming from string theory, found with the scattering analysis. The upshot is that there are many instances of regular solutions inside the family related to the stringy configurations. This investigation can be considered as a proof of existence for regular solutions for the D-brane system, but more work is required to fully understand the characteristic of these solutions and an interesting question that still remain  open is how much general are these solutions, is this family large enough to account for a significant part of the black hole entropy?

Finally, in the last section we focused on the world-volume theory of the D-brane system, that is on a supersymmetric quantum mechanics containing all the fields associated to open strings, both twisted and untwisted. This complementary point of view on the black hole system is handy to explicit count the microstates, at least for low number of branes. The counting requires only the classical approximation, that is simply looking for the supersymmetric vacua of the theory, but it is interesting to go beyond and compute the quantum ground states, for instance to study correlators of open string fields. The latter application allow us to give a precise value to the open string vevs appearing in the string scattering computations, since for generic value of the closed string moduli these vevs are not arbitrary complex values anymore. This last section is part of an ongoing research project and could  shed light on how microstates in the gravity description differ from a black hole, namely identifying quadrupoles and higher multipoles terms.

A great deal of work remain to be done in the black holes microstates field. For three charge black holes the current technology is pretty advanced and very sophisticated gravity solutions are available in literature. Possible interesting directions includes: 

\begin{itemize}
\item  What happen to an infalling observer? Is the would-be horizon soft or hard?

\item  What about the dynamic? How are fuzzball/stringy solutions formed?

\item  What is the signature of these solutions? How can an external observer distinguish them from black holes?

\end{itemize}

Regarding four charges solutions, the main topic of this thesis, it would be nice to enlarge the number of microstates geometries available and study them in parallel with the microscopic description. There are many possible tools to do this, here we have focused on SQM and string scattering amplitudes, but holography, namely $AdS_2 / CFT_1$, remain an important future tool to get new insights. 

Finally, after many years, there is the opportunity to confront models with experimental data coming from gravitational waves detections. It is still not clear if a fuzzball or other quantum gravity effects will leave a signature that is detectable just with gravitational waves, but its worth squeezing our models to extract predictions, at least by considering the implications of the universal features of the known solutions, for instance the capped space behind the would-be horizon.

 \pagebreak
 
 \section*{Acknowledgements}
 
 Let's define a variable $x = thank$. 
  
 I would like to $x$ all the people that interacted with me during this three years.  
 
 Particularly I had the great opportunity to learn a lot from my mentors Massimo Bianchi and Jose Francisco Morales. They endured my obsession for the information paradox, and that was not an easy task!
 
I $x$ Rodolfo Russo for his hospitality when I was in London and for the  illuminating conversations. 

Also, I should $x$ my collaborators Natale Zinnato for the many nice moments together and Stefano Giusto for the many enjoyable discussions.  

$x$ to the PhD guys, Rome and London!  

I $x$ so much Gino Garruti for his friendship, that I cannot forget. I really $x$ Matteo Falso with my heart.
 
 A big $x$ to my family for all the support during these years!!! You are always my favourite family! 
 
Super $x$ to 3!  Only you know the truth! Counting This one, you were on pretty many acknowledgements so fAr, eh? 
 
 It was an exciting journey, new adventures await!

 \pagebreak
 
 \appendix
 
\section{ Notations and conventions }

\label{notaz}

  The four-dimensional metric and  epsilon tensors  
 \be
 \eta_{\mu \nu} = (-,+,+,+)   \qquad \qquad   \epsilon_{0123}=-1
 \ee
  Weyl spinors and two dimensional epsilon tensors
 \bea
 \epsilon^{12}&=& \epsilon_{21}= \epsilon^{\dot 1 \dot 2 }= \epsilon_{\dot 2 \dot 1} =  1 \qquad\qquad  \epsilon^{ac}\epsilon_{cb} =\delta^a_b
  \qquad\qquad  \epsilon^{\dot a \dot c}\epsilon_{\dot c \dot b} =\delta^{\dot a}_{\dot b} \nn\\
   \mu^a &=&  \epsilon^{ab} \mu_b \;\;\;\;\;  \qquad \mu_a= \epsilon_{ab} \mu^b \;\;\; \qquad 
   \bar{\mu}_{\dot{a}}= \epsilon_{\dot{a}\dot{b}} \bar{\mu}^{\dot{b}}\;\;\;   \qquad \bar{\mu}^{\dot{a}}= \epsilon^{\dot{a}\dot{b}} \bar{\mu}_{\dot{b}}\nn\\
   \mu \chi &=& \chi \mu =  \mu^a \chi_a\qquad\qquad  \bar{\mu} \bar{\chi}  =  \bar{\chi} \bar{\mu} = \bar{\mu}_{\dot{a}} \bar{\chi}^{\dot{a}} 
   \eea
   Sigma matrices:
   \bea
  (\bar{\sigma}^\mu)^{\dot{a}b} &=& \epsilon^{\dot{a}\dot{c}}\epsilon^{b c} (\sigma^\mu)_{c \dot{c}}   \qquad\qquad 
  (\sigma^\mu)_{a \dot{b}} (\bar{\sigma}_\mu)^{\dot{c}d}=-2 \delta_a^{\textcolor{white}{b}d} \delta_{\dot{b}}^{\textcolor{white}{b}\dot{c}} \nn\\
 \sigma^\mu &=& (-1,\vec{\sigma})    \qquad \qquad        \bar\sigma^\mu = (-1,-\vec{\sigma})    \nn\\
\sigma^0 &=&  \begin{pmatrix} -1 & 0\\
0 & -1 \end{pmatrix}  \;\;\;\;\;  \sigma^1= \begin{pmatrix} 0 & 1\\
1 & 0 \end{pmatrix} \qquad
\sigma^2= \begin{pmatrix} 0 & -i\\
i & 0 \end{pmatrix}  \;\;\;\;\;  \sigma^3= \begin{pmatrix} 1 & 0\\
0 & -1 \end{pmatrix} 
\eea

\label{appN1}

The index structure is the following:
\begin{itemize}
\item  $M, N, R \dots$ for 10-dimensional vector indices.
\item  $\hat{M}, \hat{N}, \hat{R} \dots$ for 6-dimensional vector indices.
\item  $ \alpha, \beta \dots$ for left spinorial indices in directions $\lbrace  t, x_1, x_2, x_3 \rbrace$
\item  $ \dot{\alpha}, \dot{\beta} \dots$ for right spinorial indices in directions $\lbrace  t, x_1, x_2, x_3 \rbrace$
\item  $ A,B,C \dots$ for $SO(1,5)$ spinor indices. Upper position means right spinor, lower position left spinor.
\item  $\mu, \nu, \rho  \dots $ for vectorial indices in extended space-time directions $\lbrace  t, x_1, x_2, x_3 \rbrace$
\item  $i, j, k \dots $ for vectorial indices in extended spatial directions $\lbrace  x_1, x_2, x_3 \rbrace$
\item  $1,\bar{1},2,\bar{2} \dots $ for complex vectorial indices. In particular $\lbrace  1,\bar{1} \rbrace$ parametrize the first internal torus $\lbrace  y_1,\tilde{y}_1  \rbrace$, whilst $\lbrace  4,\bar{4} \rbrace \rightarrow \lbrace  t,x_3 \rbrace$, $\lbrace  5,\bar{5} \rbrace \rightarrow \lbrace  x_1,x_2 \rbrace$.
\end{itemize}

The left and right spin fields of directions $\lbrace  t, x_1, x_2, x_3 \rbrace$ are respectively bosonized by the scalar field $\varphi_I$ according to the formula:

\begin{equation}
\emph{left}: \qquad S_\alpha=e^{\frac{i}{2}(\pm \varphi_4 \pm \varphi_5)} \;\;\;  \#(-)={\rm even}
\end{equation}

\begin{equation}
\emph{right}: \qquad C_{\dot{\alpha}}=e^{\frac{i}{2}(\pm \varphi_4 \pm \varphi_5)} \;\;\;  \#(-)={\rm odd}
\end{equation}

The complex fermions $ \Psi$ are built from the real ones $\psi$ via\footnote{It's convenient to treat the torus $\lbrace  4,\bar{4} \rbrace \rightarrow \lbrace  t,x_3 \rbrace$ in a different way by defining  $\Psi^4=\frac{1}{\sqrt{2}}(\psi^{0}+\psi^{3}),  \;\;  \Psi^{\bar 4}=\frac{1}{\sqrt{2}}(\psi^{0}-\psi^{3})$ in order to use $\bar{\sigma}^{\mu}={\lbrace \bar{\sigma}^0,\bar{\sigma}^1,\bar{\sigma}^2,\bar{\sigma}^3} \rbrace$. Otherwise one can use \eqref{complefermio} and add an $i$ factor in front of $\bar{\sigma}^3$ in the definition of $\bar{\sigma}^{\mu}$.}

\begin{equation}
  \Psi^I=\frac{1}{\sqrt{2}}(\psi^{2I-1}+i\psi^{2I})  \;\;\;\;  \Psi^{\bar I}=\frac{1}{\sqrt{2}}(\psi^{2I-1}-i\psi^{2I})
  \label{complefermio}
\end{equation}

and are bosonized as:

\begin{equation}
  \Psi^I=e^{i \varphi_I}
\end{equation}

In the text we will always use the notation $\psi$, even when referring to the complex fermion, as it should be clear from the context which of the two is used. Cocycle factors, needed to implement anticommutation, will be also omitted.

  \subsection{OPEs and Correlators} 

We choose length units where  $\alpha'=2$.
         
\begin{equation}
   \langle  \psi^{\mu}(z) \psi^{\nu}(w) \rangle= \frac{\eta^{\mu \nu}}{(z-w)} 
\end{equation}

\begin{equation}
   \langle  X^{\mu}(z) X^{\nu}(w) \rangle= - \eta^{\mu \nu} \, log(z-w) 
\end{equation}

\begin{equation}
   \langle \partial_z X^{\mu}(z) e^{ikX}(w) \rangle= -\frac{ik^{\mu}}{(z-w)} 
\end{equation}

\begin{equation}
  \langle \prod^{N}_{i,j=1} e^{q_i \varphi(z_i) }e^{q_j \varphi(z_j) }\rangle=  
 \prod^{N}_{i<j} (z_i-z_j)^{ q_i \cdot q_j}
\end{equation}

\begin{equation}
  \langle \prod^{N}_{i,j=1} e^{i\lambda_i \varphi(z_i) }e^{i\lambda_j \varphi(z_j) }\rangle=  
 \prod^{N}_{i<j} (z_i-z_j)^{ \lambda_i \cdot \lambda_j}
\end{equation}

 We use the following basics OPEs in four dimensions:   
 
\begin{equation}
     C^{\dot{\alpha}} (z) S^{\beta} (w) \sim -\frac{1}{\sqrt{2 }} (\bar{\sigma}_\mu)^{ \dot{\alpha}\beta} \psi^\mu(w) 
     \label{opecs}
\end{equation}

\begin{equation}
   \psi^\mu \sim \frac{1}{\sqrt{2}} (\bar{\sigma}^\mu)^{\dot{\beta}\alpha}   C_{\dot{\beta}} S_{\alpha}
   \label{psi}
\end{equation}

\begin{equation}
  S_{\alpha}(z)  S_{\beta} (w)\sim \frac{\epsilon_{\alpha \beta}}{(z-w)^{1/2}}
  \label{opes}
\end{equation}

\begin{equation}
  C_{\dot{\alpha}}(z)  C_{\dot{\beta}}(w) \sim - \frac{\epsilon_{\dot{\alpha}\dot{\beta}}}{(z-w)^{1/2}} 
  \label{opec}
\end{equation}

\begin{equation}
   \psi^m(z) C^{\dot{\beta}} (w) \sim \frac{1}{\sqrt{2}} \frac{ (\bar{\sigma}^m)^{ \dot{\beta}\alpha} S_\alpha(w) }{(z-w)^{1/2}} 
\end{equation}

\begin{equation}
   \psi^\mu(z) S_{\alpha} (w) \sim -\frac{1}{\sqrt{2}} \frac{ (\sigma^\mu)_{\alpha \dot{\beta}} C^{\dot{\beta}}(w) }{(z-w)^{1/2}} 
\end{equation}

\begin{equation}
   \psi^\mu   \psi^\nu (z) S_{\alpha} (w) \sim  \frac{  (\sigma^{\mu\nu})_{\alpha}^{\,\,\,\,\,\beta} S_{\beta}(w) }{(z-w)} 
\end{equation}

\begin{equation}
   \psi^\mu   \psi^\nu (z) C^{\dot{\alpha}} (w) \sim  \frac{ (\bar{\sigma}^{\mu\nu})^{\dot{\alpha}}_{\,\,\,\,\,\dot{\beta}} C^{\dot{\beta}}(w) }{(z-w)} 
\end{equation}

 \subsection{How to Fix the Normalization}
 
  As an example, let's explictly fix the normalization of \eqref{opecs}.

\begin{equation}
 S^{1} =e^{\frac{i}{2} (\varphi_1+\varphi_2)}\;\;\;\;\;
S^{2}=e^{\frac{i}{2} (-\varphi_1-\varphi_2)}
\end{equation}

\begin{equation}
C^{\dot 1}= e^{\frac{i}{2} (-\varphi_1+\varphi_2)} \;\;\;\;\;
 C^{\dot 2}=e^{\frac{i}{2} (+\varphi_1-\varphi_2)}
\end{equation}

Therefore, choosing $\dot \alpha =\dot 1, \; \beta=2$ and using \eqref{complefermio}:

\begin{equation}
     C^{\dot 1} (z) S^{2} (w) \sim e^{-\varphi_1} = \Psi^{\bar 1}= \frac{1}{\sqrt{2}} (\psi^1-i \psi^2) =N (\bar{\sigma}_\mu)^{ \dot{1}2} \psi^{\mu} 
\end{equation}

So we have:

\begin{equation}
   N (\bar{\sigma}_\mu)^{ \dot{1}2} \psi^\mu = N(\psi^1  (\bar{\sigma}_1)^{ \dot{1}2}  +\psi^2  (\bar{\sigma}_2)^{ \dot{1}2} )=N(\psi^1  (-1) +\psi^2  (i) )=-N(\psi^1 -i \psi^2 )
\end{equation}

From the comparison, it follows that $N=-\frac{1}{\sqrt{2}}$.
 
 \pagebreak
 
\section{ The ten dimensional solution and its 4d reduction  }
\label{appN1}
In this appendix we collect some details on the dimensional reduction down to four dimensions of the eight harmonic family of BPS solutions describing a general system of intersecting D3-branes on $T^6$.

\subsection{The ten dimensional solution }
 
The eight harmonic family of BPS solutions associated to D3-branes intersecting  on $T^6$ is characterised by  a metric  $g_{MN}$ and a four form Ramond field $C_4$ of the form \cite{Bianchi:2016bgx}
\begin{align}   \label{ansatz}
ds^2  &= g_{\mu\nu} \,dx^\mu\, dx^\nu+\sum_{I=1}^3 h_{mn}^I \,dy_I^m\, dy_I^n\nn \\
C_4&=  C_{\mu, mnp}\,dx^\mu \wedge dy_1^m\wedge dy_2^n\wedge dy_3^p,
\end{align}
with $\mu=0,\dots 3$, $m=1,2$.   $x^\mu$ are the coordinates along the four-dimensional space time and $y^m_I=(y_I,\tilde y_I)$ span a $T^2\times T^2\times T^2$ torus with $I=1,2,3$ labelling the three two-torus. More precisely we write \footnote{
  In matrix form  
\eq
h^{I}_{mn}=  \frac{1}{Im U_I}\pmat 1 & \text{Re}\, U_I\\ \text{Re}\, U_I& | U_I|^2\fpmat,    \quad\quad 
h^{mn}_I=  \frac{1}{Im U_I} \pmat | U_I|^2 &-\text{Re}\, U_I\\-\text{Re}\, U_I & 1\fpmat.
\feq 
 }
  \begin{align}
ds^2  &=-e^{2U}\pt{dt+w}^2+e^{-2U}\sum_{i=1}^{3}dx^2_i+\sum_{I=1}^3     {1\over  \text{Im}\, U_I  } \left| dy_I+U_I d\tilde y_I \right|^2 \nn \\
C_4&=  A_{\Lambda } \gamma^\Lambda    =A_a \gamma^a + A_{\dot a}\,  \gamma^{\dot a}
\end{align}
with  $\alpha^\Lambda$ one forms in four dimensions and $\gamma_\Lambda$ three forms in the internal torus. $\Lambda=(mnp)=(a,\dot a)$ is a collective index labelling the 8 different 
three cycles $[mnp]$ on $T^6$ entering in the solution
 \bea
 \gamma^a &=(dy_1 \wedge dy_2\wedge dy_3,  d \tilde y_1 \wedge dy_2\wedge dy_3,     dy_1 \wedge d\tilde y_2\wedge dy_3,dy_1 \wedge dy_2\wedge d\tilde  y_3         )  \nn\\
\gamma^{\dot a} &=( d\tilde y_1 \wedge d\tilde y_2\wedge d\tilde y_3,  d  y_1 \wedge d\tilde y_2\wedge d\tilde y_3,     d\tilde y_1 \wedge d y_2\wedge d\tilde y_3,d\tilde y_1 \wedge d\tilde y_2\wedge d y_3         ) 
\eea
 All functions entering in the metric and four form can be written in   terms 
  of eight harmonic functions 
    \bea
  \{ V, L_I, K^I, M \} \label{harmh}
   \eea
  on $\mathbb{R}^3$ or equivalently in terms of the following combination  
  \begin{align}
Z_I&=L_I+\frac{|\ve_{IJK}|}{2}\frac{K^JK^K}{V},\nn\\
\m&=\frac{M}{2}+\frac{L_IK^I}{2V}+\frac{|\ve_{IJK}|}{6}\frac{K^IK^JK^K}{V^2}. 
\end{align}
 with $\epsilon_{IJK}$ characterizing the triple  intersections  between two cycles on $T^6$.
 One finds
  \begin{align}  
 e^{-4U}& = Z_1Z_2Z_3V-\m^2V^2, \nn\\
 U_I & =\text{Re}\, U_I+iIm U_I = -b^I+i\pt{Ve^{2U}Z_I}^{-1}   \quad\quad   b^I={K^I\over V} -\frac{\m}{Z_I},\nn\\
A_a & =( \al, \al^I-b^I\alpha)\nn \\
 A_{\dot a} & 
 =\left(\beta-b^1b^2b^3\al-\beta_I b^I+\frac{1}{2}|\ve_{IJK}|\al^I b^J b^K   , ~ \beta^I+   |\epsilon_{IJK}| \pt{\alpha b^Jb^K-2b^J\al^K}    \right)   
\end{align}
Finally the one-forms $\alpha,\alpha^I,\beta,\beta_I$ are defined in terms of the eight harmonic function via
\begin{align}
\al&=w_0-\m V^2 e^{4U}\pt{dt+w},\nn\\
\al^I&=-\frac{dt+w}{Z_I}+b^I w_0+w^I,\nn\\
\beta &=-v_0+\frac{e^{-4U}}{V^2 Z_1Z_2Z_3}\pt{dt+w}-b^Iv_I+b^1b^2b^3 w_0+\frac{|\epsilon_{IJK}|}{2}b^I b^J w^K,\nn\\
\beta_I&=-v_I+\frac{|\epsilon_{IJK}|}{2}\pg{\frac{\m \pt{dt+w}}{Z_JZ_K}+b^Jb^K w_0+2b^Jw^K} 
\end{align}
and  
 \begin{align} \label{eq: campi hodge harm}
*_3d w_0&=dV, \quad *_3dw^I=-dK^I , \quad *_3dv_0=dM, \quad 
*_3dv_I=dL_I \nn\\
*_3dw&=\frac{1}{2}\pt{VdM-MdV+K^IdL_I-L_IdK^I}
\end{align}

\subsection{The four dimensional model}

 After reduction to four dimensions, the ten dimensional solution can be viewed as a solution of a ${\cal N}=2$ truncation of maximal supersymmetric
   supergravity involving the gravity multiplet and three vector multiplets. The four dimensional model arises as a dimensional reduction of the ten dimensional lagrangian
   \eq
\lagr  = \sqrt{g_{10} }  \left( R_{10}  -\frac{1}{4 {\cdot} 5!}F_{M_1\dots M_5}F^{M_1\dots M_5} \right)  \label{l10}
\feq 
   Plugging  the ansatz (\ref{ansatz})  into (\ref{l10}) and taking all fields varying only along the four-dimensional spacetime one finds
  \eq
\lagr = \sqrt{g_4} \left(R_4-\sum_{I=1}^3 \frac{\partial_\m  U_I\partial^\m \bar{U}_I}{2\pt{Im U_{I}}^2}-\frac{1}{4{\cdot} 2!}F_{\m\n,\Lambda}F^{\m\n,\Lambda}\right)
\feq
 It is convenient to introduce a metric $\mathcal{H}^{\Lambda\Sigma}$ and its inverse to raise and lower the $\Lambda$ indices. One writes
\eq
\mathcal{H}^{abc,def}=h_1^{ad}\, h_2^{be}\, h_3^{cf}
\feq
or in matrix form
\eq
\mathcal{H}^{\La\Si}=\pmat \mathcal{H}_1& \mathcal{H}_2\\ \mathcal{H}_2^T & \mathcal{H}_3 \fpmat,
\feq
 with $\mathcal{H}_1^{ab}$,  $\mathcal{H}_2^{a\dot a}$, $\mathcal{H}_3^{\dot a \dot b}$   $4\times 4$ matrices. 
The self-duality condition of the five form field in ten dimensions  
 \eq
F_{\m\n abc}=\frac{\sqrt{g_{10}}}{2}  \epsilon_{\m\n\rho\si abc def}F^{\rho\si def},
\feq
reduces to 
\eq
F_{a bc}=\epsilon_{abc def}  \widetilde{F}^{def}   \quad \Leftrightarrow    \quad  F_\Lambda=\epsilon_{\Lambda \Sigma} \widetilde{F}^{\Sigma} 
\feq
with  $\widetilde{F}=* _4 F$ and   $\epsilon_{\Lambda \Sigma}$ an block off-diagonal antisymmetric matrix with the only non-trivial components
 \eq
\epsilon_{0\dot{0}}=-\epsilon_{I\dot{I}}=-\epsilon_{\dot{0} 0}=\epsilon_{\dot{I}I}=1  
\feq
In components
\eq\label{selfd}
F_{a}=\epsilon_{a \dot{a} }\widetilde{F}^{\dot a}   \quad\quad      F_{\dot a}=\epsilon_{\dot{a}  a }\widetilde{F}^{ a} 
\feq
These self-duality relations can be used to express the components $F_{\dot a}$ in terms of the Poincare' duals of $F_a$. Indeed, inverting the first equation in (\ref{selfd}) one finds\footnote{Here we use $*_4^2=-1$.
}
 \eq\label{selfd2}
       F_{\dot{a}}=-(\mathcal{H}_3^{-1})_{\dot{a}\dot{b}}\left( \epsilon^{\dot{b}c} \widetilde{F}_{c}+  \mathcal{H}_2^{\dot{b}c}F_c \right) 
\feq
 with $\epsilon^{\dot{a} a }= {\rm diag} (1,-1,-1,-1)$ the inverse of $\epsilon_{a \dot{a}  }$. Using these relations one  can write
 \begin{align} 
F_\Lambda F^\Lambda  & ={\cal L}_{stu}+{\cal L}_{top}
 \end{align}
 with $ {\cal L}_{top} =  -2\epsilon^{a\dot a}  \widetilde{F}_{\dot a} F_a$ a total derivative, 
\begin{align}
{\cal L}_{stu} & = 2 ( F_a\mathcal{I}^{ab}F_b+ F_a\mathcal{R}^{ab} \widetilde{F}_b)
\end{align}
and \footnote{Equivalently
\begin{align}
\mathcal{I}^{ab}&\equiv\mathcal{H}_1^{ab}+\epsilon^{a\dot{b}}(\mathcal{H}_3^{-1})_{\dot{b}\dot{c}}\epsilon^{\dot{c}b}-\mathcal{H}_2^{a\dot{b}}(\mathcal{H}_3^{-1})_{\dot{c}\dot{d}}\mathcal{H}_2^{\dot{d}b} \nn\\
\mathcal{R}^{ab}&\equiv \epsilon^{a\dot{b}} (\mathcal{H}_3^{-1})_{\dot{b}\dot{c}}\mathcal{H}_2^{\dot{c}b}+\mathcal{H}_2^{a\dot{b}}(\mathcal{H}_3^{-1})_{\dot{b}\dot{c}}\epsilon^{\dot{c}b} \nn 
\end{align}
}
\begin{align*}
\mathcal{I}^{ab}&= stu \pmat 1+\frac{\si^2}{s^2}+\frac{\tau^2}{t^2}+\frac{\nu^2}{u^2} & -\frac{\si}{s^2}&-\frac{\tau}{t^2}&-\frac{\nu}{u^2} \\
-\frac{\si}{s^2}  & \frac{1}{s^2} &0 &0 \\
-\frac{\tau}{t^2}  & 0 &\frac{1}{t^2}&0\\
-\frac{\nu}{u^2}  & 0 & 0 &\frac{1}{u^2}\fpmat		\quad
\mathcal{R}^{ab}=\pmat 2\si\tau\nu & -\tau\nu&-\sigma\nu& -\sigma\tau \\
-\tau\nu & 0 & \nu &\tau \\
-\sigma\nu  &\nu &0&\sigma\\
-\sigma\tau   &\tau &\sigma &0 \fpmat \num
\end{align*}
where
\begin{align}
U_I&=(\si+is,\tau+it,\nu+i u ) 
\end{align}
 Discarding the total derivative term, the four-dimensional  Lagrangian can then be written as
\eq
\lagr =  \sqrt{g_4}\left( R_4-\sum_{I=1}^3 \frac{\partial_\m U_I\partial^\m \bar{U}_I}{2\pt{Im U_{I}}^2}-\frac{1}{4}F_a \mathcal{I}^{ab}F_{b}-\frac{1}{4}F_{a}\mathcal{R}^{ab} \widetilde{F}_{b} \right)
\feq
The equations of motion read
\begin{align}
 & R_{\m\n} -\frac{1}{2} g_{\m\n} =   {1\over 2\pt{Im U_{I}}^2 } \left( \partial_\m U_I \partial_\n U_I - \frac{1}{2} g_{\m\n} (\partial U_I) ^2 \right) 
 + \frac{1}{2} {\cal I}^{ab} \left( F_{a\mu \sigma}  F_{b\n}^\sigma-\quar  g_{\m\n}F_a\, F_b \right) \nn \\
& \quad\quad\quad\quad   + \frac{1}{2} {\cal R}^{ab} \left( F_{a\mu \sigma}  \tilde F_{b\n}^\sigma-\quar  g_{\m\n}F_a\, \tilde F_b \right) \nn\\
 & \nabla_\m\pg{    {\cal I  }^{ab} F_b^{\m\n}+ {\cal R}^{ab} \tilde F_b^{\m\n} }=0  \nn\\
&  -\nabla_\m \frac{\nabla^\m U_I}{\pt{Im U_I}^2}= {\rm i}   \frac{\partial_\m U_I\partial^\m \bar{U}_I}{\pt{Im U_{I}}^3}+\frac{1}{2}F_a  {\partial \mathcal{I}^{ab} \over \partial \bar{U}_I} F_{b}+\frac{1}{2}F_{a}{\partial \mathcal{R}^{ab} \over \partial \bar{U}_I}   \widetilde{F}_{b}   \label{eom}
\end{align}

\subsection{The basic solutions}
\label{appN1basic}

A family of supersymmetric solutions to equations (\ref{eom}) is given in \cite{Bianchi:2016bgx}.  These solutions can be viewed as made of three different types of solutions, referred as K, L or M. In the following  we display a representative of solution in each type.

\subsubsection{L solutions}
 
 The L class of solutions can be represented by the choice 
\begin{align}
V &\equiv L\pt{x},\quad M=K^I=0,\quad L_I=1 \quad \Rightarrow \quad Z_I=1,\quad \m=0\nn\\
 \mathcal{I} &=\diag{L^{-3/2},L^{-1/2},L^{-1/2},L^{-1/2}}		\qquad
\mathcal{R}=0
\end{align}
  The solution can be written as
\begin{align}
ds^2 &=-L^{-{1\over 2}}dt^2+L^{1\over 2} \sum_{i=1}^3 dx^2_i \nn\\
 A_0  &=  w_0   \qquad   *_3 d w_0=dL \nn\\
 U_1&=U_2=U_3={\rm i}\, L^{-{1\over 2}}
\end{align}

\subsubsection{K solutions}
 
 The K solutions correspond to the choice
\begin{align}
K^3 &=-M\equiv K\pt{x},\quad L_I=V=1,\quad K^1=K^2=0  \quad \Rightarrow \quad Z_I=1,\quad \m=0 \nn\\
\mathcal{I} &=\pmat 1+K^2 & 0&0&K \\
0 & 1 &0 &0 \\
0 & 0 &1&0\\
K & 0 & 0 &1\fpmat		\qquad ~~~~~~~~~~
\mathcal{R} =\pmat 0 &0&0&0 \\
0 & 0 & -K &0 \\
0 & -K &0&0\\
0 &0 & 0 &0 \fpmat \num
\end{align}
 
The solution is given by
\begin{align}
ds^2&=- \pt{dt+w}^2+\sum_{i=1}^3 dx^2_i,  \nn\\
U_1 &=U_2={\rm i} \quad\quad   U_3=-K+{\rm i} \nn\\
A_0&=A_3=0   \quad\quad\quad \quad  A_1=A_2= -w       \quad\quad    *_3 dw =-dK 
\end{align}

\subsubsection{M solutions}

The M solutions correspond to the choice
\begin{align}
K^2 &=M\equiv M\pt{x},\qquad L_I=V=1,\qquad K^1=K^3=0  \quad \Rightarrow \quad
\mu=M \quad Z_I=1,\quad    \nn\\
\mathcal{I} &=a^{3/2}\pmat 1+2\frac{M^2}{a} & -\frac{M}{a}&0&-\frac{M}{a} \\
-\frac{M}{a} & a^{-1} &0 &0 \\
0 & 0 & a^{-1}&0\\
-\frac{M}{a} & 0 & 0  & a^{-1}\fpmat		\quad\quad
\mathcal{R}=\pmat 0 & 0&-M^2& 0 \\
0 & 0 & M&0 \\
-M^2  & M &0&M\\
0 & 0 & M  &0 \fpmat \num
\end{align}
with $a=1-M^2$.
The solution is given by
\begin{align}
ds^2&=- { dt^2 \over \sqrt{1-M^2} }+\sqrt{1-M^2} \sum_{i=1}^3 dx^2_i,  \nn\\
U_1 &=U_3=M+{\rm i} \, \sqrt{1-M^2}  \quad\quad U_2={\rm i} \, \sqrt{1-M^2}  \nn\\
A_0&=-\frac{M\, dt}{1-M^2}  \quad\quad  A_1=A_3= -\frac{dt}{1-M^2}   \quad\quad  A_2=w_2 \qquad   *_3 dw_2=-d M
\end{align}

\subsection{Sub-family of solutions}
\label{appN2}

For completeness, we list some interesting sub-families of solutions included in the eight-harmonic class. 

 \subsubsection{No scalars: IWP solution }
 
 Einstein-Maxwell theory can be embedded in four dimensional supergravity by restricting to geometries with a trivial internal square metric 
  \bea
V e^{2U} Z =1 \qquad  b^I=0 
\label{nointernal}
\eea
  These equations can be solved in terms of two harmonic functions $\text{Re}\, H$ and $ \text{Im}\, H$ via the identifications
\bea
V &=& L_1=L_2=L_3=\text{Im}\,H  \nn\\
-M&=& K^1  = K^2=K^3=\text{Re}\, H
\eea
The general solution reduces to\footnote{ In our conventions the Einstein-Maxwell lagrangian reads
\bea
{\cal L}= \sqrt{g} \left[   R-\frac{1}{4 } \, F^2 \right]
\eea
}
\bea
ds^2 &=& -  \left| H  \right|^{-2} \, (dt+w)^2 + \left| H  \right|^2 \,  d |\mbf{x}|^2 \nn\\
A_0 &=&   w_0 -{ \text{Re}\,H \over \left| H  \right|^2  } \, (dt +w)  \nn\\
    A_1 &=& A_2=A_3=  w^1 -{ \text{Im}\,H \over \left| H  \right|^2  } \, (dt +w)\label{iwp}  
\eea
with $H$ a complex harmonic function
 \be
 H=\text{Re}\, H+i\,\text{Im}\, H \qquad\quad \quad  \nabla^2 H= 0
 \ee
  and $w$ and $w_0$, $w^1$ one forms defined as
   \bea
d w&=&  {\rm i} *_3  [H  dH^*-H^* dH ] \nn\\
d w_0 &=& *_3  d \, \text{Im}\, H   \qquad    d w^1 = *_3  d \, \text{Re}\, H \label{wb}
\eea
 We notice that the contribution to the stress energy tensor of gauge fields $A_I$ exactly match that of $A_0$ , so we can replace the four gauge fields by a 
 single one given by
\be
 A =  2 w_0 -{ 2\text{Re}\,H \over \left| H  \right|^2  } \, (dt +w)  \nn\\
\ee
   The resulting solution is known in the General Relativity literature as IWP ( after Israel, Wilson and Perjes  \cite{Perjes:1971gv,Israel:1972vx} ) and
  includes very well known examples of solutions of Maxwell-Einstein gravity:
\begin{itemize}
\item{$AdS_2\times S^2$. The harmonic function $H$ reads\footnote{ Global coordinates are defined by
\be
(x_1,x_2,x_3)  = \left(  \sqrt{(\rho^2 +L^2) (1-\chi^2) }  \, \cos\phi, \sqrt{(\rho^2 +L^2)(1-\chi^2) }\, \sin\phi,  \rho\,  \chi \right)  
\ee
with $\rho \in (-\infty,\infty)$,  $\chi \in [-1, 1]$, $ \phi \in [0,2 \pi]$. These coordinates cover twice the flat space  with
 the points $(\rho,\chi)$ and $(-\rho,-\chi)$  identified.
}
\be
H={1\over \sqrt{ x_1^2+x_2^2+(x_3-{\rm i} L)^2 } }
\ee
The geometry is regular everywhere. An infinite class of regular IWP geometries, obtained as bubbling of $AdS_2\times S^2$ and parametrised by a string profile function has been recently constructed in \cite{Lunin:2015hma}. 
}

\item{ Kerr-Newman solution with  $M=Q=q$, $P=0$, $J=q L$. The harmonic function $H$ reads 
\be
H=1+{q\over \sqrt{ x_1^2+x_2^2+(x_3-{\rm i} L)^2 } }
\ee
The geometry has a naked curvature singularity at the zero of $H$.   
} 

\item{ Reissner-Nordstrom with  $M=Q=q$, $P=J=0$. The harmonic function $H$ reads 
\be
H=1+{q \over \sqrt{ x_1^2+x_2^2+x_3^2 } }
\ee
The geometry has  a curvature singularities at the zero of $H$.
   
}

\item{ Charged Taub-NUT with  $M=Q=b_1$, $P=-b_2$, $J=0$. The harmonic function $H$ reads 
\be
H=1+{b_1+{\rm i} \, b_2\over \sqrt{ x_1^2+x_2^2+x_3^2 } }
\ee
The geometry has  no curvature singularities but it has a Dirac-Misner string-like singularity.
   
}

\end{itemize}

\subsubsection{One complex scalar:  SWIP solutions}

Next, we consider a solutions with single active scalar field, let us say $U_1$,  with $U_2=U_3={\rm i}$. These conditions can be solved in terms of two 
complex harmonic functions $H_1$, $H_2$ after the identification
\begin{align}
L_1 &=V = \text{Re}\,H_2,\qquad L_2=L_3\equiv Im H_1\qquad K^1=-M = \text{Re}\, H_1\,\qquad K^2=K^3\equiv -Im H_2\label{klmswip}
\end{align}
 For this choice the general solution reduces to
\begin{align}
ds^2&=-  \pq{Im\pt{H_1\bar{H}_2}}^{-1}\pt{dt+w}^2 +Im\pt{H_1\bar{H}_2} \sum_{i=1}^3 dx^2_i, \nn \\
 U_1 &=\frac{H_1}{H_2},\qquad U_2=U_3={\rm i},  \nn\\
A_0&=w_0+\frac{Im H_2}{Im\pt{H_1\bar{H_2}}}\pt{dt+w},  \qquad  A_1=w^1-\frac{Im H_1}{Im\pt{H_1\bar{H_2}}}\pt{dt+w}  \nn \\
  A_2&=A_3=w^2-\frac{\text{Re}\, H_2}{Im\pt{H_1\bar{H_2}}}\pt{dt+w}\nn\\
  *_3 dw &=-\,\text{Re}\, \pt{H_1 d\bar{H_2}-\bar{H}_2 dH_1}\nn\\
*_3d w_0&=\text{Re}\, dH_2,\qquad *_3dw^1=-\text{Re}\, dH_1,\qquad *_3dw^2=Im dH_2.
\end{align}
  The IWP  class  corresponds to the choice $H_1={\rm i} H_2= H$. See for instance \cite{Ortin:2015hya} for more information on the SWIP solution.
 
  \subsubsection{Two complex scalars}
This solution corresponds to the choice
\begin{align}
  L_3=L_2    \qquad    K^3=K^2  \label{klmswip2}
 \end{align}
 leading to 
 \bea
 Z_1 &=&   L_1+{(K^2)^2\over V}  \qquad\qquad    Z_2 =Z_3=   L_2+{K^1 K^2\over V} \nn\\
 \mu &=&  { M \over 2} + { K^1 \,(K^2)^2   \over   V^2} + { K^1\, L_1 \over   2 V}+
 {  K^2 L_2 \over V}
 \eea
The solution reads
\begin{align}
ds^2&=-e^{-2U}\pt{ dt+w}^2+e^{2U}\sum_{i=1}^3 dx^2_i,   \qquad e^{-4 U} =Z_1 Z_2^2 V-\mu^2 V^2\nn\\
U_1 &=-b^1+{\rm i} (e^{2U} V Z_1)^{-1}   \qquad   U_2=U_3=-b^2+{\rm i} (e^{2U} V Z_2)^{-1} \nn\\
A_0&=w_0-\m V^2 e^{4U}\pt{dt+w}     \qquad      A_1 =w^1   + V e^{4U} \, \pt{dt+w}   
  \left(  Z_2^2  -K^1\, \m     \right) \nn\\
  A_2 & =A_3 =w^2  + V e^{4U} \, \pt{dt+w}   
  \left(  Z_1 Z_2  -K^I \, \m     \right)
\end{align}
with
\begin{align}
 *_3 dw &= \frac{1}{2}(V d M-M dV+K^1 dL_1 -L_1 dK^1+ 2 K^2 dL_2-2L_2 dK^2   )      \nn\\
 *_3d w_0&=dV,\qquad *_3dw^1=-dK^1 \qquad *_3dw^2= -dK^2  
\end{align}
and
\begin{align}
b^1 &={ K^1 L_1-2 K^2 L_2-M V \over 2\left[(K^2)^2+V L_1\right] }   \qquad
b^2 =b^3 =-{ K^1 L_1 +M V \over 2(K^1 K^2+V L_2 )}  
\end{align}
The SWIP solution is recovered for $L_1=V$ and $K^1=-M$ after the identifications (\ref{klmswip}).

 \pagebreak
 
  \section{T Dualities} 
  
 T-duality  acts on  SUGRA fields according to the generalized Buscher rules:
 \be
\begin{aligned}
g'_{zz} &= \frac{1}{g_{zz}}, \quad\quad\quad e^{2 \phi '} = \frac{e^{2 \phi}}{g_{zz}}, \quad\quad\quad g'_{z m} = \frac{B_{z m}}{g_{zz}}, \quad\quad\quad B'_{z m}= \frac{g_{z m}}{g_{zz}} \\[8pt]
g'_{m n} &= g_{m n}-\frac{g_{m z}\,g_{n z} - B_{m z}\,B_{n z}}{g_{zz}}, \quad B'_{m n}= B_{m n}- \frac{B_{m z}\,g_{n z}-g_{m z}\,B_{n z}}{g_{zz}}\\[8pt]
C'^{(n)}_{m_1 ... m_{n}} &= C^{(n+1)}_{m_1 ... m_{n} z} - n\,C^{(n-1)}_{[m_1 ... m_{n-1}}\,B_{m_n ] z} - n(n-1)\frac{C^{(n-1)}_{[m_1 ... m_{n-2}| z}\,B_{|m_{n-1}| z}\,g_{|m_{n} ] z}}{g_{zz}}\\[8pt]
 C'^{(n)}_{m_1 ... m_{n-1}  z} &=C^{(n-1)}_{m_1 ... m_{n-1}} - \left( n - 1 \right) \frac{C^{(n-1)}_{[ m_1 ... m_{n-2} | z}\,g_{z | m_{n-1} ]}}{g_{zz}} \label{tdual}
\end{aligned}
\ee
where by $z$ we denote the direction along with T-duality is performed and the metric is  understood in the string frame. 

 \pagebreak
 
\section{Branes at Angle}

The scattering amplitudes computed for perpendicular branes can be generalized for branes at angle \cite{Pieri:2016pdt}, here we limit ourselves to show how to build the relevant vertex operators. The rotation matrix is now generalized to:

\begin{equation}
R^{M}_{\,\,\,\,\, N}= \begin{pmatrix} 1 & 0 & 0 & 0 &0\\
0 & -\mathbf{1}_{3 \times 3} & 0 & 0 & 0 \\ 
 0 & 0 & R(\theta_1) & 0 &0\\
  0 & 0 & 0 & R(\theta_2)  &0 \\
   0 & 0 & 0 & 0& R(\theta_3)  
 \end{pmatrix}  
 \label{refl}
\end{equation}

where we have introduced the two dimensional matrices:

\begin{equation}
R(\theta_i) = \begin{pmatrix} cos(2\theta_i) & sen(2\theta_i)\\
sen(2\theta_i) & -cos(2\theta_i) \end{pmatrix} 
\end{equation}

Boundary conditions imposed by intersecting branes at angle force the open strings stretched between the branes to have non integer mode expansions, therefore leading to vertex operators   involving angle-dependent bosonic and fermionic twist field. The T-dual picture is given by open strings ending on D-branes with magnetic fields switched on along the world-volume. In the following we will construct a supersymmetric system of 4 D3 branes  at angles and then we will identify the  vertex operators corresponding to each pair of branes.

 Every Dp-brane imposes some restrictions on the spinors $\epsilon_{L,R}$ parametrizing the SUSY transformations generated by $\epsilon_L Q_L+\epsilon_R Q_R$, where $Q_{L,R}$ are the supercharges (with the same 10d chirality) of $IIB$ superstring theory. Solutions to these constraints count the number of unbroken supercharges in the system and when angles are taken into account one finds that if the branes are related by $SU(3)$ rotations some supersymmetry is preserved \cite{Berkooz:1996km}. In particular, we require that at least $\mathcal{N}=1$ is preserved between every couples of branes, leading to a condition on the (relative) angles:

\bea
\theta_1+\theta_2+\theta_3=0 \qquad mod  \,\, 2\pi
\label{angle}
\eea

where $\theta_I$ is the angle between two branes in the $\lbrace  y_I,\tilde{y}_I  \rbrace$ torus. To preserve $\mathcal{N}=2$,  \eqref{angle} must be true and  an angle must be zero, and to have $\mathcal{N}=4$ all the angles need to be zero. 

\begin{figure}
\begin{center}
\includegraphics[width=8cm]{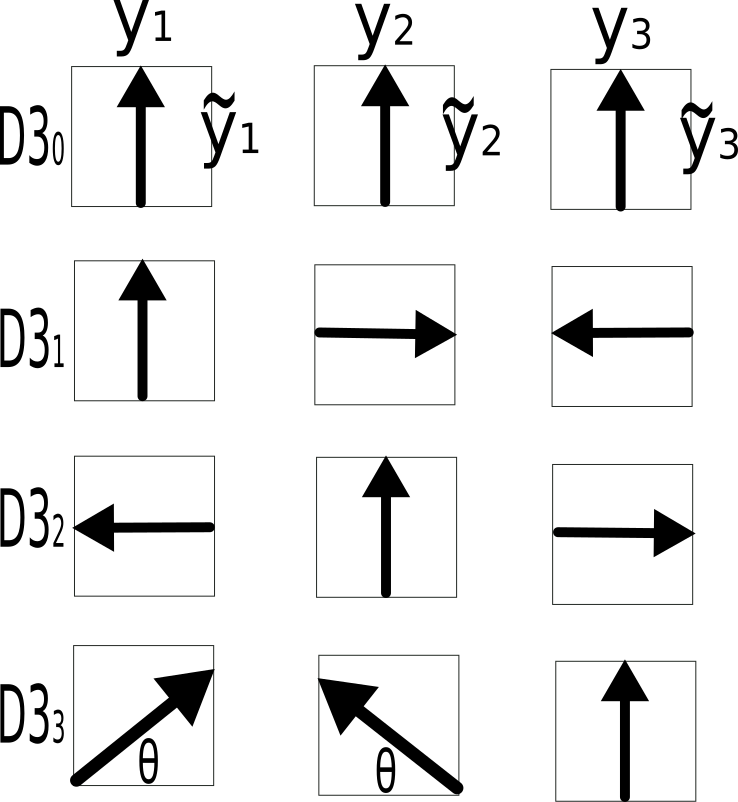} 
\end{center}
\caption{D3-brane configuration in the internal tori.}
\label{fig4}
\end{figure}

One can verify that the system of Fig.\eqref{fig4} satisfies \eqref{angle} for all the six possible pairs of branes.  The convention is that a positive angle is taken counterclockwise from the horizontal $y$ direction, so for instance a right arrow stands for $\theta=0$, a left arrow for $\theta=\pi$ and up arrow for $\theta=\pi/2$.

To be concrete, we construct the vertex operators for the fermionic open string stretched between the (ordered) pair $D3_0D3_3$  (for a more general discussion on vertex operators at angle see \cite{Cvetic:2006iz}). In the canonical $-\frac{1}{2}$ superghost picture the vertex operator is:\footnote{ The subscript $\mu$ in $V_{\mu}$ is a symbol for the polarization, not an index!}

   \begin{equation}
 V_{\mu}^{\lbrace 03 \rbrace}(z_n)=e^{-\frac{\varphi}{2}} \mu^{\beta}_{\lbrace 03 \rbrace}  S_{\beta} (\sigma^{(1)\dagger}_{1/2-\theta} e^{  i  \theta\varphi_1}) (\sigma^{(2)}_{1/2-\theta}e^{ - i \theta \varphi_2})(e^{ - i \varphi_3/2})e^{ik_{n}X}
\end{equation}

where $S_{\beta}$ is an $SO(1,3)$ spin field (see the appendix for notation and conventions) and $\mu^{\beta}$ is the fermionic polarization with Chan Paton indices  not shown explicitly. In particular every polarization $\mu^\beta$ is actually a Chan-Paton matrix, for instance $(\mu_{\lbrace 03 \rbrace}^{a \, b})^\beta$ refers to the open string stretched between the brane $a$ in the stack 0 and the brane $b$ in the stack 3 (the order is important). 

A bosonic twist $\sigma^{(I)}_{\xi}$ field must be inserted for every $I$-torus ($I=1,2,3$) in which the branes are not parallel. The twist field is angle  dependent and the angle $\xi$ is computed by rotating the first brane, the $D3_0$, to the second brane, with a dagger if the rotation is clockwise and without dagger if it is counterclockwise. Of course $\sigma_{\theta}^{\dagger}=\sigma_{1-\theta}$.  Similarly, one must add a fermionic twist $e^{i (\xi-\frac{1}{2}) \varphi_I}$ for every torus. As a check one can verify that the conformal dimension\footnote{The conformal dimension of a bosonic twist field is given by $h(\sigma_{\theta})=\frac{1}{2} \theta (1-\theta)$. Other useful formulas are $h(e^{q\varphi})= -\frac{1}{2} q^2-q$ and  $h(e^{i\lambda\varphi})= \frac{1}{2} \lambda^2$
.} of the vertex operator is equal to 1 and that the vertex is consistent with the SUSY condition \eqref{angle}. Indeed the action  of the $IIB$ positive chirality supercharge $Q^{(+++++)}=\oint dz e^{+\frac{i}{2}(\varphi_4+\varphi_5+\varphi_1+\varphi_2+\varphi_3)} e^{-\varphi /2}$  must be well defined, implying  that  no non-integer powers of $z$ can appear in the commutator $
[\epsilon Q ,V_{\mu}^{(-1/2)}]$, which gives the susy-related bosonic vertex operator.

The opposite chirality fermion, the one stretched from $D3_3$ to $D3_0$ is recovered by doing the conjugation of the previous vertex:

\begin{equation}
 V_{\bar \mu}^{\lbrace 30 \rbrace}(z_n)=e^{-\frac{\varphi}{2}} \bar \mu^{\dot{\beta}}_{\lbrace 30 \rbrace}  C_{\dot{\beta}} (\sigma^{(1)}_{1/2-\theta} e^{ - i  \theta\varphi_1}) (\sigma^{(2)\dagger}_{1/2-\theta}e^{  i \theta \varphi_2})(e^{  i \varphi_3/2})e^{ik_{n}X}
\end{equation}

The computation proceeds on the same line of the perpendicular ones, the only new non trivial correlators being the ones involving the twist fields:

\bea
 &&   
\left\langle   \sigma^{(3)}_{1/2} (z_1)    \sigma^{(3)}_{1/2} (z_3)  \right\rangle  = z_{13} ^{-\frac{1}{4} }    \nn\\
 &&   
\left\langle    \sigma^{(1)}_{1/2} (z_2)    \sigma^{(1)}_{\theta} (z_3) \sigma^{(1)}_{1/2-\theta} (z_4)   \right\rangle  =(4 \pi \Gamma_{ \lbrace\frac{1}{2}, \theta, \frac{1}{2}-\theta \rbrace})^{\frac{1}{4}} z_{23}^{-\frac{\theta}{2} }   z_{24}^{+\frac{\theta}{2}-\frac{1}{4} }   z_{34}^{-\frac{\theta}{2}+\theta^2  }   \nn\\
&&   
\left\langle    \sigma^{(2)}_{1/2} (z_1)    \sigma^{(2)}_{1/2} (z_2)    \sigma^{(2)}_{1/2-\theta} (z_3)  \sigma^{(2)}_{1/2+\theta} (z_4)   \right\rangle  = z_{12}^{-\frac{1}{4} } z_{34}^{-\frac{1}{4}+\theta^2 } \left( \frac{ z_{13} z_{24}}{ z_{14} z_{23}} \right)^{\frac{1}{4}} \mathcal{I}^{-\frac{1}{2}}[w]
\label{twist}
 \eea
 
 where we have used
 
 \bea
  \left\langle   \sigma_{\alpha} (z_1)\sigma_{\beta} (z_2)\sigma_{\gamma} (z_3)  \right\rangle  =(4 \pi \Gamma_{ \lbrace \alpha, \beta, \gamma \rbrace})^{\frac{1}{4}} z_{12}^{-\alpha \beta }  z_{13}^{-\alpha \gamma} z_{23}^{-\beta \gamma }    \nn\\
 \eea
 
\bea
\Gamma_{\alpha, \beta, \gamma}=\frac{\Gamma(1-\alpha)\Gamma(1-\beta)\Gamma(1-\gamma)}{\Gamma(\alpha)\Gamma(\beta)\Gamma(\gamma)}
\eea 

\bea
\left\langle    \sigma_{1-a} (z_1)  \sigma_{a} (z_2) \sigma_{1-b} (z_3)  \sigma_{b} (z_4)    \right\rangle  = z_{12}^{-a(1-a) } z_{34}^{-b(1-b) } \left( \frac{ z_{13} z_{24}}{ z_{14} z_{23}} \right)^{\frac{1}{2}(a+b)-ab} \mathcal{I}^{-\frac{1}{2}}(w)
\eea

\bea
\mathcal{I}[w]=\frac{1}{2\pi} (B_1(a,b) G_2(w) H_1(1-w)+B_2(a,b) G_1(w) H_2(1-w))
\eea

\bea
B_1(a,b)= \frac{\Gamma(a)\Gamma(1-b)}{\Gamma(1+a-b)} \qquad B_2(a,b)= \frac{\Gamma(b)\Gamma(1-a)}{\Gamma(1+b-a)}
\eea

\bea
G_1(w)= {}_2F_1[a,1-b; 1; w] \qquad G_2(w)= {}_2F_1[1-a,b; 1; w]
\eea

\bea
H_1(w)= {}_2F_1[a,1-b; 1+a-b; w] \qquad H_2(w)= {}_2F_1[1-a,b; 1-a+b; w]
\eea

\bea
w=\frac{ z_{12} z_{34}}{ z_{13} z_{24}}=1-x
\eea 

 where the last equality is valid after the gauge fixing 
 
  \be
  z_1=-\infty \qquad z_2=0\qquad z_3=x \qquad z_4=1 \qquad z_5=z \qquad z_6=\bar{z}
  \label{fix}
  \ee
  
   \pagebreak
   
  \section{GR: Definitions and Theorems} 
  
  \label{grDef}
  Here we alphabetically store some crucial definitions, theorems and useful properties employed throughout the thesis. Of course every definition involves the use of other terms which in turn have their own definition involving other concepts and so on, therefore a cut off must be done at some point.

\textbf{Birkhoff Theorem }: Schwarzschild metric is the unique vacuum solution outside  any localized source with spherical symmetry (the source can be even time dependent). This apply both to stars and black holes. 
There is NO Birkhoff Theorem for rotating spacetimes. It is not true that the geometry in vacuum regions outside a rotating stars is the Kerr metric, even though this is true asymptotically (higher multipoles quickly fall-off). Nonetheless  if a black hole is stationary, then is static or axially symmetric. If is axially symmetric then it must be a Kerr black hole. (Poisson pag. 204-206, Visser p.5)
In presence of charge substitute Schwarzschild with Reissner-Nordstrom and Kerr with Kerr-Newman.
 
\textbf{Black Holes}: A strongly asymptotically predictable space-time $M$ is said to contain a BH if $M$ is not contained in $J^-(\mathscr{I}^+)$. The BH region is defined as $B= \left[ M - J^-(\mathscr{I}^+)  \right]$ and the boundary is the event horizon.  To concretely locate an event horizon for the common case of stationary, asymptotically flat space-time in  spherical coordinates and in which surfaces $r=const.$ are timelike until some fixed $r=r_H$, one looks at the points at which the $r=const.$ become null, \i.e. when the normal $\partial_\mu r$ becomes null: $g^{rr}(r_H)=0$ 

 \textbf{BPS Black Holes and Rotation}: Supersymmetric black holes are special cases of stationary black holes. A stationary (asymptotically flat) black hole spacetime admits a Killing vector field  k that is timelike near spatial infinity, and unique up to normalization. However, there may be interior regions outside the horizon, called ‘ergoregions’, within which k is spacelike; in fact, an event horizon with a non-zero angular velocity necesarily lies within an ergoregion. Supersymmetric spacetimes cannot have ergoregions, however, because supersymmetry implies that k can be expressed in terms of a Killing spinor field, and this expression allows k to be timelike or null but not spacelike. It follows that the event horizon of a supersymmetric black hole must be non-rotating \cite{Townsend:2002yf}, even though angular momentum can be non zero if we consider solutions with electric-magnetic fields, see for instance \cite{Elvang:2005sa}.
 
\textbf{Closed Timelike Curve}: A closed curve with proper negative length (signature $(-+++)$). For instance, in cylindrical coordinates ${t,r,\phi,z}$ with $\phi$ periodic and spacelike, the condition for a CTC is $g_{\phi \phi}|_{t,r,z}<0$ at fixed ${t,r,z}$.
A slightly stronger requirement for the space-time is to be stably causal and for the t coordinate to be a global time function. In this way, t will then be monotonic increasing on future-directed non-space-like curves and hence there can be no CTC’s. The coordinate t is a time function if and only if $-g^{tt}>0$.
 
\textbf{Cosmic Censor Conjecture}: All physically reasonable space-time are globally hyperbolic, \i.e. apart from a possible initial singularity, no singularity is ever visible to any observer. 

 \textbf{Ergosurface}: Given a asymptotically time-translation Killing vector $K^{\mu}=\partial_t$, the ergosurface is the surface where $K^{\mu}K_{\mu}=0$. Since we can define a stationary observer as ones whose four-velocity is parallel to $K^{\mu}$, inside the ergosurface, in the ergoregion, $K^{\mu}$ becomes spacelike and the observer cannot remain stationary.   
 
 \textbf{Killing Horizon}: If a Killing vector $\chi$ field is null along some null hypersurface $\Sigma$, we say that $\Sigma$ is a Killing Horizon. Every event horizon in a stationary, asymptotically flat space-time is a Killing horizon for some Killing vector. If a space-time is static, the Killing vector is $\partial_t$.

 \textbf{Naked Singularity}: An asymptotically flat space-time $M$ which fails to be strongly asymptotically predictable is said to posses a Naked Singularity. 

\textbf{No Hair Theorem}: Stationary, asymptotically flat black hole solutions to GR coupled to electromagnetism that are non singular outside the horizon are fully characterized by mass, charge and angular momentum. Here we assume that electromagnetism is the only long range non gravitational field. This result actually depends on the matter content of the theory, so in general there will be some additional parameters, but the crucial point remain: only a small finite number of parameters fully characterize the solution.

\textbf{Strongly asymptotically predictable space-time}: Let $(M,g_{\mu \nu})$ be an asymptotically flat space-time with associated unphysical Weyl rescaled space-time $(\tilde{M},\tilde{g}_{\mu \nu})$. If in the unphysical space-time there is an open region $\tilde{V} \subset \tilde{M}$ with the closure of $\bar{M \cap J^-(\mathscr{I}^+)} \subset \tilde{V} $ such that $(\tilde{V},\tilde{g}_{\mu \nu})$ is globally hyperbolic. 
 
 \textbf{Surface Gravity}: A given Killing vector fields $\chi$ obeys the geodesic equation along the Killing horizon: $\chi^{\mu} \nabla_{\mu} \chi=-  \kappa \chi$, where $\kappa$ is a constant (in module) called surface gravity. In a static, asymptotically flat spacetime it's the acceleration of a static observer neat the horizon, as measured by a static observer at infinity.
 
\textbf{Trapped Surface}: A compact, two-dimensional, smooth spacelike submanifold $T$ having the property that the expansion $\theta$ of both ingoing and outgoing (respectively $v$ and $u$ constant) future directed null geodesics orthogonal to $T$ is everywhere negative.
It's always inside the event horizon
unless the null energy condition is violated and for stationary black holes is coincident with the event horizon.
 Extremal black holes don't have a trapped surface. 

 \pagebreak
 \section{The SQM action and supersymmetry variations}
 
 \label{sqmA}
 The action of the quantum mechanics can be written as
\bea
\mathcal{L}&=& \int\!dx_0 \,d^2\theta\,d^2\bar{\theta} \, \mathrm{tr}  \left[  Z^{\dagger \, ba}
\,e^{2gV}\,Z^{ab}  + 2 \,  e^{-2gV}\,\Phi^{a \dagger}_{\bf i}
\,e^{2gV}\,\Phi^a_{\bf i}    + \xi \, V  \right]\nn \\
&+&\!\int\!dx_0\,d^2\theta\, \mathrm{tr} \left( \frac{1 }{8 g^2}  W^\alpha W_\alpha  +  {\cal W}(\Phi,Z) \right)  
+h.c.
\eea 
 In components
 
 \begin{equation}
\begin{aligned}
\Phi^a_{\bf i}(x,\theta,\bar{\theta}) & =\phi^a_{\bf i}(x)+ \sqrt{2} \theta \mu^a_{\bf i}(x)  + \theta^2 F^a_{\bf i}(x) \\
&+i\,\theta\sigma^\mu\bar{\theta}\, D_\mu\varphi^a_{\bf i}(x)- \frac{i}{\sqrt{2}} \theta ^2 \partial_\mu \chi^a_{\bf i}(x) \sigma^\mu \bar{\theta} +\frac{1}{4}
\theta^2\,\bar{\theta}^2\,\square \, \phi^a_{\bf i}(x)  \\
Z^{ab}(x,\theta,\bar{\theta}) & =z^{ab}(x)+ \sqrt{2} \theta \mu^{ab} (x) + \theta^2 F^{ab}(x)  \\
&+i\,\theta\sigma^\mu\bar{\theta}\,D_\mu z^{ab}(x)- \frac{i}{\sqrt{2}} \theta ^2 D_\mu \chi^{ab}(x) \sigma^\mu \bar{\theta}+\frac{1}{4}
\theta^2\,\bar{\theta}^2\,\square \, z^{ab}(x) \\
V(x,\theta,\bar{\theta})  & =-\theta\sigma^\mu\bar{\theta}\,x_\mu+i\,\theta^2\,\bar{\theta}\,\bar{\lambda}
-i\,\bar{\theta}^2\,\theta\,\lambda+\frac{1}{2}\theta^2\bar{\theta}^2 \,D       \\
W_\alpha (x,\theta) &=2g\Big(\!\!-i\,\lambda_\alpha-\frac{i}{2}
(\sigma^\mu\bar{\sigma}^\nu)_{\alpha}^{~\beta}\,\theta_\beta\,F_{\mu\nu} +\theta_\alpha D+
\theta^2\,\sigma^\mu_{\alpha\dot{\alpha}}\,D_\mu\bar{\lambda}^{\dot{\alpha}}\Big) \\
\end{aligned}
\end{equation}
with
\be 
D_{\mu}=\left(   \partial_{t}  + {\rm i}\,  A_t   , {\rm i}\, x_{i} \right)   \qquad  F_{t i}=D_t x_i  \qquad  F_{ij}={\rm i}\,  [ x_i , x_j ] 
\ee
 The action is invariant under the supersymmetry variations
$\delta= \sqrt{2} (\epsilon^{\alpha}\delta_\alpha+ \bar\epsilon_{\dot\alpha} \delta^{\dot \alpha})$
  with  
\bea
\delta_\alpha z^{(ab)}&=& \mu_\alpha^{(ab)}         \qquad \qquad\qquad~~~~~~~~~~~    \delta_{\dot \alpha} \bar z^{(ab)}= \bar\mu_{\dot \alpha}^{(ab)} \nn\\
\delta_\alpha \mu_{\beta}^{(ab)} &=&- {\partial \overline{\mathcal W} \over \partial \bar{z}^{(ba)} }  \epsilon_{\alpha\beta}\qquad \qquad  ~~~~~~~~~  \delta_{\dot \alpha} \bar\mu_{\dot \beta}^{(ba)}
= - {\partial {\mathcal W} \over \partial z^{(ab)} } \epsilon_{\dot \alpha\dot \beta} \nn\\
\delta_\alpha \bar\mu_{\dot\alpha}^{(ba)} &=& - i \sigma^0_{\alpha\dot\alpha} \dot{\bar z}^{(ba)} - \sigma^i_{\alpha\dot\alpha}\,  x_i^{(ba)}\, \bar z^{(ab)}  \qquad
\delta_{\dot\alpha} \mu_{\alpha}^{(ab)} =i \sigma^0_{\alpha\dot\alpha} \dot{z}^{(ab)} - \sigma^i_{\alpha\dot\alpha}\,  x_i^{(ab)}\,  z^{(ab)}   \label{susy1}
\eea

\bea 
\delta_\alpha \phi^{(a)}_{\bf{i}}&=& \chi_{\alpha\bf{i}}^{(a)}         \qquad \qquad\qquad~~~~~~~~~~~    \delta_{\dot \alpha} \bar \phi^{(a)}_{\bf{i}}= \bar\chi_{\dot \alpha \bf{i}}^{(a)} \nn\\
\delta_\alpha \chi_{\beta \bf{i}}^{(a)} &=&- {\partial \overline{\mathcal W} \over \partial \bar \phi^{(a)}_{\bf{i}} }  \epsilon_{\alpha\beta}\qquad \qquad  ~~~~~~~~~  \delta_{\dot \alpha} \bar\chi_{\dot \beta \bf{i}}^{(a)}
= - {\partial {\mathcal W} \over \partial \phi^{(a)}_{\bf{i}} } \epsilon_{\dot \alpha\dot \beta } \nn\\
\delta_\alpha \bar\chi_{\dot\alpha \bf{i}}^{(a)} &=& - i \sigma^0_{\alpha\dot\alpha} \dot{\bar \phi}_{\bf{i}}^{(a)} - \sigma^i_{\alpha\dot\alpha}\,  x_i^{(a)}\, \bar \phi^{(a)}_{\bf{i}}  \qquad
\delta_{\dot\alpha} \chi_{\alpha \bf{i}}^{(a)} =i \sigma^0_{\alpha\dot\alpha} \dot{\phi}_{\bf{i}}^{(a)} - \sigma^i_{\alpha\dot\alpha}\,  x_i^{(a)}\,  \phi^{(a)}_{\bf{i}}
\eea

\bea
\delta_\alpha x^{(a)}_{\mu}&=& -i \bar \lambda^{(a)}_{\dot \alpha}    (\bar{\sigma}^{\mu})^{\dot \alpha \beta}  \epsilon_{\alpha \beta}   \qquad \qquad\qquad~~~~~~~~~~~    \delta_{\dot \alpha} x^{(a)}_{\mu}= i    \epsilon_{\dot \alpha\dot \beta}   (\bar{\sigma}^{\mu})^{\dot  \beta \alpha}  \lambda^{(a)}_{\alpha} \nn\\
\delta_\alpha \lambda_{\beta}^{(a)} &=& (\sigma^{\mu \nu} F_{\mu \nu}^{(a)}  + i D^{(a)})_{\alpha}^{\, \, \,\gamma}\epsilon_{\gamma\beta}\qquad \qquad   \delta_{\dot \alpha} \bar\lambda_{\dot \beta}^{(a)}
= \epsilon_{\dot \alpha \dot \gamma}  (\bar \sigma^{\mu \nu} F_{\mu \nu}^{(a)}  - i D^{(a)})_{\, \, \, \dot \beta}^{\dot \gamma}    \label{susy3}
\eea

where $x_i^{(ab)} = x_{i}^{(a)}-x_{i}^{(b)}$.
%
%
%
%
%

  To compute the supercharges we need only the time derivatives 
 coming from the kinetic terms. Specifying to $Q_a=1$ the various contributions are
\bea
\frac{1}{8g^2}\bigg(\!\int  \,d^2\theta\,\mathrm{tr}\big(W^\alpha W_\alpha\big)+ \mbox{c.c.}\bigg)\! 
 &=&    \sum_{a=1}^4   \left[\frac{1}{2}  \sum_{i=1}^3 \left(\partial_t x_i^{(a)}\right)^2 + i \bar{\lambda}\, \bar{\sigma}_0 \,   \partial_t \, \lambda^{(a)}  \right]
  +\ldots \nn\\
   \int  \,d^2\theta\,d^2\bar{\theta} \, \mathrm{tr}  Z^{\dagger \, ba}
\,e^{2gV}\,Z^{ab} &=& \sum_{(ab)} \bigg[ \partial_t z^{(ab)} \partial_t \bar{z}^{(ba)}   + i \bar{\mu}^{(ba)} \bar{\sigma}^{0}  \partial_t \mu^{(ab)} \bigg] +\ldots\nn\\
 2 \,  \int  \,d^2\theta\,d^2\bar{\theta} \, \mathrm{tr}  \,  e^{-2gV}\,\Phi^{a \dagger}_{\bf i}
\,e^{2gV}\,\Phi^a_{\bf i}   &=& \sum_{a=1}^4 \sum_{\mathbf{i} = 1}^3  \left(  \partial_t \phi_{\mathbf{i}}^{(a)} \partial_t \bar{\phi}_{\mathbf{i}}^{(a)}     + i \bar{\chi}_{\mathbf{i}}^{(ba)} \bar{\sigma}_0   \partial_t\chi_{\mathbf{i}}^{(ab)} \right) +\ldots 
\eea

The whole lagrangian in field components is:

 \bea
\mathcal{L}=\mathcal{L}_X + \mathcal{L}_{\Phi}+ \mathcal{L}_{Z}  + \mathcal{L}_{\Phi Z} + \mathcal{L}_{X Z}
\eea

with

  \bea     
    \mathcal{L}_Z  =\sum_{(ab)} \bigg[ D_{t} z^{(ab)} D_{t} \bar{z}^{(ba)}  - F^{(ab)} \bar{F}^{(ba)}    + i \bar{\mu}_{\dot \alpha}^{(ba)} (\bar{\sigma}^{\mu})^{ \dot\alpha \alpha}  D_{t}\mu_\alpha^{(ab)} \bigg]
   \eea     
   
   \bea 
D_t z^{(ab)} = \dot{z}^{(ab)} + i z^{(ab)} A_{t}^{(ab)}
\eea


\bea
\mathcal{L}_X = \sum_{a=1}^4 \left[\frac{1}{2}  \sum_{i=1}^3 ( \partial_t x_i^{(a)})^2 + i \bar{\lambda}_{\dot \alpha} \bar{\sigma}_0^{ \dot\alpha \alpha} \partial_t \lambda^{(a)}_{ \alpha}- \frac{1}{2} D^{(a)}D^{(a)} - c^{(a)} D^{(a)} \right]
\eea

 \bea
\mathcal{L}_{\Phi} = \sum_{a=1}^4 \sum_{\mathbf{i} = 1}^3  \left(  \partial_t \phi_{\mathbf{i}}^{(a)} \partial_t \bar{\phi}_{\mathbf{i}}^{(a)}  - F^{(a)}_{\mathbf{i}} \bar{F}^{(a)}_{\mathbf{i}}    + i \bar{\chi}_{\dot \alpha, \mathbf{i}}^{(ba)} \bar{\sigma}_0^{ \dot\alpha \alpha}  \partial_t\chi_{\alpha,\mathbf{i}}^{(ab)} \right)
\eea
   
   \bea     
\mathcal{L}_{XZ}  = - \sum_{a,b} \bigg[   ((x_i^{(ab)})^2 + D^{(ab)})  \bar{ z}^{(ba)}z^{(ab)} +   x_i^{(ab)}  \bar{\mu}_{\dot \alpha}^{(ba)} (\bar{\sigma}^{i})^{ \dot\alpha \alpha}  \mu_\alpha^{(ab)} \nn \\
+i \sqrt{2}  \left(  \bar{ z}^{(ba)} \epsilon^{ \alpha \beta}\lambda_{\beta}^{(ab)} \mu^{(ab)}_{\alpha} -\epsilon^{\dot \beta \dot \alpha}\bar{ \mu}^{(ba)}_{\dot \beta} \lambda_{\dot \alpha}^{(ab)} z^{(ab)}\right) 	
    \bigg]
   \eea  
   
   where $D^{(ab)} = D^{(a)}-D^{(b)} $ and $\lambda_{\dot \alpha}^{(ab)} = \lambda_{\dot \alpha}^{(a)}-\lambda^{(b)}_{\dot \alpha}$.
   

 \bea
\mathcal{L}_{\Phi Z} =\sum_{a=1}^4 \left[ \frac{\partial W}{\partial z^{(ab)}} F^{(ab)} + \frac{\partial W}{\partial \phi^{(a)}_{\mathbf{i}} } F^{(a)}_{\mathbf{i}} + h.c. \right] + \nn \\
+ \frac{1}{2} \left[ \sum_{(ab)} \sum_{(cd)} \left( \frac{\partial^2 W}{\partial z^{(ab)}\partial z^{(cd)}}  \epsilon^{\beta \alpha} \mu_{\alpha}^{(ab)} \mu_{\beta}^{(ab)} \right) + 
 \sum_{a} \sum_{b} \left( \frac{\partial^2 W}{\partial \phi^{(a)}_\mathbf{i} \partial \phi^{(b)}_\mathbf{i}}  \epsilon^{\beta \alpha} \chi_{\alpha,\mathbf{i}}^{(ab)} \chi_{\beta,\mathbf{i}}^{(ab)} \right)+h.c. \right] 
\eea

 where we assumed that the fields do not depend on spatial coordinates and the stack of branes contain actually only a single brane.

 \pagebreak
\section{Bifundamental Vertex Operators}
 \label{bifVer}

We focus on bifundamental fields. The bosonic modes of massless open strings connecting two different D-branes are 
described by the following NS vertex operator in the $-1$ picture

\bea
   \label{eq:zvo14}
  z^{(ab)} : \qquad   V_z=z^{(ab)}  e^{-\varphi} \,\prod_{\mathbf{j}\not= \mathbf{j}_{ab}  }  \left( \Delta_{\mathbf{j} } \, e^{-\frac{i}{2} h_{\mathbf{j}}}\right)~,
  \eea
\bea
   \label{eq:zvo14b}
 \bar{ z}^{(ab)} : \qquad   V_{\bar{z}}=\bar{z}^{(ab)}  e^{-\varphi} \,\prod_{\mathbf{j}\not= \mathbf{j}_{ab}  }  \left( \Delta_{\mathbf{j} } \, e^{\frac{i}{2} h_{\mathbf{j}}}\right)~,
  \eea
with $\mathbf{j}_{ab}$ labelling the plane along which branes of type $a$ and $b$ are parallel,  $\Delta_{\mathbf{j}}$ is the bosonic $Z_2$ twist field in the $\mathbf{j}^{\rm th}$ plane in the $T^6$, and $z$ is the Chan-Paton matrix. Notice that the conjugate variable to ${z}^{(ab)}$ is $\bar{z}^{(ba)}$, while ${z}^{(ab)}$ and $\bar{z}^{(ab)}$ are completely independent. The fermionic states in the same multiplet correspond to the following R vertex operators 
\bea
  \label{eq:Vpsi14}
   \mu_\alpha^{(ab)} :\qquad  V_{\mu_\alpha}=\mu_{\alpha}^{(ab)} \,  e^{-\frac{\varphi}{2}}  \, S^{\alpha} \, e^{\frac{i}{2} h_{\mathbf{j}_{ab}  }} \,\prod_{\mathbf{j}\not=\mathbf{j}_{ab} } \Delta_{\mathbf{j}} \nn\\
   \bar\mu_{\dot \alpha}^{(ab)} :\qquad  V_{\mu_{\dot \alpha}}=\bar \mu_{\dot\alpha}^{(ab)} \,  e^{-\frac{\varphi}{2}}  \, S^{\dot\alpha} \, e^{-\frac{i}{2} h_{\mathbf{j}_{ab}  }} \,\prod_{\mathbf{j}\not=\mathbf{j}_{ab} } \Delta_{\mathbf{j}}
\eea

where $S^\alpha=(S^{++},S^{--})$ are the two chiral spin fields in $\mathbf{R}^{1,3}$. Finally the auxiliary field can be described be the following vertex operator in the zero picture
\bea
  \label{eq:VF14}
   F^{(ab )} :\qquad  V_F &=& F^{(ab)}  \, e^{i h_{\mathbf{j}_{ab}  }} \,\prod_{\mathbf{j}\not=\mathbf{j}_{ab} } \left(\Delta_{\mathbf{j}} \, e^{\frac{i}{2} h_{\mathbf{j}}} 
   \right)~ \nn\\
   \bar F^{(ab )} :\qquad  V_{\bar F} &=& \bar F^{(ab)}  \, e^{-i h_{\mathbf{j}_{ab}  }} \,\prod_{\mathbf{j}\not=\mathbf{j}_{ab} } \left(\Delta_{\mathbf{j}} \, e^{-\frac{i}{2} h_{\mathbf{j}}} 
   \right)~ 
\eea
 
 \pagebreak

\bibliography{references_D3Phd}
\bibliographystyle{abe}



\end{document}